\documentclass[11pt]{article}
\pdfoutput=1

\usepackage{setspace}
\usepackage[letterpaper,top=25mm,left=31mm,bottom=28mm,right=31mm]{geometry}
\usepackage{titlesec}
\titleformat*{\section}{\Large\bfseries}
\titleformat*{\subsection}{\large\bfseries}
\titleformat*{\subsubsection}{\bfseries}
\titleformat*{\paragraph}{\small\bfseries}

\usepackage{mathtools}
 \usepackage{tcolorbox}
    \tcbuselibrary{listings,theorems}
\newcommand\ebox[2]{\begin{equation}\label{#1}
\tcboxmath[colback=white,boxrule=.25mm]{\begin{split} #2\end{split}}
\end{equation}}

\usepackage{upgreek}
\def\doublestroke#1{\pdfliteral{1 Tr .2 w}#1\pdfliteral{0 Tr 0 w}}

\def\nua{{\doublestroke{\nu}}}

\usepackage{empheq}
\usepackage{microtype}
\usepackage{relsize}
\usepackage[dvipsnames]{xcolor}
\usepackage{cite}
\usepackage{enumitem}
\usepackage{ifpdf}
\usepackage{lmodern}
\usepackage[T1]{fontenc}
\usepackage[ansinew]{inputenc}
\usepackage{amsmath}
\usepackage{amsthm}
\usepackage{relsize}
\usepackage{amssymb}
\usepackage{comment}
\usepackage{tikz}
\usepackage{graphicx}
\usepackage{color}
\definecolor{darkblue}{cmyk}{0.9,0.9,0,0}
\definecolor{darkgreen}{rgb}{0,0.55,0}
\usepackage[colorlinks=true,linkcolor=darkblue,citecolor=darkblue,urlcolor=darkblue]{hyperref}
\usepackage{epsfig}
\usepackage{epstopdf}
\usepackage{graphicx}

\makeatletter
\long\def\@makecaption#1#2{
  \vskip\abovecaptionskip
  \sbox\@tempboxa{{\captionfonts #1: #2}}
  \ifdim \wd\@tempboxa >\hsize
    {\captionfonts #1: #2\par}
  \else
    \hbox to\hsize{\hfil\box\@tempboxa\hfil}
  \fi
  \vskip\belowcaptionskip}
\makeatother

\def\sfS{\mathsf{S}}

\definecolor{greencirc}{rgb}{0, 0.73, 0.5}
\def\thalf{{\tfrac12}}
\newcommand\result[2]{\noindent\textbf{#1}~\textit{#2}}
\def\enumcom{\begin{enumerate}[label*=  \sf (\roman*),wide, labelwidth=!, labelindent=0pt]}

\def\cc{\text{(c.c)}}
\def\cV{\mathcal{V}}
\def\sl{SL(2,\ZZ)}
\def\ni{\noindent}
\newcommand\supr[1]{\underset{#1}{\text{sup}}}

\def\vol{{\rm vol}}
\def\HH{\mathbb{H}}
\def\CC{\mathbb{C}}
\def\ZZ{\mathbb{Z}}

\def\RR{\mathbb{R}}
\def\QQ{\mathbb{Q}}
\def\qz{\mathrm{Q}_\cZ}
\def\Q{\mathrm{Q}}
\def\lz{L_{\cZ}}
\def\zz{\z_\cZ}
\def\Lamz{\L_{\cZ}}

\def\cS{\mathcal{S}}
\def\max{{\rm max}}

\def\ab{\overline\alpha}
\def\Pb{\bar P}

\def\vareps{\varepsilon}
\def\rhob{\bar\rho}
\def\E{\mathcal{E}}

\def\sfa{\mathsf{a}}
\def\sfb{\mathsf{b}}

\def\RMT{{\rm RMT}}

\def\cF{\mathcal{F}}

\def\L{\Lambda}
\def\half{{1\o2}}

\def\R{\mathbb{R}}
\def\qb{\overline{q}}
\def\S{\mathcal{S}}
\def\Re{\text{Re}}
\def\Im{\text{Im}}

\def\x{\times}

\DeclareMathOperator*{\Res}{Res}
\def\cD{\mathcal{D}}

\def\eps{\epsilon}

\def\bul{$\bullet$~}

\def\hb{\overline h}
\def\Z{\mathbb{Z}}

\def\1{{\rm 1-loop}}

\def\cZ{\mathcal{Z}}
\def\h{{1\o2}}

\def\c{\cite}

\def\cA{\mathcal{A}}

\def\cM{\mathcal{M}}

\def\zb{\bar z}

\def\cA{\mathcal{A}}

\def\c{\cite}

\def\vs{\vskip .1 in}

\def\G{\Gamma}
\def\p{\partial}
\def\o{\over}
\def\g{\gamma}
\def\D{\Delta}
\def\rar{\rightarrow}

\def\eqr{\eqref}

\def\cO{{\cal O}}

\def\i{\infty}

\def\ssec{\subsection}
\def\sssec{\subsubsection}
\def\sec{\section}

\def\foot{\footnote}
\newcommand{\es}[2] {\begin{equation} \label{#1} \begin{split} #2 \end{split} \end{equation}}
\newcommand{\e}[2] {\begin{equation} \label{#1} #2 \end{equation}}

\newcommand{\beq}{\begin{equation}}
\newcommand{\eeq}{\end{equation}}
\newcommand{\beqy} {\begin{eqnarray}}
\newcommand{\eeqy} {\end{eqnarray}}
\newcommand{\bsmat}{\begin{smallmatrix}}
\newcommand{\esmat}{\end{smallmatrix}}
\newcommand{\bmat}{\begin{matrix}}
\newcommand{\emat}{\end{matrix}}

\def\({\left(}
\def\){\right)}
\def\[{\left[}
\def\]{\right]}

\def\<{\langle}
\def\>{\rangle}

\def\a{\alpha}
\def\b{\beta}
\def\g{\gamma}
\def\G{\Gamma}
\def\d{\delta}
\def\D{\Delta}
\def\z{\zeta}

\def\k{\kappa}
\def\l{\lambda}

\def\t{\tau}
\def\s{\sigma}
\def\vs{\vskip .1 in}

\def\fb{\bar{f}}

\usepackage{varioref}
\usepackage{makeidx}
\makeindex

\usepackage[english]{babel}
\usepackage{caption}
\usepackage{subfig}

\numberwithin{equation}{section}

\usepackage{tabularx}
 
\begin{document}

\begin{spacing}{1.15}

\begin{titlepage}

\vspace*{2cm}
\begin{center}
{\Large \bfseries An $L$-function Approach to\\ \vs Two-Dimensional Conformal Field Theory}

\vspace*{1cm}

Eric Perlmutter

\vspace*{2mm}

\textit{\small Institut de Physique Th\'eorique, Universit\'e Paris-Saclay, CEA, CNRS\\Orme des Merisiers, 91191, Gif-sur-Yvette Cedex, France\\
\vskip .2 in
Institut des Hautes \'Etudes Scientifiques\\ 
35 Route de Chartres, 91440, Bures-sur-Yvette, France}

\vspace{2mm}

{\tt \small perl@ipht.fr}

\vspace*{0.5cm}


\end{center}
\begin{abstract}

We introduce a framework for two-dimensional conformal field theory (CFT) in the language of analytic number theory. Attached to the torus partition function of every two-dimensional CFT is a self-dual, degree-4 $L$-function of root number $\varepsilon=1$, with a universal gamma factor determined by $SL(2,\mathbb{Z})$ and local conformal invariance. Due to the richness of CFT operator spectra, these are not, in general, standard $L$-functions. We explicate their analytic structure, exploring the interplay of the Hadamard product over non-trivial zeros with the generalized Dirichlet series over CFT scalar primary conformal dimensions. We derive a zero sum rule in terms of the spectrum, and a global zero density bound in terms of the spectral gap. Convergence of the series representation implies square root cancellation of the degeneracies; we relate this to random matrix behavior of high-energy level spacings. Random matrix universality of the CFT implies ``Riemann zeta universality'' of the $L$-function: an average relation between the $L$-function on the critical line and Riemann zeta on the 1-line. This in turn yields a subconvexity bound. For a compact free boson, the $L$-function is a product of Riemann zeta functions times an analytic factor. Extensions to correlator $L$-functions and spinning spectra are briefly discussed. 

In the course of this work, we are led to sharpen the notion of random matrix universality in two-dimensional CFTs. We formulate a precise version of the following standalone conjecture, logically independent of $L$-functions: in unitary, compact Virasoro CFTs with central charge $c>1$, fixed-spin primary spectra at high energy are asymptotically simple, with random extreme gap statistics.

\end{abstract}

\end{titlepage}
\end{spacing}

\pagenumbering{roman}
\begin{spacing}{0.7}

\setcounter{tocdepth}{2}
\tableofcontents
\end{spacing}

\newpage

\pagenumbering{arabic}
\setcounter{page}{1}

\begin{spacing}{1.15}

\sec{Introduction}

This work proposes a novel language for two-dimensional conformal field theories (CFTs) and their gravity duals, namely, the language of analytic number theory, and specifically of $L$-functions. 

One motivation (hardly original) is the resolution of black hole microstate spectra in quantum gravity in anti-de Sitter (AdS) spacetime, dual to high-energy spectra in CFT\thinspace---\thinspace a central open problem since the advent of AdS/CFT. On the CFT side, rigorous bootstrap-type computations at finite central charge $c$ have addressed two main categories of related questions. The first are questions of asymptotics, in which universal behaviors are derived at the cost of washing out microscopics via coarse-graining. The second are questions of the spectral edge, which are agnostic about the bulk of the spectrum altogether. There are also bootstrap methods in an expansion around large central charge, dual to semiclassical AdS gravity, where quantization is obstructed by a wall of complicated non-perturbative effects in $1/c$. Having evolved from what one might call the ``Cardy Era'' (asymptotics) through the ``Tauberian Era'' (asymptotics with rigorous spectral averaging) and now into the ``Random Era'' (asymptotic statistical correlations), we appear to be moving closer to a ``Resolution Era'', in which fine structure of high-energy spectra may be genuinely resolved\thinspace---\thinspace sans statistics or averaging.  

To make the evolutionary leap, it seems that we should answer the following conceptual question, which remains mysterious: in what sense do bootstrap principles (broadly defined) constrain high-energy spectra microscopically, or at least their all-point autocorrelations? This framing in turn suggests a key objective: {\it to find and solve constraints that act on high-energy states alone.}

A recent surge of remarkable progress on the three-dimensional quantum gravitational path integral and its boundary interpretation makes AdS$_3$/CFT$_2$ an obviously fertile setting in which to pursue these ideas. The discovery of JT/RMT duality \cite{SaadShenkerStanford2019} led directly to certain avatars of ensemble constructions in one dimension higher \cite{ChandraCollierHartmanMaloney2022} and, though rather more subtle, insights into random dynamics of BTZ black hole states \cite{CotlerJensen2021,Maxfield:2020ale}. But still here, the question of resolution has met significant challenges: in order to probe spectral fine structure even statistically, one apparently must include gravitational contributions that cannot be understood according to known rules of semiclassical path integration. Current developments in AdS$_3$/CFT$_2$ involve several complementary physical perspectives in response to these challenges, notable for the diversity of formal mathematical approaches involved: topological \cite{CollierEberhardtZhang2023,CollierEberhardtZhang2024,deBoer:2025rct,Hartman2025CTV,BaoHungJiangLiu2024}, geometric \cite{Maldacena:2004rf,ChandraCollierHartmanMaloney2022,Abajian:2023bqv,Hartman2025Triangulation}, statistical \cite{BelinDeBoerJafferisNayakSonner2024,Jafferis:2025vyp,Jafferis:2025yxt,Pelliconi:2024aqw,DeBoerLiskaPostSasieta2024}, spectral-automorphic \cite{2307,Haehl:2023tkr,HaehlReevesRozali2023JHEP,HaehlReevesRozali2023PRD,2503}, and conformal representation-theoretic \c{ChandraCollierHartmanMaloney2022,deBoer:2024mqg,Post:2024itb}.

However, a number-theoretic approach to AdS$_3$/CFT$_2$ is conspicuously absent, and it is natural to ask what role one might play in the resolution of microstates.

It is, from a broad point of view on physical mathematics, a bit surprising that analytic number theory has not yet made deeper inroads to quantum gravity and CFT. To be clear, there is a long and decorated history of number-theoretic \textit{objects} appearing within these subjects, most often in the setting of BPS-protected quantities, rational CFTs, vertex operator algebras, or other rigid (and computable) formal structures \c{Bah:2022wot}. What we are seeking is something which applies more fundamentally and universally, to the \textit{architecture} of generic CFTs without extra symmetries at all\thinspace---\thinspace a connection between the intricacy expected of operator spectra in field theory (black hole spectra in gravity), and the famously intricate properties seen in the well-studied world of $L$-functions. See \c{WittenOoguri2015} for some motivational remarks.

One reason for optimism in this regard is that the richness of $L$-functions appears to transcend symmetry: while many $L$-functions can ultimately be related to automorphic forms, they exhibit profound, seemingly magical structures that are not fixed by symmetry in any known way.

Moreover, these structures are often of a ``physical'' nature: as we hope to convey below, many problems pertaining to $L$-functions have a flavor that resonates with physics intuition. 

More pragmatically, $L$-functions give rise to useful, semi-soluble toy models that mimic various challenging problems of interacting physical systems. Their solutions often even use physics techniques themselves: a shining example is the demonstration of quantum chaotic dynamics and random spectral statistics of non-trivial Riemann zeta zeros \c{Montgomery1973,Odlyzko1987,BerryKeating1999}. Combined with the interconnectedness of $L$-functions with fields across mathematics, this makes it plausible that, beyond providing mere analogies, the analytic language of $L$-functions could be co-opted to address physics problems \textit{directly}\thinspace---\thinspace including, in particular, the resolution of microscopics in quantum gravity and CFT. (See \c{Hartnoll:2025hly,DeClerck:2025mem} for a recent step in this direction.)

Driven by that belief, we develop a number-theoretic framework for general two-dimensional CFTs, leaving the holographic translation to three-dimensional gravity language mostly implicit. In short, to every torus partition function, one can attach an $L$-function with universal transformation properties. These CFT $L$-functions are not ``standard'' $L$-functions, neither in the technical nor the colloquial sense of the word; but their complexity is precisely what is needed to reflect the chaotic dynamics of irrational CFT spectra, just as the $SL(2,\ZZ)$-invariant partition functions of such theories are (highly) non-standard modular functions indeed. On the other hand, in perhaps the simplest compact two-dimensional CFT\thinspace---\thinspace a single free boson on a circle\thinspace---\thinspace the attached $L$-function is the product of two Riemann zeta functions. This provides an encouraging sign that the space and properties of two-dimensional CFTs may be effectively stratified through the lens of $L$-functions.

In the long term, our hope is that this framework inspires new strategies for old questions about two-dimensional CFTs and AdS$_3$ quantum gravity. Indeed, as completed $L$-functions are constructible from their non-trivial zeros, one can imagine a possible future (audacious as it may be) in which AdS$_3$ black hole \textit{spectra} are regarded as secondary, their reconstruction from the black hole zeros left as an exercise for the reader. 

\sssec*{Summary}
~~~\,In {\bf Section \ref{s2}} we provide a friendly introduction to $L$-functions, squarely intended for physicists. We present some rudimentary formal properties, describe the large scale structure of the space of $L$-functions, and summarize a selection of important research milestones which seem synergistic for the physics questions at hand. We point the reader to expert mathematical reviews and references (of which we give many).

In {\bf Section \ref{s3}}, we introduce the construction of $L$-functions for two-dimensional CFTs. The $L$-function may be defined by a Rankin-Selberg modular integral, placing it within an established formal setting of automorphic $L$-functions.\foot{An essential difference with $L$-functions constructed from classical modular forms (e.g. Maass or holomorphic cusp forms) is that a given Fourier mode of a CFT partition function contains an infinite set of data\thinspace---\thinspace the local operator dimensions of fixed spin\thinspace---\thinspace whereas a given mode of a classical modular form is characterized by a single number.} Given a CFT with (two copies of) an extended chiral algebra $\mathcal{A}$, the $L$-function is attached to the $\mathcal{A}$-primary torus partition function\thinspace---\thinspace more precisely, a regularized version thereof, $\cZ(\t)$, which removes low-energy states while preserving modularity \c{2107} and (if present) discreteness of the spectrum. We call the $L$-function $\lz(s)$, and its completion $\Lamz(s)$. We focus for most of the paper on \textit{Virasoro CFTs}, defined as those CFTs without extra currents, with easy generalization to arbitrary $\mathcal{A}$ (see Section \ref{s101}). We also introduce, in Section \ref{s33}, an ``$L$-quotient'' $\qz(s)$, defined as a certain ratio of $\lz(s)$ with Riemann zeta functions that removes universal symmetry information of the $L$-function. At first this seems innocuous, but we will notice some intriguing features of $\qz(s)$ throughout the paper. 

\begin{figure}[t]
\centering
{
\subfloat{\includegraphics[scale=.9]{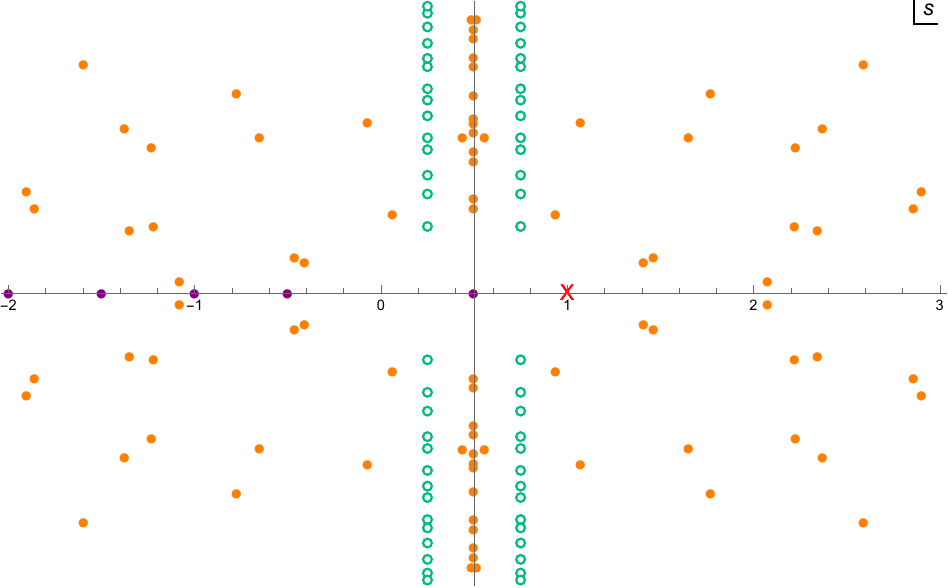}}
}
\caption{The poles and zeros of $\lz(s)$. The ${\color{red} \x}$ is a simple pole at $s=1$. The ${\color{violet} \bullet}$ are trivial zeros and the central zero. The ${\color{orange} \bullet}$ are non-trivial zeros, coming in quartets $\{\rho,\rhob, 1-\rho,1-\rhob\}$, infinite in number but with locations \`a priori unknown; their density and locations are constrained by the zero density bound \eqr{gdb} and sum rules \eqr{BLreln2} and \eqr{sumrule3}. The ${\color{greencirc} \boldsymbol\circ}$ are zeros at the locations of the non-trivial zeros of $\z(2s-1)$ and $\z(2s)$, whose status is somewhat different from the others; see the main text.}
\label{fig:poleszeros}
\end{figure}

{\bf Sections \ref{s4} and \ref{s5}} are devoted to presenting the analytic structure of $\lz(s)$\thinspace---\thinspace see Figure \ref{fig:poleszeros}\thinspace---\thinspace and to developing two distinct representations of $\lz(s)$. The first is the Weierstrass factorization over zeros and poles. The second is the series representation, in this case a generalized Dirichlet series over the scalar primary dimensions of $\cZ(\t)$, all of which all lie above the ``heavy'' threshold $\D={c-1\o 12}$ for Virasoro CFTs (dual to the semiclassical black hole threshold in AdS$_3$ gravity). We initiate an exploration of the interplay of these two representations. This yields two sum rules for the zeros (whose locations are \`a priori unknown): one in terms of a modular integral, and one directly in terms of the CFT spectrum. We also derive what we call a ``global zero density bound'', a constraint on the total number of non-trivial zeros of $\lz(s)$ in terms of the lightest scalar operator of $\cZ(\t)$.

In our construction, unlike the Hilbert-P\'olya picture, it is the energies of the CFT which are the frequencies of the series, not the zeros. The latter are instead ``dual'' to the energies in the canonically conjugate sense. Cardy asymptotics of CFT operator spectra make the question of series convergence more nuanced; we carefully present the mathematical conditions for convergence in Section \ref{s53}, and a physical interpretation for them in terms of random matrix behavior\thinspace---\thinspace specifically, the random statistics of \textit{extreme gaps} \c{BAB}\thinspace---\thinspace in Section \ref{s54}. We hope this connection will provide a useful lever for future bootstrap studies of generic CFTs, as it has so far not been understood how to inject the presence of chaos or randomness into bootstrap computations.

In {\bf Section \ref{s6}}, we step back and interpret our results from the point of view of UV-IR relations, AdS$_3$ effective field theory and the swampland. We explain a quantitative sense in which the low-energy data nearly fix the high-energy data by crossing: after subtracting these classical contributions, the quantum fluctuations that remain exhibit square root cancellation. Conversely, the existence of a landscape of UV finite AdS$_3$ quantum gravity theories predicts the existence of a landscape of $L$-functions obeying the stated formal properties of $\lz(s)$.

In {\bf Section \ref{s8}}, we return to formalism, deriving an approximate functional equation for $\lz(s)$. We first explain what this is, and review its derivation for standard $L$-functions. We then derive an approximate functional equation for $\lz(s)$ from scratch. This leads to our best, manifestly convergent form of the aforementioned zero sum rule in \eqr{sumrule3}. 

In {\bf Section \ref{s9}}, we ask what special properties of the $L$-function are implied by randomness in the CFT spectrum. The answer is appealing: it controls the behavior of $\lz(s)$ high on the \textit{critical line}, $s=\half+it$. We observe a phenomenon that we call \textit{Riemann zeta universality}\thinspace---\thinspace see Figure \ref{fig:zetauniv}\thinspace---\thinspace which follows from imposing a linear ramp, a hallmark of long-range spectral rigidity, on the (scalar) spectral form factor of the CFT. From this we extract a subconvexity bound (another concept that we will review) for $\lz(\half+it)$ in terms of its behavior on the 1-line ($s=1+it$), and set up the computation of its moments.

In {\bf Section \ref{s10}}, we present some generalizations and explicit examples: we define $L$-functions for CFTs with arbitrary extended chiral algebras $\mathcal{A}$; we retreat from the irrational CFT setting to the world of $c$ free bosons compactified on tori, i.e. Narain CFTs, giving explicit details and examples of their $L$-functions using results of \c{2107}; and we extend the construction to correlator $L$-functions, noting that any modular-covariant torus observable admits an $L$-function.

In {\bf Section \ref{s11}}, we make some initial remarks about the extension of the $L$-function construction to the spinning primary sectors of CFTs. We do not attempt a complete analysis, but part of the answer is inherited from $\lz(s)$. We observe a nifty simplification that happens only for the Virasoro case. This appears to have the intriguing mathematical implication, to be investigated elsewhere, that the set of non-trivial Riemann zeta zeros and the set of parity-even $\sl$ Maass cusp form eigenvalues are not mutually independent.

In {\bf Section \ref{s12}}, we outline conceptual directions for future work. 

Appendices provide mathematical background and proofs of results from the main text. 

Mathematicians may have noticed that we skipped {\bf Section \ref{s7}} in our summary. We did so to emphasize that Section \ref{s7} stands logically apart from the rest of the paper, with no derivations nor any direct relation to $L$-functions, and may be read independently. Its presence originates from Section \ref{s54}, where our analysis turns out to graze the question of what random matrix universality in two-dimensional CFTs should mean. This is a rather open question, moreso than in ordinary non-continuum quantum systems. We take the opportunity, in Section \ref{s7}, to frame this question with some precision, after which we formulate two-dimensional CFT versions of random extreme gap statistics. This leads, in Section \ref{s71}, to the formulation and discussion of the conjecture mentioned in the Abstract, which we dub the Simple Extreme Conjecture.

\sssec*{Notation and nomenclature}
We write our modular parameter as $\t=x+iy$ with $x,y\in\RR$. We use $\cF$ to denote the fundamental domain for the action of $\sl$ on $\mathbb{H}$, not including the cusp at $i\i$. Unless otherwise noted, a function like $\cZ(\t)$ depends non-holomorphically on $\t$. We write Petersson inner products with the $\sl$ non-holomorphic Eisenstein series, $E_s(\t)$, as
\e{}{(\cZ,E_{1-s}) := \int_\cF{dx dy\o y^2} \cZ(\t) E_s(\t)}
This notation arises from $\sl$ spectral decomposition, in which $s=\half+it$ and the inner product complex conjugates the second entry, but as is conventional we will continue to use this notation for all $s\in\CC$. In general, we write $s=\s+it$ with $\s,t\in\R$.

We write two-dimensional CFT torus partition functions as sums over discrete spectra of local operators, as in \eqr{Zdef}. Discreteness is not necessary for our construction, but we will use that language because compact CFTs are ultimately our main interest. One could generalize what follows to accommodate continua, trading sums for integrals and degeneracies for densities of states\thinspace---\thinspace and on the $L$-function side, introducing integrals over frequencies (in the series representation) and continuous measures over non-trivial zeros (in the Weierstrass representation)\thinspace---\thinspace but to avoid being overly pedantic, we just present the formulas once using the language of discreteness.

We use the term ``standard $L$-function'' in a generalized, colloquial sense, not just to refer (as mathematicians do) to automorphic $L$-functions attached to standard representations of $GL(n)$, but also to the $L$-functions most explicitly studied by mathematicians, obeying some set of nice properties. In particular, our use of ``standard $L$-functions'' includes Selberg class $L$-functions.

We use $f(T) = O(T)$ to mean $|f| \leq C T$ for some constant $C>0$, where dependence on other parameters is (usually) suppressed. We often use $\eps$ to denote a small positive constant, not necessarily the same at each occurrence. We use $\approx$ to indicate an asymptotic including the coefficient ($f\approx g$ means $\lim f/g = 1$), and $\sim $ to indicate asymptotic scaling alone ($f\sim g$ means $\lim f/g = O(1)$).

We use the following shorthand for integrals parallel to the imaginary axis:
\e{}{\int\limits_{~(\b)} := \int_{\b-i\i}^{\b+i\i}}
We use the following shorthands for sums:
\e{}{\sum_\l := \sum_{\l\,\in\,\{\l_n\}}\,,\qquad \sum_n := \sum_{n\,\in\,\ZZ_+}}
In general, indices $n$ run over positive integers unless otherwise noted. For generalized Dirichlet series, we pass back and forth between writing
 \e{}{\sum_n {a_n\o \l_n^{s-\half}}\qquad\text{and}\qquad \sum_{\l} {a_\l\o \l^{s-\half}}}
depending on context, where $a_n := a_{\l_n}$ is understood.

We use $\rho$ to denote the non-trivial zeros of $\lz(s)$, and occasionally those of a general $L$-function, with clarity from context. We use the shorthands
\e{}{\prod_\rho := \prod_{\rho\,\in\,\sfS_\rho(\lz)}\,,\qquad \sum_\rho := \sum_{\rho\,\in\,\sfS_\rho(\lz)}}
where $\sfS_\rho(\lz)$ is the non-trivial zero set of $\lz(s)$, 
\e{}{\sfS_\rho(\lz) := \{\rho:\Lamz(\rho)=0\}}
At the same time, we will use the standard notation $\rho_n$ for non-trivial Riemann zeta zeros, defined by $\z(\rho_n)=0$ with $\Im(\rho_n)\neq 0$; if zeros of $\lz(s)$ and $\z(s)$ ever appear in the same expression, we take care to disambiguate, for example by referring to $\sfS_\rho(\lz)$ explicitly. 

We do not assume the Riemann Hypothesis.

\sec{A breezy primer on $L$-functions}\label{s2}
This section is intended as a gentle introduction to the subject of $L$-functions from a modern high-energy theorist's perspective. Sections \ref{s21} lays groundwork for the remainder of the paper, whereas Sections \ref{s22} and \ref{s23} serve mainly to provide wider context. The reader is strongly encouraged to consult Chapter 5 of the textbook by Iwaniec and Kowalski \c{IKtext}, a standard reference in the field, for an eminently readable mathematical introduction. We also recommend \c{Davenport2000, Iwaniec1997} for their treatments of classical $L$-functions; \c{Titchmarsh1986} for a clear and comprehensive text on the Riemann zeta function; the more concise Riemann zeta text \c{Patterson1988}; and \c{Titchmarsh1939,Goldfeld2006} for related material. 

\ssec{What are $L$-functions?}\label{s21}

Perhaps surprisingly (and intriguingly), the obvious first question does not have a simple answer. Well-studied, axiomatic subclasses of $L$-functions possess extra structure that is not indicative of the general case, but admit proof of certain features that may otherwise remain conjectural. Rather than immediately diving into taxonomy, we present some of the core features that $L$-functions tend to have and return to the landscape of $L$-functions afterwards.\foot{The use of ``landscape'' here is intentional: physicists might find useful a faithful analogy between the space of $L$-functions and our state of knowledge thereof, and the landscape of (say) consistent quantum gravitational vacua. In this analogy, standard/Selberg class $L$-functions lie beneath the lamppost.} 

Let us say that an $L$-function is a function which admits a formal generalized Dirichlet series representation that meromorphically continues to the entire complex plane under a specific type of functional equation, perhaps with other arithmetic properties. $L$-functions are often denoted $L_f(s)$, where $s\in\CC$ and the subscript indicates some auxiliary object, often an automorphic form or group representation, to which $L_f(s)$ is ``attached''. A generalized Dirichlet series takes the form
\e{}{L_f(s) = \sum_n {a_f(n)\o \l_n^s}\,,\quad a_f(n) \in \mathbb{C}}
where $\{\l_n\}\subseteq\RR_+$ is a unbounded set. In the simplest, most familiar $L$-functions, the series is an ordinary Dirichlet series, for which $\l_n=n$, which converges absolutely in some right-half-plane $\s > \s_a$. The functional equation is best understood in terms of a \textit{completed $L$-function}, $\L_f(s)$, constructed as follows:
\e{}{\L_f(s) := q^{s/2} \g_f(s) L_f(s)\,.}
$q\in\ZZ_+$ is the \textit{conductor}, and $\g_f(s)$ is the \textit{gamma factor}, of the specific form
\e{}{\g_f(s) := \pi^{-{ds/2}}\prod_{i=1}^d \G\({s+\kappa_i\o 2}\)\,.}
The number $d \in \ZZ_+$ is the \textit{degree} of the $L$-function. The parameters $\kappa_i\in\CC$ are known as \textit{spectral parameters} (or \textit{Archimedean parameters}) of the $L$-function. With this form, the functional equation is
\e{}{\L_f(s) = \vareps \L_{\bar f}(1-s)\,.}
The phase $\vareps$ is called the \textit{root number}. In general, the functional equation relates $L_f(s)$ to its ``dual'' $L$-function, $L_{\bar f}(s)$; in the series representation, this corresponds to complex conjugation of the $a_f(n)$. When the functional equation relates $L_f(s)$ to itself, the $L$-function is called \textit{self-dual}, and $\vareps=\pm 1$. We have used the \textit{analytic normalization} here, in which $\s=\half$ is the axis of reflection. Typically, $\L_f(s)$ is meromorphic with poles only at $s=0,1$, and the ``trivial zeros'' of $L_f(s)$ are the poles of $\g_f(s)$.

$L$-functions may also be endowed with an Euler product,
\e{}{L_f(s) = \prod_{i=1}^d\prod_{p} (1-\a_i(p) p^{-s})^{-1}\,,\quad \a_i(p)\in\CC}
which also converges in a right half-plane. The $\a_i(p)$ are known as \textit{local roots}. The product runs over some set of primitives $p$\thinspace---\thinspace often, but not necessarily, the primes. The presence of an Euler product is intimately tied with a multiplicative, Hecke algebraic structure for the series coefficients $a_f(n)$; this is guaranteed to exist if $f(\t)$ is a Hecke eigenform. These coefficients are conjectured to obey various fascinating bounds on their magnitude (e.g. Ramanujan-Petersson) and distribution (e.g. Sato-Tate).

The numbers $(d,q,\{\kappa_i\},\vareps)$ are fundamental data that partially characterize the $L$-function (known as the \textit{Selberg data} in the context of the Selberg class of $L$-functions). We emphasize the word ``partially'': distinct $L$-functions can and do possess identical Selberg data. These data may be combined to form certain useful objects that appear in $L$-function phenomenology. One is the so-called \textit{analytic conductor,}  
\e{aconddef}{\mathfrak{q}(s) := q\prod_{i=1}^d (|s+\kappa_i|+3)\,.}
This controls, in a unified manner, the growth of various quantities as some of these data become large in various \textit{aspects}: e.g. the $q$-aspect (growth within a family of $L$-functions indexed by $q$) or $t$-aspect (growth on vertical lines $s=\s+it)$.\foot{In certain applications to e.g. subconvexity bounds, one often sees assumptions of boundedness on the ratios $\kappa_i/\kappa_j$, such that no parameter can become too large or small. These are known as \textit{non-conductor-dropping conditions}.} 

Needless to say, the most famous $L$-function is the Riemann zeta function. Its series representation
\e{}{\z(s) = \sum_{n=1}^\i {1\o n^s}\,,}
absolutely convergent for $\s>1$, analytically continues to $s\in\CC$ via a functional equation
\e{Lamdef}{\L\({s\o2}\) := \pi^{-s/2}\G\({s\o2}\) \z(s) = \L\({1-s\o 2}\)\,.}
So $\z(s)$ is a self-dual, degree-1 $L$-function with root number $\vareps=1$, $q=1$ and $\kappa_1=0$. 

\ssec{Where are $L$-functions?}\label{s22}

We now turn to the $L$-function landscape. What is known about the space of $L$-functions is highly incomplete, even at the level of definitions. One must emphasize the \textit{generality} of the notion of an $L$-function. Axiomatization allows for definite results, but with the downside risk of viewing such results as overly representative of the general case. The most common class is the \textit{Selberg class}, obeying several simplifying assumptions: an Euler product; entirety of $(s-1)^m L_f(s)$ for some $m\in\Z_{\geq 0}$; and an absolutely convergent ordinary Dirichlet series expansion, with coefficients obeying Ramanujan-type, sub-polynomial bounds $|a_f(n)| = O(n^\eps)$ for every $\eps>0$. This should be viewed as a limited working proxy for a definition of $L$-functions: there are myriad examples of interest that fall outside of this class, including both those that have not yet been proven to satisfy the axioms and those that are known definitively to violate them. The situation is well-summarized in the opening pages of the Iwaniec-Kowalski chapter on $L$-functions (p.96 of \c{IKtext}):\foot{For another more recent view on the topic of axiomatization and related issues, see \c{Booker}.}

\begin{quotation}
\textit{It is important to note that many interesting L-series exist beyond our context,
either because of lack of strength in our fingers to prove the required assumptions
(for instance Artin L-functions for non-trivial irreducible characters are not known
to be entire in general, and general L-functions of varieties are not even known to
be meromorphic...) or because some of the conditions
fail. For instance, in the second class come Dirichlet L-functions of non-primitive
characters (they have Euler product but the functional equation has extra factors),
L-functions of general modular forms... which have
no Euler products and a functional equation relating $L_f(s)$ with some function not
directly related to $L_{\bar f}(s)$.}
\end{quotation}

\ni Indeed, for physicists the ``general modular forms'' category is especially germane and hardly exotic in our day-to-day lives; $L$-functions attached to cusp forms for noncongruence subgroups of $SL(2,\ZZ)$, which lack a Hecke theory and hence an Euler product, furnish a good example.  We also note the $m$-fold symmetric power $L$-functions, whose analyticity/automorphy have not been proven in general.\foot{Symmetric square $L$-functions for $GL(2)$ modular forms are known to be automorphic for $GL(3)$ \c{GelbartJacquet1978}\thinspace---\thinspace this is the Gelbart-Jacquet lift to $GL(3)$, well-reviewed in \c{Goldfeld2006}\thinspace---\thinspace but the general proof of automorphy of $L_{\text{Sym}^m f}(s)$ for $GL(m+1)$ is still lacking.} 

Not wishing to let the pendulum swing too far in either direction, we also emphasize that a hugely influential set of results have come from study of Selberg class $L$-functions, and ``standard'' $L$-functions, $L_\pi(s)$. These are $L$-functions attached to automorphic representations $\pi$ of $GL(n)$. The word ``standard'' has both narrow and broad usages: in the former (most common), $\pi$ is the defining (standard) representation, while in the latter, $\pi$ is any finite-dimensional representation.  It is in this setting that many rigorous proofs are made. The famous conjectures of Langlands posit that ``all $L$-functions'' are standard $L$-functions in the broad sense\thinspace---\thinspace the need for the quotation marks being universally understood. We will elaborate upon properties of standard $L$-functions at various points throughout the paper. The reader is encouraged to peruse \href{http://lmfdb.org}{lmfdb.org} for a user-friendly catalog of known $L$-functions. 

\ssec{Why (are) $L$-functions?}\label{s23} 

So, why should a high-energy theorist care about $L$-functions? The short answer (in our view) is that the study of $L$-functions has a ``physical'' flavor, resonant with modern perspectives in theoretical physics, that suggests possible deeper connections. To give a sense of this, we describe a selection of major areas of current research on $L$-functions. It would be pointless to aim for completeness; our goal is to whet the appetite in under two pages, and to give some references. 

Much of the action for standard $L$-functions takes place within the critical strip $\s\in[0,1]$, and especially on the critical line $s=\half+it$. The questions here pertain to the value distribution of $L$-functions. Standard (primitive) $L$-functions are believed to obey the \textit{(Generalized) Riemann Hypothesis (RH)}: all non-trivial zeros have $\s=\half$.\foot{See \c{conrey03-rh,conrey19-rh} for two useful overviews (among many) of varying degrees of formality.} This in turn implies two logically independent conjectures of note. The first is the \textit{(Generalized) Lindel\"of Hypothesis}: the suprema of the $L$-function grow slower than any power of $|t|$ as $|t|\rar\i$. The second is the \textit{(Generalized) Density Hypothesis}: within the rectangle $\s\in[\half,1-\a]$ and $|t|\leq T$, the cardinality of zeros $N(\a,T)$ obeys $N(\a,T)= O(T^{2(1-\a)+\eps})$ for every $\eps>0$. The latter two conjectures are strictly weaker than RH. Many works address yet more refined questions, such as the value distribution of $\z(\half+it)$ within intervals (e.g. \c{fyodorov, arguin, harper}). 

While proving the RH seems beyond any known techniques, progress toward proving the Lindel\"of Hypothesis and Density Hypothesis in certain contexts is closer at hand. Progress toward Lindel\"of falls under the umbrella of what are known as subconvexity bounds, for reasons that will be reviewed in Section \ref{s91}: their goal is to establish the tightest possible upper bound on the asymptotic growth of suprema.\foot{See \c{Michel2022Bourbaki} for a recent overview with useful references; \c{Bourgain2017} for the best known result for the Riemann zeta function; and \c{Nelson2021} for the best known result in the $GL(n)$ case.} Progress toward the Density Hypothesis comes in the form of zero density theorems, which upper-bound $N(\a,T)$ as a function of $\a$.\foot{See \c{GuthMaynard2024} for a recent advance.} These problems and their offspring have a clear bootstrap sensibility, even without having to commit to any interpretation of the zeros as eigenenergies of a mysterious physical system \`a la Hilbert-P\'olya. A more literal example of an $L$-function question of a bootstrap nature is the determination of the ``highest lowest zero'', a spectral maximin problem \c{Miller02,BFCetc,TangMiller}: letting $L(\half+i\g_n)=0$ define the non-trivial zeros with $\g_n\in\RR_+$ and $\g_1 := \min(\{\g_n\})$, how large can $\g_1$ be? (Note that Riemann zeta is \textit{not} the ``extremal $L$-function'' here.)

The physical nature of critical line phenomena is strongly reinforced by the appearance of random matrix theory (RMT) in the subject, of which there are increasingly many. Non-trivial zeros of the Riemann zeta function high on the critical line were observed decades ago to exhibit random matrix autocorrelations to extreme numerical precision \c{Montgomery1973,Odlyzko1987}: this is a paradigmatic example of a mathematical avatar of establishing RMT dynamics in a complex system. This led to later use of RMT heuristics to motivate computations and conjectures concerning the asymptotics of arbitrary critical moments of Riemann zeta \c{keatingsnaith1, conreyghosh, conreygonek}, culminating in the Conrey-Farmer-Keating-Rubenstein-Snaith (CFKRS) conjecture for primitive $L$-functions \c{cfkrs}, which even includes the overall coefficients.\foot{See \c{Ng} for a useful overview of Riemann zeta moments.} The conjecture of CFKRS implies the Generalized Lindel\"of Hypothesis (cf. e.g. Section 5.3 of \c{Patterson1988}).

The appearance of RMT heuristics in the formulation of the CFKRS conjecture was informed by yet another intervening development, which touches on a question familiar from modern conformal field theory and the swampland program. In \c{katzsarnak}, Katz and Sarnak introduced the notion of a `family'' of $L$-functions. The strategy was to embed a given $L$-function $L_f(s)$ within a class of similar objects by defining an infinite family $\mathsf{F}$, such that upon summing $L_f(s)$ over all $f\in\mathsf{F}$, one can prove a result that applies on average to every member and, potentially, to individual members. Their subsequent computations concerned the distribution of zeros in the neighborhood of the \textit{central point,} $s=\half$, instead of high on the critical line; without entering into details, they showed that the low-lying zeros of families of $L$-functions obey statistics of a Gaussian random matrix ensemble, the particular choice of ensemble (GUE, GOE, GSE) depending on the family. Actually, the result informs what a ``family'' ought to mean, as $L$-functions are not manifestly equipped with a notion of a family to which they belong: on the contrary, that notion is partially characterized by the symmetry class of central zeros, and the definition of a family continues to be refined. See e.g. \c{ks2, conreyfarmer,IwaniecLuoSarnak2000,rudnick-sarnak-zeros,SarnakShinTemplier2016,ShinTemplier2016,radz2023,blomer2024,cheek2025} for works in these directions.\foot{A footnote for any wayward mathematicians: the question of what constitutes a ``family of CFTs'' is an urgent one for non-perturbative conformal field theory and, by way of the AdS/CFT Correspondence, quantum gravity. We return to this in the Discussion.}

The question of whether a given $L_f(s)$ vanishes \textit{exactly} at the central point is of interest for a few reasons. In general, it is believed that $L_f(\half)\neq 0$ unless there is a hidden ``reason''; if it does vanish, the order of vanishing becomes an observable. This motivates an analysis of the proportion of $L$-functions that vanish within a given family $\mathsf{F}$, e.g. \c{IwaniecSarnak2000}. It also stars in the Birch-Swinnerton-Dyer (BSD) conjecture \cite{BirchSwinnertonDyer1965, BirchSwinnertonDyer1965II,WilesBSD}: for an $L$-function attached to an elliptic curve $E$, the BSD conjecture equates the order of vanishing at the central point with the rank of $E$, and gives a formula for the leading nonzero coefficient. 

We should mention that $L$-functions also appear in the mathematical-physical study of quantum chaos (logically distinct from RMT). A famous example is the Quantum Unique Ergodicity conjecture \c{rudnick-sarnak94,lindenstrauss06,soundararajan10,sarnak11-queprogress}: given a Maass cusp form $\phi_n(\t)$ for $SL(2,\ZZ)$ of unit $L^2$-norm, the integral of $|\phi_n(\t)|^2$ against the standard hyperbolic measure over a compact subregion of the modular surface $\cF$ reduces to the volume of the region in the large eigenvalue limit $n\rar\i$. The proof of this conjecture and questions about the approach to ergodicity can be related to the subconvexity of triple-product $L$-functions of modular forms, as reviewed in e.g. \c{bisain24,nelson2025}.

In closing let us just note that, of course, verification of the many conjectural automorphic and analytic properties of $L$-functions (and discovery of new ones, e.g. \c{bober2023,He03072025,zubrilina2025murmurations}) is vigorously pursued, as are applications to the distribution of prime numbers and other arithmetic questions that we have omitted here. 
 
Further details on some of the above topics will appear throughout the paper as we seek their avatars for CFT $L$-functions, to which we now turn. 

\sec{$L$-functions for two-dimensional CFT}\label{s3}

Our construction of $L$-functions for 2d CFTs will follow an approach familiar to those with some experience with standard $L$-functions, or with the spectral approach to 2d CFTs initiated in \c{2107}. In particular, the $L$-functions are straightforwardly defined by a Rankin-Selberg integral for $\sl$. We first present the construction of the $L$-functions using a  direct route, circling back to general automorphic formalism afterwards. 

We start with the torus partition function of a compact Virasoro CFT,
\e{Zdef}{Z(\t) = {\sum_{h,\hb}}' d_{h,\hb} \chi_{h,\hb}(q,\qb) = {1\o |\eta(\t)|^2}{\sum_{h,\hb}}' d_{h,\hb} q^{h-{c-1\o 24}}\qb^{\hb-{c-1\o 24}}}
where $q:=e^{2\pi i\tau}$, $c$ is the central charge, $\chi_{h,\hb}(q,\qb)$ is a Verma module character of Virasoro $\x \overline{\text{Virasoro}}$, $\eta(\tau)$ is the Dedekind eta function, and $(h,\hb)$ are eigenvalues under Virasoro generators $(L_0,\bar L_0)$. The sum is over Virasoro primaries, as indicated (quite naturally indeed) by the prime.\foot{A side quest of this paper is to propagate the ``prime for primary'' summation convention far and wide.} We have in mind that $c>1$ and $Z_p(\t)$ is unitary: this implies positive conformal weights $h,\hb>0$ and degeneracies $d_{h,\hb} > 0$, with the further condition $d_{h,\hb}\in \Z_+$ should one impose integrality. The \textit{primary partition function} is
\e{Zpdef}{Z_p(\t) := \sqrt{y}|\eta(\t)|^2 Z(\t)}
defined so that, for $\g\in SL(2,\Z)$ here and below,
\e{}{Z(\t) = Z(\g \t)\,,\quad Z_p(\t) = Z_p(\g \t)\,.}
$T$-invariance allows us to grade the Hilbert space by spacetime spin $j\in\ZZ$ and to write the partition function as 
\e{}{Z_p(\t) = \sum_{j\,\in\,\Z} e^{2\pi i j x} Z_p^{(j)}(y)\,.}
A primary state of conformal weights $(h,\hb)$ has conformal dimension $\D=h+\hb$ and spin $j=h-\hb$.

Next, from $Z_p(\t)$, construct a subtracted partition function $\cZ(\t) := Z_p(\t) - \tilde Z_p(\t)$ which preserves modular invariance and is square-integrable: 
\e{}{\qquad \cZ(\g\t)=\cZ(\t) \in L^2(\cF)\,.}
The object $\tilde Z_p(\t)$ is a fiducial object designed to enforce these constraints \c{2107}: it could, but need not, be a bona fide unitary CFT partition function.\foot{In such a case, unitarity is sufficient to guarantee square-integrability of the resulting $\cZ(\t)$, because unitary partition functions are finite for $\t\in\cF$ (this follows from Cauchy-Schwarz which implies $Z(\t) \leq Z(iy)$). Note that compared to \c{2107,2307} we call the regularized partition function $\cZ(\t)$ instead of $Z_{\rm spec}(\t)$ for a few reasons: besides emphasizing the new setting (we are not spectrally decomposing), and avoiding clumsy notation for the $L$-functions, we make the physics distinction that whereas $Z_{\rm spec}(\t)$ was previously used with modular completion via Poincar\'e summation, we will not commit to that choice here.} We give but one canonical algorithm for constructing $\cZ(\t)$ in Section \ref{s33}, valid for any $Z_p(\t)$. One can view this as a regularization, generalizing Zagier's approach \c{zagier} for power law growth to the exponential case. The square integrability of $\cZ(\t)$ means that, as a partition sum, $\cZ(\t)$ only has support only on heavy states, i.e. those states with $\D>{c-1\o 12}$.

Consider now the weight-zero Petersson pairing
\e{peterspeter}{(\cZ,E_{1-s}) := \int_\cF{dx dy\o y^2} \cZ(\t) E_s(\t)}
where $E_s(\t)$ is the non-holomorphic Eisenstein series for $\sl$. We review some salient features of $E_s(\t)$ in Appendix \ref{appa}. The integral is absolutely convergent. $E_s(\t)$ obeys a functional equation most easily stated in terms of its completion $E^*_s(\t)$,
\e{}{E^*_s(\t) := \L(s)E_s(\t) = E^*_{1-s}(\t)\,,}
where $\L(s)=\L(\thalf - s)$ is the completed Riemann zeta function defined in \eqr{Lamdef}. This implies a corresponding functional equation for $(\cZ,E_{1-s})$. We observe that upon defining
\e{olap}{(\cZ,E_{1-s}) := \(2\pi\)^{\half-s} \G\big(s-\thalf\big){\lz(s)\o \z(2s)}}
the factor $\lz(s)$ transforms as an $L$-function. More precisely, it obeys a specific functional equation of characteristic form, most easily stated in terms of the completed $L$-function 
\e{Lzdef}{\Lamz(s) := 2^s\g_{\cZ}(s) \lz(s)\,,}
with gamma factor
\e{gcft}{\g_\cZ(s) = \pi^{-2s} \prod_{i=1}^4 \G\({s+\k_i\o 2}\)\,,\quad \{\k_i\} = \{0,1,-\thalf,\thalf\}\,,}
which satisfies
\e{LZFE}{\Lamz(s)=\Lamz(1-s)\,.}
This is the functional equation of a self-dual $L$-function of degree-4, gamma factor $\g_\cZ(s)$, and root number $\vareps=1$.

This definition, and in particular the values of the spectral parameters $\{\kappa_i\}$, are dictated by the symmetry structure of $\cZ(\t)$. To see this, it is instructive to present the completed $L$-function $\Lamz(s)$ as a modular integral with the \textit{completed} Eisenstein series:
\e{Lamzint}{{\Lamz(s)\o4\sqrt{\pi}} =\int_\cF{dx dy\o y^2} \cZ(\t) E^*_s(\t)\,.}
Writing the gamma factor using the duplication formula as
\e{gprop}{\g_\cZ(s) \propto \G(s)\G(s-\thalf)}
up to exponential factors, the origin of each factor becomes completely clear from \eqr{Lamzint}: the $\G(s)$ is an \textit{$\sl$ factor} coming from the completed Eisenstein series, while the $\G(s-\thalf)$ is a \textit{Virasoro factor} coming from $\cZ(\t)$, namely, from the $\sqrt{y}$ scaling of the large $y$ expansion of the scalar partition function $\cZ^{(0)}(y)$ (i.e. the constant Fourier coefficient of $\cZ(\tau)$). To see the latter explicitly, ``unfold'' the inner product using the Eisenstein series' Poincar\'e sum representation \eqr{Espoincare} to obtain the Mellin representation
\e{unfold}{(\cZ,E_{1-s}) = \int_0^\i dy \, y^{s-2}\cZ^{(0)}(y)\,.}
At large $y$, 
\e{Z0y}{\cZ^{(0)}(y\rar\i) \approx \sqrt{y} \sum_{\l>0}a_\l e^{-2\pi \l y}\,.}
By Mellin inversion, matching \eqr{unfold} to this expansion and using the definition \eqr{olap} gives the formal series
\e{zzdef}{\zz(s) := {\lz(s)\o \z(2s)} = \sum_\l {a_\l\o \l^{s-\half}}\,.}
This is the spectral zeta function attached to $\cZ^{(0)}(y)$, a generalized Dirichlet series over the scalar data of $\cZ(\t) = Z_p(\t) - \tilde Z_p(\t)$. The set $\{\l_n\}$ comprises the scalar dimensions of $\cZ(\t)$, with respective multiplicities $a_\l$ (suppressing a ``(0)'' superscript for tidiness): 
\e{laldef}{\l = \D-{c-1\o 12}\,,\qquad a_\l = d_\D-\tilde d_\D\,.}
That $\zz(s)$ expanded like so gives the requisite form \eqr{Z0y} for a Virasoro CFT is directly tied to the specific gamma factor $\G(s-\thalf)$ in \eqr{olap}. 

The previous paragraph shows that the construction generalizes straightforwardly to CFTs with extra conserved currents generating an extended chiral algebra $\mathcal{A}\supset \text{Virasoro}$. This is clear from \eqr{Lamzint}, with the following implication: to the $\mathcal{A}$-primary partition function of a CFT with $N$ generating currents is attached an $L$-function with the same properties as the Virasoro case ($N=1$), except now with spectral parameters 
\e{kappaN}{\{\k_i\}  = \{0,1,\tfrac{N}{2}-1,\tfrac{N}{2}\}\,.}
The former two again make a $\G(s)$ factor from $\sl$ invariance, while the latter two make a $\G(s-\tfrac N2)$ factor from $\mathcal{A}$ symmetry via unfolding. (We collect some formulas more explicitly in Section \ref{s101}.)

Note that the fundamental strip of the Mellin transform is to the right of $s=1$, due to the pole of the integrand:
\e{E1exp}{E_{s\rar 1}(\t) \approx {{\rm vol}^{-1}(\cF)\o s-1} + \widehat E_1(\t) + \ldots\,,\qquad  {\rm vol}(\cF)={\pi\o3}\,.}
This pole, and its reflection at $s=0$, are the only poles of $E_s^*(\t)$. The residue of the $L$-function at $s=1$ is thus controlled by the modular average of $\cZ(\t)$,
\e{modavdef}{\Res_{s=1}\,(\cZ,E_{1-s}) = \<\cZ\> := {\rm vol}^{-1}(\cF)\int_\cF{dx dy\o y^2} \cZ(\t) \,.}
This means that $s(1-s)\Lamz(s)$ is entire, analytically continuing $\lz(s)$ to the whole complex plane. 

\sssec*{Rankin-Selberg convolution context}

It is edifying to view this construction against the backdrop of Rankin-Selberg convolution $L$-functions. 

Given two weight-$k$ automorphic functions $f(\t)$ and $g(\t)$, the modular integral
\e{}{\L_{f \otimes g}(s) := \int_{\cD} {dx dy \o y^2} y^k f(\t) \overline{g(\t)} E^*_s(\t)}
defines a completed $L$-function, $\L_{f \otimes g}(s) = \L_{f \otimes g}(1-s)$, whose series representation is the Dirichlet convolution of the respective series for the individual automorphic $L$-functions $L_f(s)$ and $L_g(s)$. The domain $\cD$ is a suitable modular domain $\Gamma\backslash \HH$, where $y^k f(\t)\overline{g(\t)}$ is $\Gamma$-invariant; in common cases, $\G = \sl$. An elementary version of the above is the Rankin-Selberg convolution of two cusp forms. Consider a holomorphic cusp form $f(\t)$ of weight $k\in\Z_+$ under $\G_0(N)$ (``level $N$'') and trivial multiplier, with Fourier expansion
\e{holfourier}{f(\t) = \sum_{n=1}^\i a_f(n) q^n\,,\quad a_f(n)\in\mathbb{C}\,.}
Then the $L$-series
\e{Lffser}{L_{f \otimes \fb}(s) = \z(2s) \sum_{n=1}^\i{|a_f(n)|^2\o n^{s+k-1}}}
admits an analytic continuation to $s\in\CC$ via a Rankin-Selberg integral
\e{}{(4\pi)^{1-k-s}\G(s+k-1) {L_{f \otimes \fb}(s) \o \z(2s)} = \int_{\cF} {dx dy \o y^2} \,y^k|f(\t)|^2 E_s(\t)\,.}
$L_{f \otimes \fb}(s)$ is the Rankin-Selberg convolution $L$-function, self-dual of root number $\vareps=1$, with a degree-4 gamma factor with spectral parameters
\e{}{\{\k_i\} = \{0,1,k-1,k\}\,.}
At $s=1$,
\e{ffres1}{\Res_{s=1}L_{f \otimes \fb}(s) = {(4\pi)^k\o \G(k)} {\<f,f\>\o \text{vol}(\cF(N))}}
where 
\e{}{\<f,f\>=\int_{\cF(N)}{dx dy\o y^2}\,y^k |f(\t)|^2}
is the weight-$k$ Petersson inner product on $\cF(N):= \G_0(N)\backslash \mathbb{H}$  \c{IKtext}.%

It is plain to see that our $L$-function is an instance of this construction, generalized to a non-holomorphically-factorized integrand. The correspondence is\foot{Some factors of $2$ and $\pi$ differ on account of our conventions, namely our preference for the frequencies of $\zz(s)$ to be $\l := \D-{c-1\o 12}$ and the degeneracies $a_\l$ to be the physical ones, not rescaled. In the Rankin-Selberg convolution context in which the $a_f(n)$ are non-integer, one commonly rescales by numerical factors for simplicity.}
\es{Lcorresp}{\lz(s) \qquad &\longleftrightarrow \qquad L_{f \otimes \fb}(s)\\
\zz(s)  \qquad &\longleftrightarrow \qquad\sum_{n=1}^\i{|a_f(n)|^2\o n^{s-\half}}}
In particular, $\lz(s)$ has exactly the same functional equation and gamma factor as $L_{f\otimes \fb}(s)$ upon formally identifying $k=\half$. Indeed, the structure of the respective integrals is literally identical when $k=\half$\thinspace---\thinspace including, crucially, at the level of the formal structure of the $q$-expansion that enters the unfolding.\foot{In contrast, the Rankin-Selberg convolution of two Maass cusp forms generates a degree-4 $L$-function with a different gamma factor. This happens because, while the coefficients of the $L$-series are still constructed from the Fourier coefficients, the Maass cusp form Fourier modes carry Bessel functions of $y$ (see \eqr{Ephifourier}), which generate an infinite large $y$ expansion mode-by-mode (unlike \eqr{holfourier}).} Both $\Lamz(s)$ and $\L_{f\otimes \fb}(s)$ are entire apart from poles at $s=0,1$, with the residues at $s=1$ given by modular averages.

Similar remarks apply to the $N$-current case, where the identification with Rankin-Selberg convolution becomes $k={N\o 2}$. It is notable that the Virasoro case, with $k=\half$, involves an analytic continuation in the modular weight of $f(\t)$: there are no weight-$\half$ holomorphic cusp forms for $\G_0(N)$. 

The relation \eqr{Lcorresp} is enlightening from the CFT point of view, as it casts $\lz(s)$ as a non-holomorphic generalization of a familiar factorized object $L_{f \otimes \fb}(s)$. We can view $f(\t)$ as akin to a chiral CFT partition function times an eta factor $\eta(\t)$. A generic CFT does not holomorphically\foot{See Section 4 of \c{mcgady19} for an $L$-function formula for central charges in holomorphic CFTs.} factorize; nevertheless, we can attach $L$-functions\thinspace---\thinspace with universal analyticity and transformation properties\thinspace---\thinspace to generic CFTs in a direct generalization of the Rankin-Selberg convolution construction, in spite of the inherent complexity of their spectral data. 

Of course, such complexity does give rise to interesting differences between $\lz(s)$ and $L_{f\otimes \fb}(s)$. At the analytical level, $L_{f\otimes f}(s)$ is divisible by $\z(s)$\thinspace---\thinspace that is, their quotient is entire\thinspace---\thinspace whereas the same is not true of $\lz(s)$; the residue at $s=1$ of $L_{f \otimes \fb}(s)$ is strictly positive, unlike $\lz(s)$; and $\lz(s)$ vanishes at $s=\half$, unlike $L_{f\otimes f}(s)$. Other differences lie in the nature of the series expansions: $\lz(s)$ has a \textit{generalized} Dirichlet series expansion and $a_\l \in \ZZ$, whereas $L_{f \otimes \fb}(s)$ has an ordinary Dirichlet series expansion and $|a_f(n)| \in \RR_+$. We will elucidate these features of $\lz(s)$ throughout the rest of this work.

\ssec{Comments}\label{s31}

\enumcom

\item The functional equation shows that $\lz(s)$ has conductor $q=4$. However, since the $L$-series is a generalized Dirichlet series, with leading frequency $\l_1$ not normalized to unity, this is not an invariant notion in our context. The more precise statement is that, in our convention in which the frequency spectrum $\{\l_n\}$ is comprised of the (shifted) conformal dimensions, $q=4$. Alternatively, one could regard $q=4\l_1^{-2}$ as a normalized conductor. 

\item One could append a $s-\thalf$ to the gamma factor \eqr{gcft} without spoiling the existence of a functional equation for $\lz(s)$. Doing so would render $\lz(s)$ odd, flipping the sign of the root number $\vareps$. This seems contrived in the sense explained above: the $\G(s-\thalf)$ in the functional equation is a direct descendant of the Virasoro symmetry, needing no further adornment. Also, such a freedom is \textit{not} present for the $L$-functions for CFTs with $N>1$ conserved currents, with spectral parameters \eqr{kappaN}.

\item For standard $L$-functions, their Dirichlet series frequencies $n\in\ZZ$ are ``simple'' and their coefficients are ``complicated'', obeying various remarkable (conjectural) statistical properties. For $\zz(s)$, the nature of the series data is essentially inverted: its frequencies $\l\in \R_+$ are complicated, whereas its coefficients $a_\l \in \ZZ$ are simple. Indeed, the frequencies are expected to be densely and chaotically distributed with Gaussian random autocorrelations, and $a_\l = \pm 1$ modulo accidental degeneracies. We will discuss these features and their implications for series convergence in later sections. 

\ni It is highly non-trivial that $\Lamz(s)$ analytically continues the formal series \eqr{zzdef} to all $s\in\CC$ at all. Whereas ordinarily, analytic continuation reflects some structure in the series coefficients\thinspace---\thinspace e.g. arithmeticity, reciprocity laws or group structure\thinspace---\thinspace for CFT $L$-functions this cannot be the explanation. 

\item $\cZ(\t)$ is not a Hecke eigenform, so we do not expect $\lz(s)$ to admit an Euler product. In this respect, $\lz(s)$ is rather more complicated then even the Beurling prime zeta functions \c{hilberdink-lapidus06}, whose frequencies are constructed from an arbitrary infinite set of real primitives.

\item The previous two remarks make plain that the CFT $L$-functions are rather exotic compared to standard $L$-functions, lying outside the most well-studied classes of $L$-functions (Selberg, $GL(n)$ automorphic) under the lamppost. This is as it should be, in order to reflect the rich structures present in the spectral data of irrational CFTs. Morally, this is a Mellin transform of the same fact in $\tau$-space: standard modular functions $f(\t)$ sorely lack the intricate structure needed to serve as candidate partition functions of irrational CFTs.

\item A natural question from the mathematical point of view, notwithstanding the above remarks, is whether there are physically relevant CFTs (albeit simple theories with highly regular spectra) where $\lz(s)$ \textit{is} comprised of standard $L$-functions. We observe a happy synergy here: up to an analytic prefactor, a product of two Riemann zeta functions is the $L$-function of a single compact free boson! (See Section \ref{s102}.)

\item In CFTs with conformal manifolds, their $L$-functions are continuous functions of the exactly marginal couplings. We note the contrast with Sarnak's rigidity conjecture for Selberg class $L$-functions: the only continuous family of $L$-functions are primitive $L$-functions with shifted arguments, $L(s+iy)$. If the conformal manifold is supersymmetric, the $L$-function contains a BPS subsector that does not depend on couplings. 

\end{enumerate}

\ssec{Subtraction}\label{s32}

We now return to the matter of regularization, or ``cuspification''. Given a partition function $Z_p(\t)$ with a discrete spectrum, our goal is to construct $\cZ(\t)$ by subtracting the light states $\D\leq {c-1\o 12}$ in a manner preserving the following three conditions: 

{\sf i)} ~~modularity,

{\sf ii)} ~\,discreteness,

{\sf iii)} ~regularity for all $\t\in\cF$. 

\ni The resulting $\cZ(\t)\in L^2(\cF)$. We now present one canonical way to do so, that we call \textit{$J$-subtraction}. The name refers to the modular $J$-function, call it $J(q)$, a meromorphic modular function with $q$-expansion $J(q) \approx q^{-1} + 196884q+\ldots$, where we have set the constant term to zero for simplicity. $J(q)$ is regular for all $\t\in\cF$, with a zero on the unit $\t$-circle. 

First, we define a certain non-holomorphic Hecke operator,
\e{}{T'_{\a,\ab}(q,\qb) := J(q)^{\a}J(\qb)^{\ab}\sum_{\substack{m,n=0\\m+n\leq \lfloor\a+\ab\rfloor}}\g_{\a,\ab}^{(m,n)}J(q)^{-m} J(\qb)^{-n}}
where the $\g_{\a,\ab}^{(m,n)}=\g_{\a,\ab}^{(n,m)}$ are fixed uniquely by the condition
\e{Thhb}{T'_{\a,\ab}(q,\qb) \stackrel{!}{=} {q^{-\a}\qb^{-\ab}\o |\eta(q)|^2} + O(|q|^{\eps})~~\forall\,\eps>0}
where $|q|^\eps$ means a monomial in $q,\qb$ of total power $\eps$. In other words, given a character whose leading term in the $q$-expansion is $q^{-\a}\qb^{-\ab}$, upon replacing that term with $J$-functions we iteratively remove their ``spurious'' light descendants by subtracting more $J$-functions, whose own light descendants we remove likewise, proceeding until none remain. Consequently, $T'_{\a,\ab}(q,\qb)$ equals the character $\chi_{h,\bar h}(q,\qb)$ with $(h,\bar h) = (\a+{c-1\o 24},\ab+{c-1\o 24})$, up to heavy states which are uniquely fixed by modularity under this $J$-function prescription. 

We next define a fiducial partition function and its primary counterpart
\e{ZJsum}{Z^{(J)}(\t) = \mathop{{\sum}'}_{\substack{h,\hb\\h+\hb \leq {c-1\o 12}}} d_{h,\hb} T'_{{c-1\o 24}-h,{c-1\o 24}-\hb}(q,\qb)\,,\qquad Z_p^{(J)}(\t) = \sqrt{y}|\eta(q)|^2Z^{(J)}(\t) }
where the degeneracies $d_{h,\hb}$ are those of $Z_p(\t)$: this is \textit{$J$-completion}. Then the difference
\e{}{\cZ(\t) = Z_p(\t) - Z_p^{(J)}(\t)}
is a modular-invariant, square-integrable regularization of $Z_p(\t)$, with support only on heavy states: this is \textit{$J$-subtraction}. 

Let us verify the three conditions on $\cZ(\t)$. First, for $\a-\ab \in \Z$, $T'_{\a,\ab}(q,\qb)$ is $\sl$-invariant; therefore, $Z_p(\t)$ being a sum over integer-spin states, so is $Z_p^{(J)}(\t)$. Second, and in contrast to Poincar\'e summation, $J$-subtraction also manifestly preserves discreteness. Finally, $Z_p^{(J)}(\t)$ is non-singular for all $\t\in\cF$. To see this, we can write
\e{}{T'_{\a,\ab}(q,\qb) = J(q)^{\a}J(\qb)^{\ab}\sum_{p=0}^{\lfloor\a+\ab\rfloor}|J(q)|^{-p}\sum_{n=0}^{\left\lfloor {p \o2}\right\rfloor} \g^{(n,p-n)}_{\a,\ab}{2\cos\big((2n-p)\arg(J)\big) \o 1+\delta_{n,{p\o 2}}}\,.}
and note that $\a+\ab\geq 0$ for every term in the sum \eqr{ZJsum}. Note that $\arg(J) \approx -2\pi x$ over most of $\cF$, where $J(q) \approx q^{-1}$ is a good approximation. 

Meromorphy of $J(q)$ implies that the heavy spectrum of $Z_p^{(J)}(\t)$ is comprised of towers of integer-spaced states, one tower for every light operator. In addition, their degeneracies have a certain relic of the Monster symmetry of the $J$-function: they are real linear combinations of Monster group representations determined by binomial expansion. Unitarity of $Z_p(\t)$ implies $\a,\ab\in\RR$ and hence $\g^{(m,n)}_{\a,\ab}\in\mathbb{R}$, but for generic values of $c$ one has $\g^{(m,n)}_{\a,\ab}\notin\Z$. For all of these reasons, the ``spurious'' heavy states of $Z^{(J)}_p(\t)$ can be filtered out from the frequency spectrum of $\cZ(\t)$, distinguished from the genuine (presumably erratically-distributed) states of $Z_p(\t)$ of positive integer degeneracy.

$J$-subtraction is a canonical way to build the $\cZ(\t)$ associated to any $Z_p(\t)$, and hence an $L$-function attached to $Z_p(\t)$; we certainly make no claims of uniqueness of $J$-subtraction within the space of possible regularizations. Indeed, $J$-subtraction is but one instantiation of what one might call \textit{extremal subtraction}: for every light primary of dimension $(h,\hb)$, subtract a partition function with central charges $(c',\bar c') = (c-24h,c-24\hb)$ and a gap to threshold.

We dubbed $T'_{\a,\ab}(q,\qb)$ a non-holomorphic Hecke operator because it mimics the action of the usual $\sl$ Hecke operator $T_k$ on the $J$-function. Let us review the latter. For $k\in\Z_+$, 
\e{}{T_k J(q) = \sum_{ad=k}\sum_{b=0}^{d-1} J\(e^{2\pi i\({a\t+b\o d}\)}\) = q^{-k} + O(q)}
where the $O(q)$ terms and beyond are heavy states determined uniquely by modularity and meromorphy. As $T_k J(q)$ is a polynomial in the $J$-function, it is non-singular for all $\t \in \cF$. The operator $T'_{\a,\ab}(q,\qb)$ does essentially the same thing by minimally generalizing this construction to the non-holomorphic setting: it modular completes a power $q^{-\a}\qb^{-\ab}$ to the unique $\sl$-invariant object built purely out of $J$-functions, without adding any new singularities nor light Virasoro primaries.\foot{The Virasoro element of our construction is by design: removing the $|\eta(q)|^2$ in \eqr{Thhb} would have been closer to the literal form of $T_k J(q)$, but at the cost of having to clumsily subtract Virasoro descendants by hand. We thank Yiannis Tsiares for discussions on this topic.} This is not unlike Witten's construction \c{Witten:2007kt} of extremal CFTs, only that we are turning the logic on its head somewhat by viewing $Z^{(J)}(\t)$ not \textit{constructively}, as a bona fide CFT partition function obeying all bootstrap constraints, but rather \textit{destructively}, as a modular completion to be subtracted from one. However, the $J$-completion point of view is interesting in its own right, and we return to that in the Discussion.

\ssec{A useful $L$-quotient}\label{s33}
Our discussion around \eqr{gprop} evinced the symmetry-based origins of the gamma factor $\g_\cZ(s)$ in the functional equation for $\Lamz(s)$, being determined solely by Virasoro and $\sl$ invariance. This suggests the construction of an object that strips off this universal information. To do so, we introduce 
\e{qdef1}{\qz(s) := {1\o 4\sqrt{\pi}}{\L_\cZ(s)\o \L(s)\L(s-\half)} \,.}
Cancelling the gamma factors, we see that $\qz(s)$ is a ratio of $L$-functions, i.e. an \textit{$L$-quotient}:
\e{qdef2}{\qz(s) = 2^{\half-s}{\lz(s)\o \z(2s)\z(2s-1)}\,.}
As such, $\qz(s)$ admits a generalized Dirichlet series representation. Moreover, it is reflection-symmetric:
\e{}{\qz(s) = \qz(1-s)\,.}
This is manifest from \eqr{qdef1} upon using the functional equation for $\L(s)$. 

Having quotiented by the symmetry-based information, one might wonder whether $\qz(s)$ captures the ``core'' of the $L$-function. Versions of this idea will indeed be ratified in what follows. At any rate, $\qz(s)$ will sometimes be a useful object in terms of which to phrase certain results.

\sec{Analytic structure I: Poles and zeros}\label{s4}

The analytic structure of $\lz(s)$, depicted in Figure \ref{fig:poleszeros}, is strongly constrained by its origin as a Rankin-Selberg integral. 

\subsubsection*{Poles}

The inner product $(\cZ,E_{1-s})$ converges absolutely away from poles of $E_s(\t)$:
\e{}{(\cZ,E_{1-s}) < \i ~~\forall\,s\notin\left\{1,{s_n\o 2}:\z(s_n)=0\right\}}
The pole at $s=1$ is simple, cf. \eqr{modavdef}. By the definition \eqr{olap}, the zeta poles are not present in $\lz(s)$, so there is just one pole:
\e{}{\lz(s) < \i ~~\forall\,s\neq 1}
with
\e{}{\Res_{s=1}\lz(s) = \sqrt{2}\z(2) \<\cZ\>\,.}
We record for future use that, by reflection, this implies a value at the origin
\e{lz0}{\lz(0) = {\<\cZ\>\o 12\sqrt{2}}\,.}

\subsubsection*{Zeros}

Because $s(1-s)\Lamz(s)$ is entire, the trivial zeros of $\lz(s)$ are the poles of $s(1-s)\g_\cZ(s)$. This implies the set
\e{}{\qquad \qquad \qquad \qquad \lz(s) = 0 ~~\forall\,s \in \half\, \Z_- \qquad~~ (\textsf{trivial zeros})}
as well as a zero at the distinguished \textit{central point,} $s=\half$:
\e{centralzero}{\qquad\qquad \quad~ \lz\(\half\)=0\,,\quad \underset{s=\half}{\text{ord}}\,\lz(s) \geq 1\qquad (\textsf{central zero})}
This zero is generically simple, with coefficient
\e{bsd}{L_\cZ'\(\half\) = \half\int_\cF{dx dy\o y^2} \cZ(\t) E'_{\half}(\t)}
where $E'_{\half}(\t) = \p_s E_s(\t)|_{s=\half}$ has Fourier expansion
\e{}{E'_{\half}(\t) = 2\sqrt{y}\(\g_E - \log(4\pi) + \log y + 4\sum_{j=0}^\i \cos(2\pi j x)\s_0(j) K_0(2\pi j y)\)}
where $\g_E$ is Euler's constant. A higher-order zero at the central point could occur in principle if $\cZ(\t)$ has finely-tuned, resonant behavior with $E'_{\half}(\t)$ that forces \eqr{bsd} to vanish.\foot{We remark that $L$-functions constructed by Rankin-Selberg should have bounded order of vanishing: if all derivatives vanish at $s=\half$, the Eisenstein integral vanishes in a neighborhood of $s=\half$ and, by the identity theorem for meromorphic functions, for all $s\in\CC$.} 

The central point $s=\thalf$ is special in number theory. By the BSD conjecture, for $L$-functions attached to elliptic curves the order of vanishing at the central point equals the rank $r$ of the elliptic curve, and the leading nonzero Taylor coefficient is conjecturally positive (given by a complicated formula). The region near the central point also serves as a playground for various explorations of the ``complexity'' of $L$-functions, their organization into families, and bounds on the locations of zeros.\foot{Standard $L$-functions obey rigorous upper bounds on their order of vanishing in terms of the conductor $q$ \c{blomer2024}. On the Generalized Riemann Hypothesis, the order $r$ is essentially $r \sim \log q/\log\log q$, which is a measure of the average spacing of zeros near the central point: i.e. in an segment of the critical line near the origin of order one size, there are approximately $r$ zeros. This undergirds a general correlation between the complexity of an $L$-function, its conductor $q$, and its order of vanishing $r$ at the central point.} Given that standard self-dual $L$-functions with root number $\vareps=1$ are generally believed to be nonzero at $s=\thalf$ unless there is a special arithmetic reason, it is natural to ask what the reason is for \eqr{centralzero}, and whether there is a bias for a positive sign of \eqr{bsd}.

Finally, there are non-trivial zeros of $\lz(s)$, which are the zeros of $\Lamz(s)$:
\e{}{\qquad \qquad \qquad \qquad \lz(\rho)=\Lamz(\rho)=0 \quad \qquad (\textsf{non-trivial zeros})}
There is an infinite number of non-trivial zeros. This follows from the fact that, as we will prove in the next subsection, $\Lamz(s)$ is a complex-analytic function of \textit{order one}, a property defined in \eqref{oo}, and a basic fact about order one functions (e.g. \c{Davenport2000}, Chapter 11) is that
\e{orderonezerosum}{\sum_\rho {1\o |\rho|}=\i\,,\quad \sum_\rho {1\o |\rho|^{1+\eps}}<\i~~\forall\,\eps>0\,.}

\sssec*{$\zz(s)$ and $\qz(s)$}

A few lines of algebra translate these analytic properties of $\lz(s)$ into those of $\zz(s)$ and $\qz(s)$. For future use we explicitly record a few of them:

\begin{itemize}

\item $\zz(s)$ has trivial and central zeros
\e{}{\qquad\qquad\qquad\qquad\qquad\zz(s) = 0 ~~\forall\,s\in\half-\Z_{\geq 0}\quad \quad (\textsf{trivial/central zeros})}
where now the central zero is a double zero, 
\e{}{\underset{s=\half}{\text{ord}}\,\zz(s) \geq 2\,.}

\item $\zz(s)$ is regular for all $\s>\half$ except for the pole at $s=1$. This builds in compatibility with a discrete Virasoro primary scalar spectrum (or average thereof): by spectral decomposition
\e{}{\cZ^{(0)}(y)-\<\cZ\> = {1\o 2\pi i} \int\limits_{~(\half)}ds (\cZ,E_{1-s})y^{1-s}}
and the relation \eqr{modavdef}, regularity in the right-half-plane guarantees that $\cZ^{(0)}(y)$ takes the requisite form \eqr{Z0y} at $y\rar\i$, with no inverse powers of $y$ inside the sum.

\item $\qz(s)$ is regular at $s=1$, because the pole of $\z(2s-1)$ cancels that of $\lz(s)$. Its only possible poles lie at the locations of non-trivial zeros of $\z(2s)$ and $\z(2s-1)$:
\e{qpoles}{\qz(s)<\i~~\forall\, s\notin\left\{{\rho_n\o 2},{1+\rho_n\o 2}\right\}\,.}
On RH, these poles are simple for all $n$. 

\item We stress that \eqr{qpoles} comprise the \textit{maximal} pole set of $\qz(s)$: some of them could be absent. We do not know how to establish their existence on the basis of modularity alone. As we explain in Appendix \ref{appd}, these poles require certain cancellations among spin sectors of the partition function $\cZ(\t)$, and imply intriguing ``conspiracies'' among Riemann zeta and Maass cusp form data. Nevertheless, we prove in Appendix \ref{appd}, conditional on a weak assumption, that at least some of these poles are always present in a $c>1$ Virasoro CFT. At any rate, in what follows we remain agnostic about their presence and allow them throughout our analysis.

\end{itemize}

\ssec{Order one}\label{s41}

An important property of $\Lamz(s)$, shared by all standard $L$-functions, is that it is of \textit{order one}: 
\e{oo}{\L_\cZ(s) \ll e^{|s|^{1+\eps}}~~\forall\,\eps>0}
as $|s|\rar\i$. It is so important that we give two independent proofs. The first, completely general proof is given below.\foot{We prove this anew because the usual proofs for standard $L$-functions tend to rely on a half-plane of absolute convergence.} A second, conditional proof is given in Appendix \ref{appe}.

The idea of the proof is as follows. Consider $s(1-s)\Lamz(s)$, which is entire. To prove \eqr{oo}, it suffices to establish its uniform boundedness on the critical line: then by analyticity, and its factorial growth at $s\rar\i$, Phragmen-Lindel\"of for strips (see Appendix \ref{appb}) implies that $s(1-s)\Lamz(s)$, and hence $\Lamz(s)$, is order one for all $\s\geq \half$. The $s\rar 1-s$ invariance of $s(1-s)\L_\cZ(s)$ finally extends this to the entire complex plane. 

We start with the Rankin-Selberg integral $(\cZ,E_{1-s})$. This converges everywhere the Eisenstein series $E_s(\t)$ has no pole, with uniform convergence in $s$ over compact sets. To establish boundedness it remains only to study the behavior along vertical lines. We sit on the critical line, $s=\half+it$. By construction, $\cZ(\t)$ is of rapid decay in $y$. This actually implies the ``dual'' statement that \textit{$(\cZ,E_{\half-it})$ is of rapid decay in $t$}. One hands-on way to prove this\foot{See Section 9 of \c{EM} for another proof.} is to consider $(\cZ,\D_\t^m E_{\half-it})$, where $\D_\t = -y^2(\p_x^2+\p_y^2)$ is the hyperbolic Laplacian. On the one hand, acting directly on the Eisenstein series gives
\e{}{(\cZ,\D_\t^m E_{\half-it}) =\({1\o4}+t^2\)^m (\cZ,E_{\half-it})\,,\quad m\in\Z_{\geq 0}}
via the Laplace equation \eqr{Eslaplace}. On the other hand, by self-adjointness,
\e{}{(\cZ,\D_\t^m E_{\half-it}) = (\D_\t^m \cZ,E_{\half-it})\,,}
and by Cauchy-Schwarz,
\e{}{(\D_\t^m \cZ,E_{\half-it}) \leq ||\D_\t^m \cZ||_{_2} ||E_{\half-it}||_{_2}}
where $||f(\t)||_{_2} := \sqrt{\int_\cF{dx dy\o y^2}f(\t)\overline{f(\t)}}$ is the $L^2$-norm. Putting things together,
\e{oorhs}{ (\cZ,E_{\half-it}) \leq \({1\o4}+t^2\)^{-m}||\D_\t^m \cZ||_{_2} ||E_{\half-it}||_{_2}~~\forall\,m\in\Z_{\geq 0}\,.}
We now bound both terms on the RHS. First, $\D_\t^m \cZ(\t)$ is still of rapid decay for any $m$, so $||\D_\t^m \cZ||_{_2}<\i$ for all $m$ and is independent of $t$. Next, we use that (cf. e.g. \c{Terras})
\e{}{\qquad \qquad\big(||E_{\half-it}||_{_2}\big)^2 \leq C\log(2+|t|)~~\qquad (|t|\rar\i)}
for some fixed constant $C$. So the RHS of \eqr{oorhs} is finite and decays, giving the desired result
\e{}{\qquad~~(\cZ,E_{\half-it}) = O(|t|^{-m})~~\forall\,m\in\Z_{\geq 0}\qquad (|t|\rar\i)\,.}
With this in hand, the proof proceeds quickly according to our earlier logic. We have the entire function
\e{}{s(1-s)\L_\cZ(s) = s(1-s)\L(s) (\cZ,E_{1-s})\,.}
$\L(s)$ is order one, $(\cZ,E_{1-s})$ is of rapid decay on the critical line, and both grow factorially as $s\rar\i$. Phragmen-Lindel\"of then implies that $s(1-s)\L_\cZ(s)$ is order one for all $\s\geq \half$. By the functional equation, this extends to all $s\in\CC$.  \qed

\ssec{Hadamard product}\label{s42}

Because $s(1-s)\Lamz(s)$ is an entire function, it admits a Hadamard product representation as a factorization over its zeros, i.e. the non-trivial zeros of $\lz(s)$. The order one condition \eqr{oo}, together with the functional equation \eqr{LZFE}, implies an especially simple form:
\ebox{Lamzprod}{s(1-s)\L_\cZ(s) = e^A\prod_{\rho} \(1-{s\o \rho}\)\,,\quad e^A = -{2\pi^{3/2}\o 3}\<\cZ\>}
where the formula for the residue $e^A$ follows from \eqr{lz0}.

Normally, Hadamard products for entire functions $f(s)$ of order one are written in terms of the so-called canonical factors, which include exponentials in $s$:
\e{}{f(s) = f(0)\,e^{Bs}\prod_\rho \(1-{s\o \rho}\)e^{s/\rho}}
where the exponent $B$ is determined by Taylor expansion. But a functional equation $f(s)=f(1-s)$ allows us to conveniently eliminate the exponentials, because it implies
\e{Beq}{B = -\sum_\rho  {1\o\rho}\,.}
Deriving this is elementary, but we include it to demonstrate the power of the functional equation and the order one condition.
 
Given a zero $\L_f(\rho)=0$, there is also a complex conjugate zero $\L_f(\bar\rho) = 0$. A general functional equation relates zeros of $f(s)$ to those of its dual. However, for a self-dual $L$-function like $\lz(s)$, one has $\L_f(\rho) = \L_f(1-\rho)$ too. So zeros come in $\ZZ_2 \x \ZZ_2$-symmetric \textit{quartets,}
\e{}{\L_f(\rho)=\L_f(\bar\rho)=\L_f(1-\rho)=\L_f(1-\bar\rho)=0}
save for those zeros with $\s_n=\half$ or $t_n=0$, which live in doublets. Accordingly we can restrict the product to those zeros in the first quadrant of the complex $s-\half$ plane,
\e{++def}{\sfS_\rho^{(++)}(f) := \left\{\rho = \s+it:\L_f(\rho)=0\,,~\s-\thalf \in \RR_+\,,~ t\in \RR_+ \right\}\,.}
We can thus write $\L_f(s)$ in terms of canonical factors as 
\e{fs}{\L_f(s) = {e^{A+Bs}\o s(1-s)} \prod_{\rho\,\in\,\sfS_\rho^{(++)}(f)} \(1-{s\o \rho}\)\(1-{s\o \bar\rho}\)\(1-{s\o 1-\rho}\)\(1-{s\o 1-\bar\rho}\)e^{s(\rho^{-1} + \bar\rho^{-1} + (1-\rho)^{-1} + (1-\bar\rho)^{-1})}}
Evaluating at $s=0,1$ and using the functional equation, the polynomial factor drops out, yielding\foot{Note that, for a non-self-dual $L$-function, $B = -\sum_\rho \Re(\rho)^{-1}$ instead. Nevertheless, one can still write the Hadamard product in the form \eqr{Lamzprod}, although the proof of that fact takes more ingenuity and is not standard material: we are aware only of a MathOverflow post from 2020, containing an apparently novel proof in the eyes of experts \c{370538}.}
\e{}{B = -\sum_{\rho\,\in\,\sfS_\rho^{(++)}(f)}\( \rho^{-1} + \bar\rho^{-1} + (1-\rho)^{-1} + (1-\bar\rho)^{-1}\)}
which reduces precisely to \eqr{Beq}. Applying this to $\Lamz(s)$ implies \eqr{Lamzprod}. \qed 

\ssec{A sum rule for the non-trivial zeros}\label{s43}

We now derive a sum rule for the non-trivial zeros of the $L$-function. The computation is textbook material for standard $L$-functions. 

From Hadamard, the coefficient $B$ can be extracted from the expansion near $s=0$ or $s=1$: 
\e{}{s(1-s)\L_\cZ(s) \approx e^A(1+sB+\ldots)}
For our $L$-function $\lz(s)$,
\e{}{B = {d\o ds}\log(s(1-s)\L_\cZ(s))\big|_{s=0} = {\lz'\o \lz}(0)-\varphi}
where
\e{varphidef}{\varphi := - {d\o ds}\log\(s(1-s)2^s\g_\cZ(s)\) = \log(8\pi^2)+2\g_E-1 \approx 4.52\,.}
We can also use the functional equation to expand around $s=1$, which will be useful later. Developing the expansion for both $\lz(s)$ and $\Lamz(s)$ as
\es{Ls1}{\lz(s\rar 1) &\approx \ell_0\({1\o s-1} + {\ell_1\o \ell_0} + O(s-1)\)\\
\L_\cZ(s \rar 1) &\approx r_0\({1\o s-1} +\({\ell_1\o \ell_0} - \varphi-1\)+O(s-1)\)}
for some constant $\ell_1$, where the residues are related to the modular average $\<\cZ\>$ as
\es{l0r0}{\ell_0 &:= \Res_{s=1}\lz(s) = \sqrt{2}\z(2)\<\cZ\>\\
r_0 &:= \Res_{s=1}\L_\cZ(s) = {2\pi^{3/2}\o 3}\<\cZ\>}
the relation is
\e{}{\varphi-{\ell_1\o\ell_0} = {\lz'\o \lz}(0)-\varphi\,.}
On the other hand, recall from \eqr{Beq} that $B$ is the sum over the reciprocal of all zeros of $\Lamz(s)$. We thus have a sum rule for the non-trivial zeros of $\lz(s)$ in terms of its behavior at the edge of the critical strip:
\ebox{BLreln}{{\sum_\rho {1\o \rho} = {\ell_1\o\ell_0}-\varphi}}

Typically \c{Patterson1988}, the zero sum representation for $B$ is stated to be convergent under the prescription that one pairs the zeros $\rho$ and $\bar\rho$, 
whereupon a sufficient condition for convergence is
\e{rebound}{\underset{\rho}{\text{max}}\,(\Re(\rho))<\i}
thanks to \eqr{orderonezerosum}. Non-trivial zeros of standard $L$-functions are confined to the critical strip, but in our case, we do not know for certain whether this boundedness condition holds.  Nevertheless, $B<\i$. One way to see this is to note that self-duality buys us one order at large $|\rho|$, due to the presence of the dual zeros. In particular, we have
\e{Bquartet}{B = -2\sum_{\rho \,\in\, \sfS_\rho^{(++)}(\lz)} \({\Re(\rho)\o |\rho|^2}+{\Re(1-\rho)\o |1-\rho|^2}\)\,.}
where $\sfS_\rho^{(++)}(\lz)$ is the set of first-quadrant zeros of $\lz(s)$, cf. \eqr{++def}. Convergence of $B$ is controlled by the large $|\rho|$ tail of the sum; in this regime, $|\rho-1| \approx |\rho|$, so actually $B <\i$ for any self-dual $L$-function of order one, regardless of whether \eqr{rebound} holds, because $\sum\limits_\rho |\rho|^{-2}<\i$ by \eqr{orderonezerosum}.

Finiteness of $B$ follows from another, more constructive point of view: namely, from a representation of $\ell_1/\ell_0$ as a Rankin-Selberg integral. Recall the definition \eqr{olap} of $\lz(s)$ via the Rankin-Selberg transform \eqr{peterspeter}. The key point is that since $\cZ(\t)$ has rapid decay as $y\rar\i$, we can expand the Eisenstein series around (say) $s=1$, insert this expansion into the modular integral, and obtain convergent results order-by-order. The expansion of $E_s(\t)$ near $s=1$ was given in \eqr{E1exp}. Carrying this out, the zeroth-order match yields \eqr{l0r0}, while the first-order match yields an expression for $\ell_1/\ell_0$ which, upon plugging into \eqr{BLreln}, gives the following sum rule:
\ebox{BLreln2}{\sum_\rho {1\o\rho} = \<\cZ\>^{-1}\int_\cF{dxdy\o y^2} \cZ(\t) \widehat E_1(\t)+\widehat\varphi}
where 
\e{}{\widehat\varphi := 2{\z'\o\z}(2)+\log8\pi+\g_E-\varphi \approx -1.86\,.}
The integral is finite due to the rapid decay of $\cZ(\t)$ and the regularity of $\widehat E_1(\t)$ for all $\t\in\cF$ away from the cusp (where it grows only linearly). 

\sssec{Where are the zeros?}
We do not know the sign of the zero sum \eqr{BLreln2}, since $\cZ(\t)$ is not \`a priori sign-definite. This turns the sum rule into a useful tool: it can be used to probe the locations of the zeros. 

\begin{figure}[t]
\centering
{
\subfloat{\includegraphics[scale=.6]{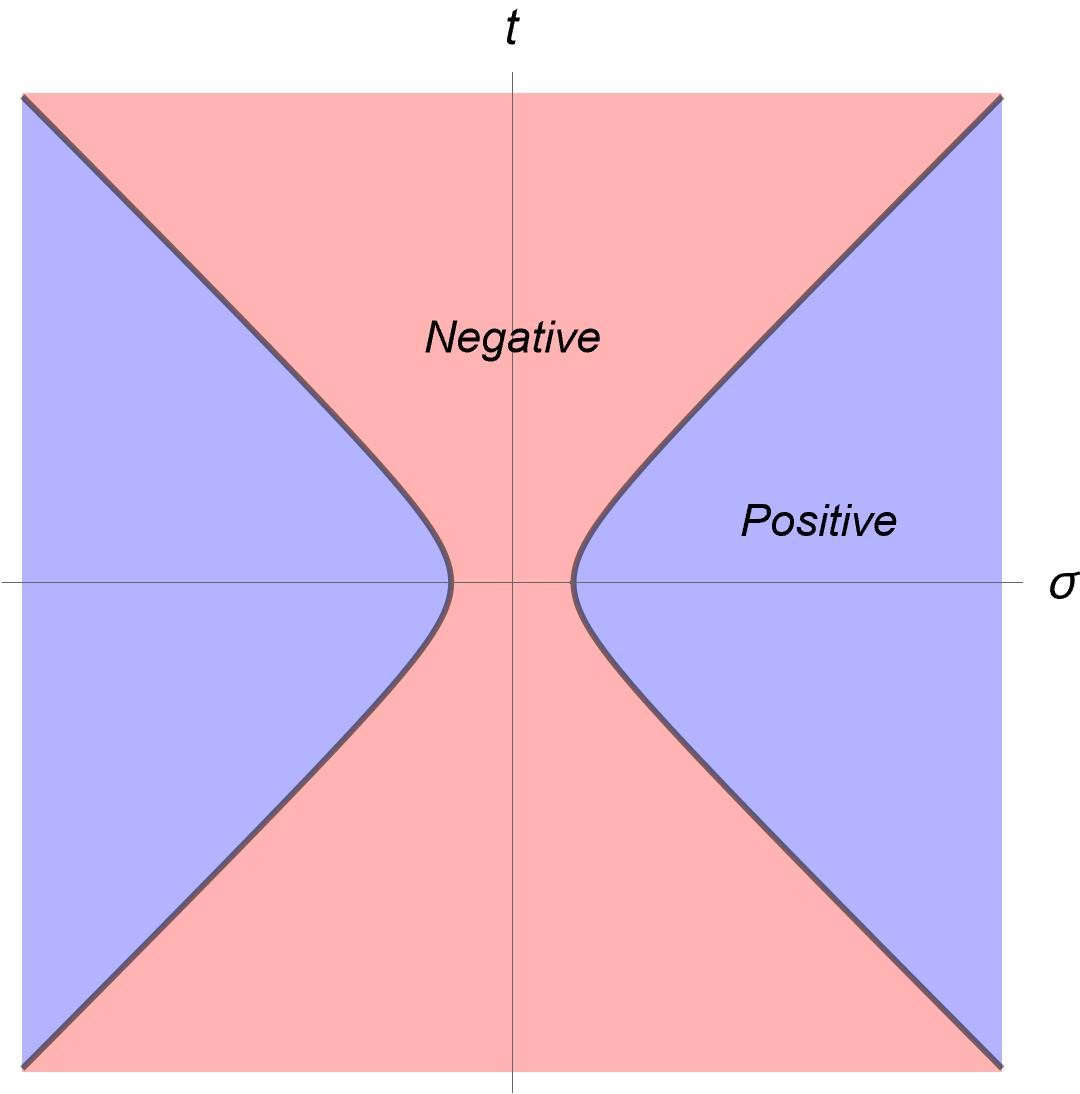}}
}
\caption{The sign of a contribution to $B$ of a full quartet of zeros, determined by the zero locus of the summand of \eqr{Bquartet}. The origin is at $(\s,t)=(\half,0)$. The critical strip lies in the negative region.}
\label{fig:B}
\end{figure}

A quartet of zeros contributes to $B$ as in \eqr{Bquartet}. Writing $\rho=\s+it$, we see that all zeros in the critical strip $\s\in[\half,1]$, and those with small enough $\s$, give negative contributions to $B$, i.e. positive contributions the sum rule (see Figure \ref{fig:B}).\foot{For context, we note that for the Riemann zeta function, $B_\z = \half\log4\pi-\half\g_E-1\approx -0.023$.} Defining a positive region (for $B$)
\e{}{\mathcal{R}_+ := \{s=\s+it : t^2 > |\s(1-\s)|\}}
we have the following result: \textit{if $\rho \notin \mathcal{R}_+~\forall\,\rho$, then $B<0$.} As a condition on the data of the $L$-function,
\e{}{B<0 \quad \Leftrightarrow \quad {\ell_1\o\ell_0} > \varphi}
Equivalently, in terms of the modular integral,
\e{ZBbound}{B<0 \quad \Leftrightarrow \quad{\int_{\cF} {dx dy\o y^2} \cZ(\t) \widehat E_1(\t)}
\gtrsim 1.78  \int_{\cF} {dx dy\o y^2} \cZ(\t)} 
where we have approximated $-\vol^{-1}(\cF)\widehat\varphi\approx 1.78$. This provides a diagnostic for the presence of zeros outside the critical strip, indeed in the entire positive region $\mathcal{R}_+$: \textit{if these inequalities are (or perhaps must always be) violated, then $\exists ~\rho \in \mathcal{R}_+$.} This is an interesting avenue for future work.

To that end, we note a salient feature of the integrand: 
\e{E1hvalues}{\widehat E_1(\t) \geq \widehat E_1\(e^{i\pi/3}\) \approx 1.85\,.}
If $\cZ(\t)$ \textit{were} sign-definite for all $\t\in\cF$ then \eqr{ZBbound} would follow. Given that $\cZ(\t)$ is a difference of partition functions, there is no reason to expect sign-definiteness of $\cZ(\t)$ locally. However, the sign of $\<\cZ\>$ is computing the \textit{average} sign of $\cZ(\t)$: combined with \eqr{E1hvalues}, that mildly suggests a bias in favor of $B<0$.\foot{Let us also note that $\D_\t \widehat E_1(\t) = \vol^{-1}(\cF)$. This means that the modular average may be written as
\e{}{\<\cZ\> = \int_\cF{dxdy\o y^2} \cZ(\t)\D_\t\widehat E_1(\t) = \int_\cF{dxdy\o y^2} \D_\t\cZ(\t)\widehat E_1(\t)}
So the ratio in question is the ratio of modular integrals of $\cZ(\t)$ against $\widehat E_1(\t)$ with and without an insertion of $\D_\t$. Equivalently, it is the ratio $\<\D_\t^{-1}\cZ\>/ \<\cZ\>$, with $\D_\t^{-1}$ defined by formal inversion. Perhaps this could be useful.}

\sec{Analytic structure II: CFT spectrum}\label{s5}

We now turn to analysis of the generalized Dirichlet series (GDS) representation of CFT $L$-functions. We will present some aspects of this series, make contact with the poles and zeros discussed in the previous section, and extract some physical consequences for the CFT data. 

We will have already presented the formal series for $\zz(s)$ in \eqr{zzdef}. It will be useful to introduce distinct notations for $\zz(s)$, $\lz(s)$ and $\qz(s)$,

\e{zlqser}{\zz(s) = \sum_\l {a_\l\o\l^{s-\half}}\,,\qquad \lz(s) = \sum_\nu {a_\nu\o\nu^{s-\half}}\,,\qquad \qz(s) = \sum_\mu {a_\mu\o\mu^{s-\half}}}
with the various quantities obviously related by convolution through the relations \eqr{zzdef} and \eqr{qdef2}.\foot{We have defined the frequencies of $\qz(s)$ such that these relations imply $\mu_1 = 2\l_1$.} In this section we phrase things in terms of the frequencies $\{\l_n\}=\{\D_n-{c-1\o 12}\}$, the shifted scalar conformal dimensions in the spectrum of $\cZ(\t)$. Recall that $\{\l_n\} = \{\l_1,\l_2,...\}$ is an infinite set, where we order $0<\l_1 < \l_2 < \ldots$ and $\l_{n\rar\i} \rar\i$.  

We begin with a general remark. The rigorous formulation of the spin-graded Cardy formula \c{Cardy:1986ie,baursridip} is as follows: for spin-$j$ primaries,
\e{cardydef}{ \rho_{\rm Cardy}^{(j)}(\D) = {2\o 1+\d_{j,0}}{1\o\D}e^{2\pi \sqrt{{c-1\o 3}\D}} + (\text{non-perturbative})}
is the coarse-grained asymptotic density as $\D\rar\i$, 
\e{coarsecardy}{\int_{\D-{\d\o 2}}^{\D+{\d\o 2}} dx \,\rho^{(j)}(x) \approx \Omega\,\rho_{\rm Cardy}^{(j)}(\D)\,,\quad \Omega \in [\d-1,\d+1]}
where $\rho^{(j)}(x)$ is the primary spin-$j$ density. This is derived by $S$-modular transform of the vacuum state, with other light states giving non-perturbative corrections in $\D$. Now, recall that we are studying a unitary CFT partition function $Z_p(\t)$, regularized into $\cZ(\t)$. Where is the central charge hiding in $\cZ(\t)$? Naively, $\cZ(\t)$ does not know about the central charge: the vacuum and all light states, and hence their Cardy modular images at high frequency, have been removed by subtraction. Correspondingly, the frequency support of $\zz(s)$ starts at $\l_1>0$. However, because \eqr{coarsecardy} is coarse-grained, the Cardy density still sets the \textit{average inverse level spacing} at $\D\rar\i$\thinspace---\thinspace even though the integrated density no longer grows exponentially. Focusing on scalars, define the \textit{nearest-neighbor spacing} between scalar dimensions as
\e{deldef}{\d_n := \D_{n+1}-\D_n}
where $n\in\ZZ_+$. Then Cardy implies
\e{}{\qquad\qquad\qquad\quad\<\d_{n}\>^{-1} \approx \rho_{\rm Cardy}^{(0)}(\D_n)\qquad \quad (n\rar\i)}
where $\<\cdot\>$ may be defined as a statistical average within $O(1)$ windows. Therefore, the central charge may be extracted from the asymptotic level spacings as
\e{}{c = 1+{3\o(2\pi)^2}\lim_{n\rar\i}{\log^2\<\d_n\>\o \D_n}\,.}
This alternative, \textit{UV definition} of central charge\thinspace---\thinspace different from the usual, \textit{IR definition} as a ground state energy on the cylinder\thinspace---\thinspace is valid for any unitary compact CFT. But it becomes especially pertinent for subtracted quantities like $\cZ(\t)$, and in turn $\zz(s)$, where the ground state has been removed. 

\ssec{Dirichlet vs. Hadamard: A global zero density bound}\label{s51}

As we have already seen from equating the Rankin-Selberg and Hadamard representations of $\lz(s)$, it is useful to equate the different representations of our $L$-function with one another. Pitting the series representation of $\lz(s)$ against the Hadamard representation probes the interplay of the spectrum with the zeros:
\e{}{\textsf{(Dirichlet)}\qquad \l \quad \longleftrightarrow \quad \rho \qquad \textsf{(Hadamard)}}
We take the opportunity to (re-)emphasize an overarching point of this work: \textit{the spectrum $\{\l_n\}$ is constructible from the zeros.} And as noted in Section \ref{s32}, if one chooses to define $\cZ(\t)$ by $J$-subtraction, one can filter the actual CFT frequencies of $Z_p(\t)$ from the spurious frequencies of $Z_p^{(J)}(\t)$ that enter $\{\l_n\}$ as byproducts of regularization. 

One instance of this interplay allows us to derive a quantitative bound on the \textit{number} of non-trivial zeros in terms of the \textit{spectral gap} $\l_1$.

\result{Global zero density bound:}{Let $N_{\lz}$ and $N_\z$ be the respective cardinalities of non-trivial zeros of $\lz(s)$ and $\z(s)$ inside a disc of radius $T\gg1$ centered at $s=\thalf$. Then the spectral gap $\l_1$ is correlated with the number of non-trivial zeros as
\ebox{gdb}{\log\bigg({N_{\lz} \o N_\z}\bigg) > 0 \quad \Longleftrightarrow \quad \l_1 < \half}
}

To show this\foot{The name ``global zero density bound'' originates from zero density theorems for standard $L$-functions, which bound the number of zeros within the critical strip as a function of $\s$. The result \eqr{gdb} bounds the number zeros of $\lz(s)$ globally, without input (nor output) as to their locations. For progress in zero density theorems for the Riemann zeta function, see \c{GuthMaynard2024}.}, it is convenient to use the $L$-quotient $\qz(s)$. Its Weierstrass representation follows from its definition combined with the Hadamard factorizations of $\Lamz(s)$ and $\L(s)$. The former was given in \eqr{Lamzprod}, with zeros grouped in quartets eliminating the exponentials in the canonical factors. Dropping an overall constant,
\e{eq59}{\qz(s) \propto  \(s-\thalf\)^2\prod\limits_{\rho\,\in\,\sfS_\rho(\lz)}\(1-{s\o\rho}\)\prod\limits_n \bigg(1-{s\o{\rho_n\o2}}\bigg)^{-1}\bigg(1-{s\o1-{\rho_n\o2}}\bigg)^{-1}\,.}
The first factor runs over the non-trivial zeros of $\lz(s)$, while the second, recpirocal factors run over the non-trivial zeros of $\z(2s)$ and $\z(2s-1)$, respectively. 

We now take $s\rar\i$. Dropping an overall constant,
\e{qasymp}{\qz(s) \sim \mu_1^{-s} \qquad (s\rar\i)\,.}
On the other hand, from its Weierstrass representation, the $s\rar\i$ limit gives a signed count of the number of terms in \eqr{eq59} for which $s$ is much larger than the magnitude of the zero. This means that the sign of $\log\mu_1$ is correlated with the relative number of zeros of $\lz(s)$ and $\z(s)$: if $\qz(s)$ grows (say) at large $s$, there must be more terms in the numerator than the denominator that give a non-negligible contribution. We can define the number of zeros following standard definitions in number theory as follows. Take $N_L(T)$ to be the cardinality of non-trivial zeros of an $L$-function $L(s)$ inside a disc of radius $T$,
\e{}{N_L(T) := \#\{\rho: \L_L(\rho)=0\,,~ |\rho-\thalf|\leq T\}\,.}
The global cardinality of zeros is defined by the large $T$ asymptotic, here implemented by taking large $s$.\foot{From Jensen's theorem, order one functions obey $N(T) \sim \a T\log T$ to leading order in large $T$, for some constant $\a\in\RR_+$. Since the $s\rar\i$ asymptotic \eqr{qasymp} is exponential, not factorial, on account of the cancelling gamma factors, the spectral gap controls a subleading correction to this. The correction can be shown to be of the form $N_{\lz}(T) - N_\z(T) \sim -T\log \mu_1$ times a positive constant.} Noting that $\mu_1 = 2\l_1$ yields \eqr{gdb}. \qed

A sanity check may be done via an analogous computation for a ratio of (completed) standard $L$-functions $\L_i(s)$ which have the same gamma factor $\g_i(s)$, but different conductors $q_i$. On the Dirichlet side, the role of $\mu_1$ is played by $q_2/q_1$. On the zeros side, it follows from standard zero counts (e.g. \c{IKtext}, Theorem 5.8) that $N_{L_1}(T) - N_{L_2}(T) \approx T\log ({q_1/ q_2})$ to leading order in large $T$. (For standard $L$-functions one usually defines the cardinality of zeros inside a long rectangle $|t|\leq T, |\s|\leq\s_*$ for some $\s_*$ of $O(1)$, which is equivalent to the disc count when the zeros are confined to the critical strip.)

\ssec{Dirichlet vs. Hadamard II: Zero sum rule reconsidered}\label{s52}

Recall \eqr{BLreln}, our sum rule for the non-trivial zeros of $\lz(s)$ in terms of its expansion near $s=1$. In a moment, we will show that when the GDS for $\lz(s)$ converges, $s=1$ sets the boundary of convergence. This means that \eqr{BLreln} may be expressed in terms of a partial sum over the frequencies $\{\nu_n\}$ of $\lz(s)$: the expression for a general series is
\e{}{\ell_1 = \lim_{N\rar\i} \(-\ell_0\log \nu_N + \sum_{n=1}^N {a_n\o \sqrt{\nu_n}}\)\,.}
Therefore,
\e{BLreln4}{\sum_\rho {1\o \rho}  = \lim_{N\rar\i} \({1\o\ell_0}\sum_{n=1}^N {a_n\o \sqrt{\nu_n}}-\log \nu_N \)-\varphi}
This is a sum rule for the zeros in terms of the CFT spectrum, with the divergence subtracted. One can study this relation numerically. However, in Section \ref{s8} we will derive a better, manifestly convergent, version of this sum rule. 

\ssec{Convergence}\label{s53}

We now turn to the topic of convergence. The main obvious questions here are whether, where, and to what degree the series converges; and whether the answers to those questions are subject to certain necessary or sufficient conditions on the underlying CFT data.\foot{See \c{hrtext} for a classic reference on generalized Dirichlet series.}

Any GDS admits an abscissa of simple convergence, denoted $\s_c$, defined as the smallest $\Re(s)$ to the right of which the series converges: if $f(s)$ admits a convergent series
\e{fser}{f(s) = \sum_n {a_f(n)\o \l_n^{s}}<\i ~~\forall\,\s>\s_c(f),}
then
\e{scdef}{\s_c(f) := \min\(\s : \sum_n {a_f(n)\o \l_n^{\s+it}}<\i~~\forall\,t\)\,.}
It is a general theorem about GDS (see e.g. Chapter II of the useful reference \c{hrtext}) that $\s_c(f)$ is equal to the real part of the location of the right-most pole. Let us also note another useful property of convergent series, namely, a \textit{uniform growth bound} high on vertical lines within the half-plane of convergence: as $|t|\rar\i$ for fixed $\s>\s_c(f)$, there exists some $C(\s)<\i$ such that
\e{bound}{|f(\s+it)| \leq C(\s)(1+|t|) ~~\forall\,|t|\geq 0\,.}
We prove this result in Appendix \ref{appc}. 

We now turn to $\zz(s)$. Recall from Section \ref{s4} that the Rankin-Selberg representation of $\zz(s)$ dictates that its right-most pole is at $s=1$. Therefore, if the GDS for $\zz(s)$ converges for \textit{any} finite $\s$, it converges all the way down to $\s=1$: in other words,
\e{sczz}{\s_c(\zz) = 1~\text{or}~\i\,.}
And by $\lz(s) = \z(2s)\zz(s)$ and absolute convergence of $\z(2s)$ for $\s>\thalf$,\foot{Recall that, given a product $f(s) = L(s) g(s)$ for which $g(s)$ converges conditionally and $L(s)$ converges absolutely, $f(s)$ converges conditionally with $\s_c(f) = \max(\s_c(g),\s_a(L))$.}
\e{}{\s_c(\lz) = \s_c(\zz)\,.}
To be clear, $\lz(s)<\i$ for all $\s>1$ regardless of the value of $\s_c(\lz)$, cf. Sections \ref{s3}--\ref{s4}. 

We can now address the question of whether $\zz(s)$ converges. The first, preliminary remark is that $\zz(s)$ does not, in general, converge \textit{absolutely}: that is, $\s_a(\zz)=\i$, where $\s_a$ is the abscissa of absolute convergence, defined as in \eqr{scdef} but with absolute values inside the summand. This is because $a_\l = d_\D - \tilde d_\D$ flips sign very rapidly, but each degeneracy individually has coarse-grained Cardy growth as $\D\rar\i$: if we wash out the signs, the exponential growth of the sum (barring exceptional cancellation in the difference) prevents absolute convergence for any finite $\s$. (The picture to have in mind is when $\{\D_n\}$ and $\{\tilde\D_n\}$ are non-overlapping up to some small possible subset, such that $|a_\l| = d_\D + \tilde d_\D$ for most states, giving twice Cardy growth.)

On first pass one would expect conditional convergence of the series representation of $\zz(s)$, precisely because the Cardy growth of $Z_p(\t)$ and $\tilde Z_p(\t)$ cancel one another: the Cardy density \eqr{cardydef} is derived precisely by $S$-modular transform of the vacuum state, but we have removed all light states in the difference $\cZ(\t)$. However, this intuition is too quick: the Cardy growth is only guaranteed to cancel \textit{on average}, but this is not a sufficient condition on the $\{a_\l\}$ to guarantee conditional convergence of the series. \textit{What is the mathematical condition on the series coefficients $\{a_\l\}$ for convergence, and how does it translate into a physical condition on the spectrum of the underlying CFT?}

Luckily, we can answer this by again appealing to well-known formulas in the general theory of GDS \c{hrtext}. A positive abscissa of convergence $\s_c(f)$ of a series \eqr{fser} is determined by the growth of partial sums over expansion coefficients as \c{hrtext, Titchmarsh1939}
\e{abscissa}{\s_c(f) = \underset{N\rar\i}{\lim\sup}\,{\log A_f(N)\o \log\l_N}\,,\qquad A_f(N) := \left|\sum_{n=1}^N a_f(n)\right|\,.}
Adapted to our series $\zz(s)$, which includes a shift by $-\thalf$ in its definition \eqr{zlqser}, \eqr{sczz} implies that if the series representation of $\zz(s)$ converges, then $\s_c(\zz)=1$ and
\e{ANsqrt}{A_{\zz}(N) \sim \sqrt{\l_N}}
as $N\rar\i$. Thus, convergence holds if and only if there are no oscillations outside a square root envelope out to arbitrarily large $N$. If instead the series does not converge, then $\s_c(\zz)=\i$ and $A_{\zz}(N)$ has oscillations whose magnitude grows faster than any polynomial. 

This has two important implications. First, the square root \eqr{ANsqrt} admits a tidy CFT interpretation. If we treat $A_{\zz}(N)$ in an approximate sense by replacing the sum by an integral and treating $a_\l \rar a(\l)$ as a smooth density, \eqr{ANsqrt} implies the coarse-grained behavior
\e{liouenvelope}{a(\l) \sim {1\o \sqrt{\l}}}
as $\l\rar\i$. This is, up to an overall constant, the \textit{Liouville} CFT primary density. So we learn that assuming convergence, the coarse-grained scalar spectra of two CFTs with identical light spectra differ by a multiple of Liouville at high energies. 

The second, more important consequence of \eqr{abscissa}--\eqr{ANsqrt} is the \textit{spectral} condition underlying convergence, which we give a nickname: 
\vs
\centerline{\textit{$\zz(s)$ converges if and only if it obeys No Clustering.}}
\vs
\ni By \textit{clustering} we mean that many frequencies bunch together with the same sign of $a_\l$, such that $A_{\zz}(N)$ has a large ``spike'' as we increase $N$. Convergence is the condition that spikes of magnitude $\gg \sqrt{\l}$ do not persist out to $\l\rar\i$. 

\begin{figure}[t]
  \centering

  \begin{minipage}{\textwidth}
    \centering
 \includegraphics[width=.53\textwidth]{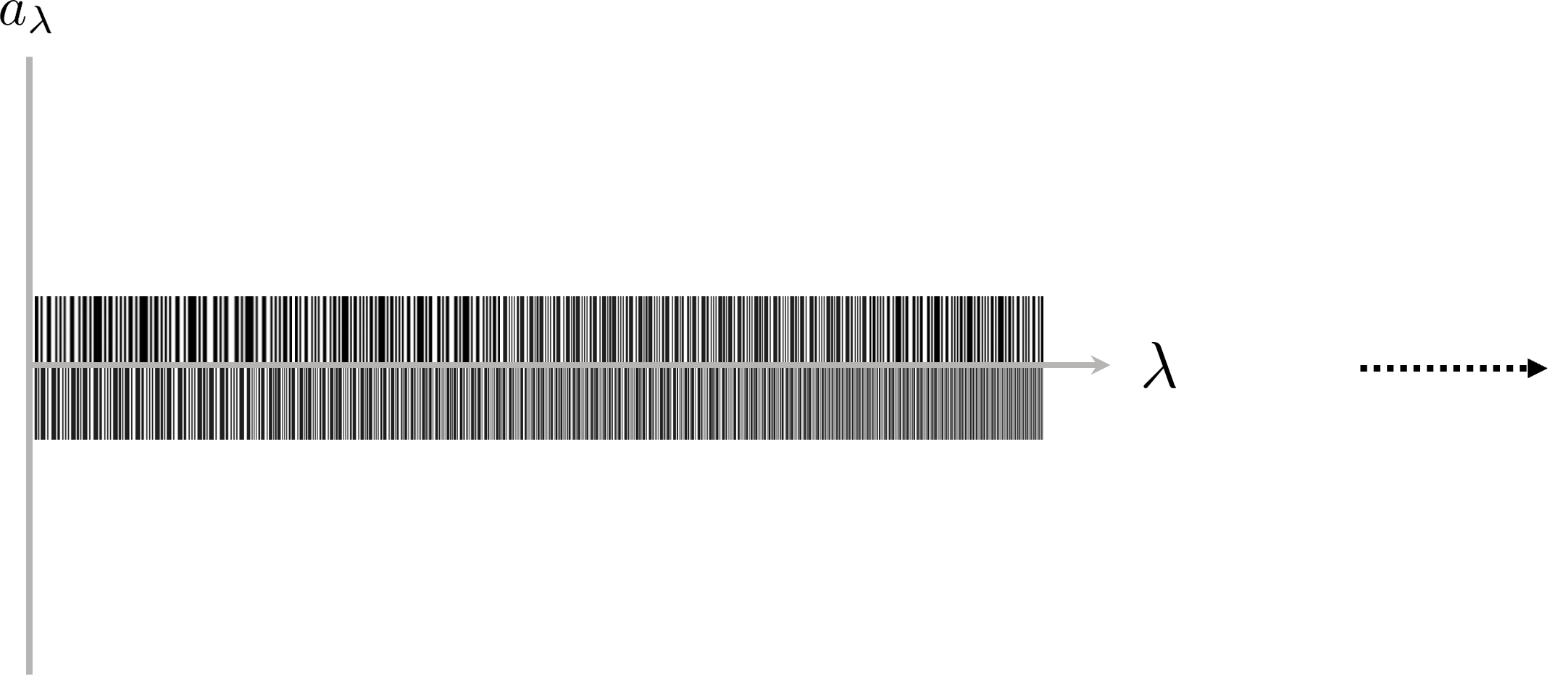}  \hfill
     \raisebox{-0.25cm}{\includegraphics[width=.42\textwidth]{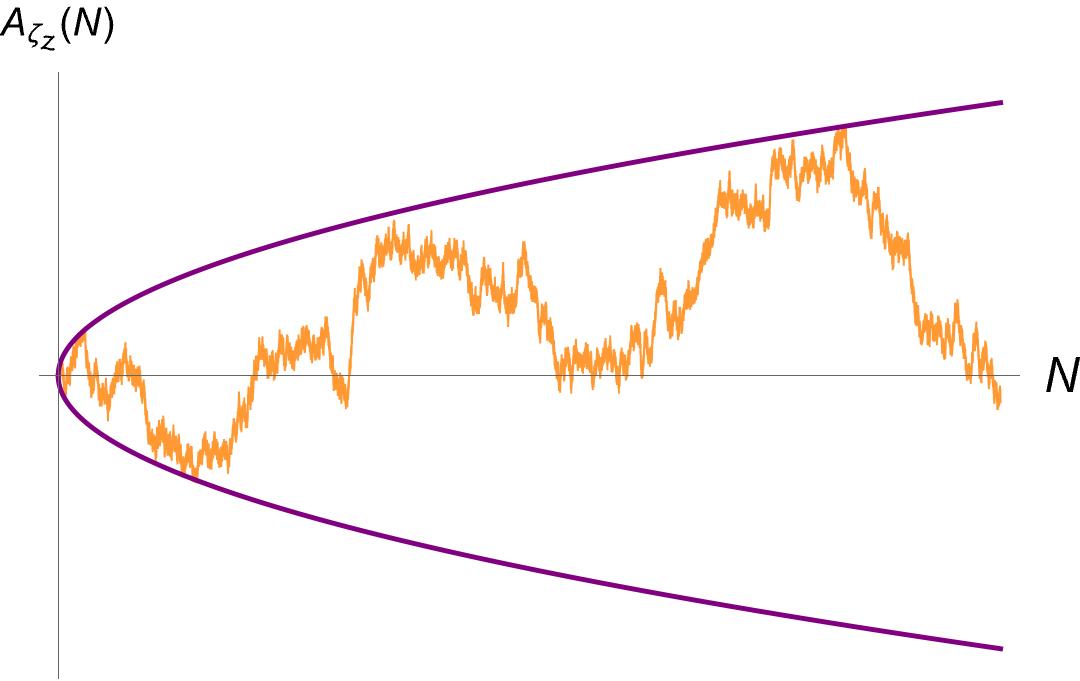}}

    \par\vspace{0.5ex}
    \parbox[t]{\textwidth}{\textit{No Clustering:} erratic oscillations generate pseudorandom behavior, the growth of partial sums is confined to a square root envelope, and $\s_c(\zz)=1$.}
  \end{minipage}

  \vspace{2ex}

  \begin{minipage}{\textwidth}
    \centering

       \includegraphics[width=.53\textwidth]{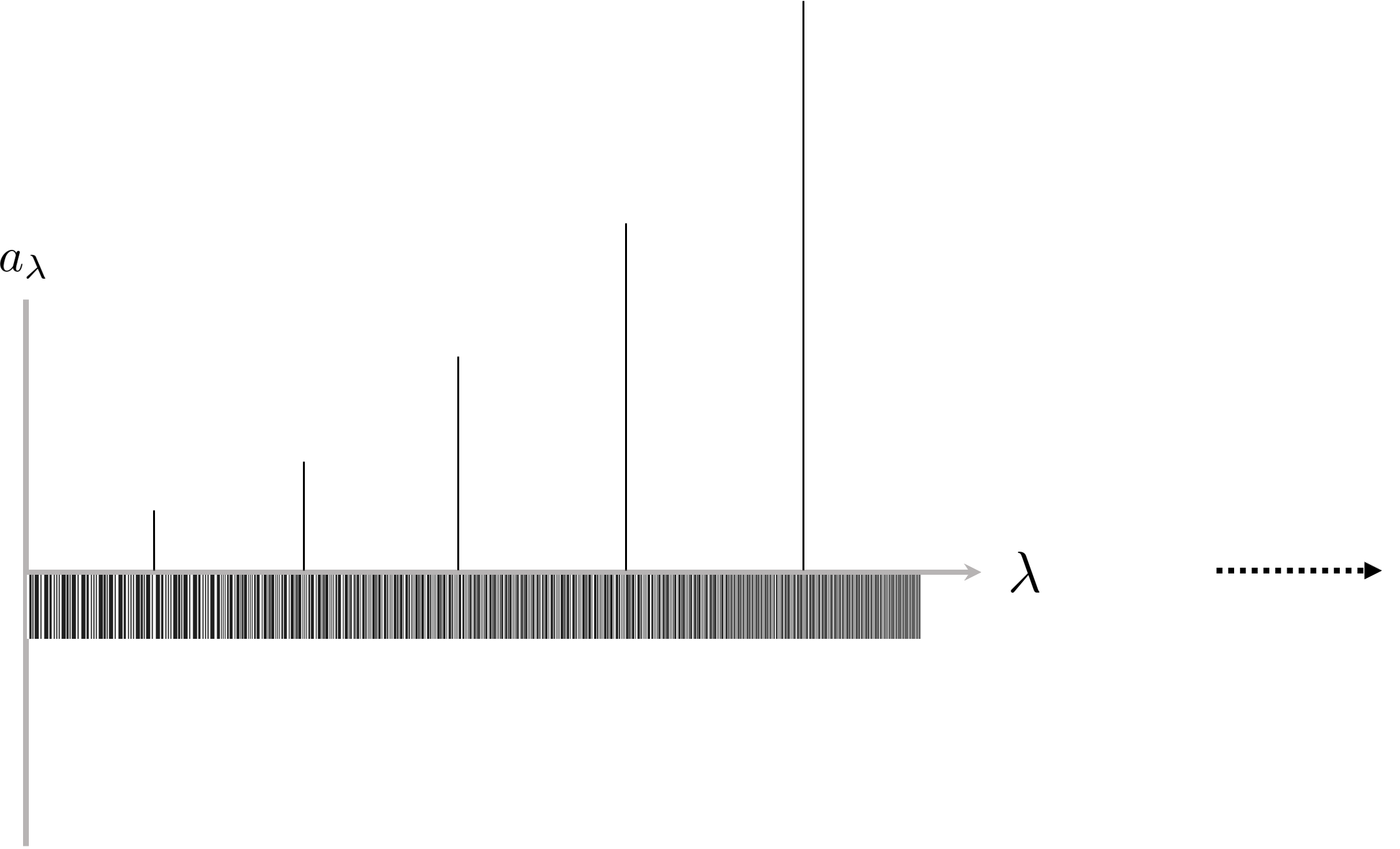} \hfill
   \raisebox{-0.25cm}{\includegraphics[width=.42\textwidth]{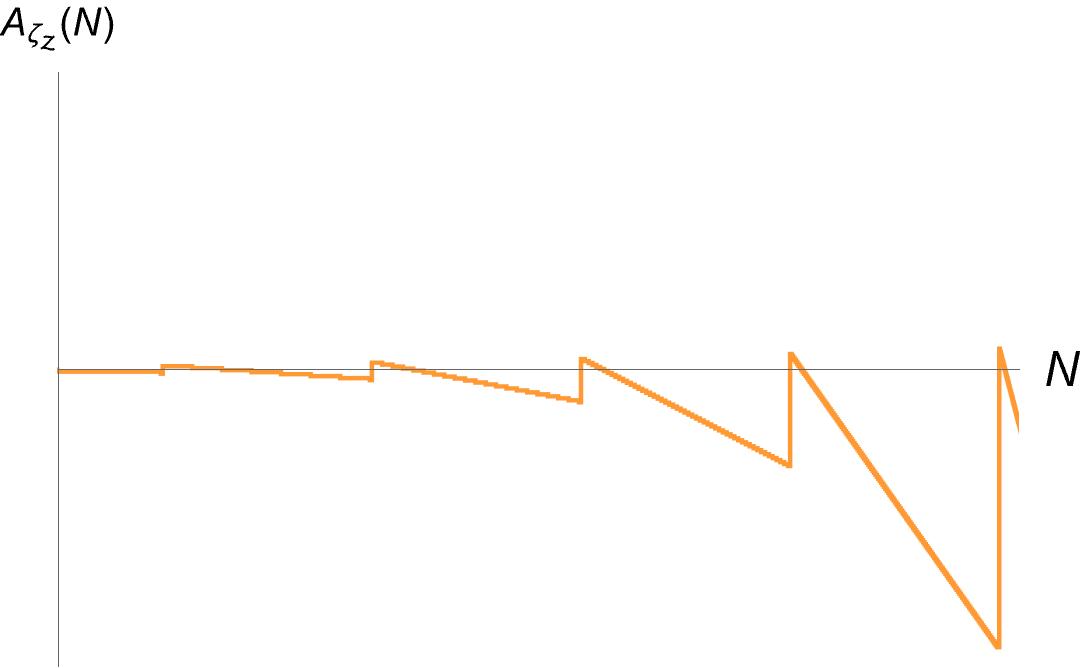}}

    \par\vspace{0.5ex}
        \parbox[t]{\textwidth}{\textit{Clustering:} partial sums oscillate exponentially due to persistent unbounded suprema as $\l\rar\i$, and $\s_c(\zz)=\i$. (Exponential spikes on the left not drawn to scale.)\vskip .05in}

  \end{minipage}

  \caption{Two scenarios for frequency spectra of $\zz(s)$.}
  \label{fig:nc}
\end{figure}

The CFT interpretation of this is transparent. In forming $\cZ(\t) = Z_p(\t) - \tilde Z_p(\t) \in L^2(\cF)$, there are both positive and negative coefficients $\{a_\l\}$. By the absence of light states in $\cZ(\t)$, we have guaranteed that averaging over any frequency window of size $\d=O(1)$ centered upon some $\l\rar\i$ generates large cancellations between positive and negative Cardy densities; however, the density of, say, $Z_p(\t)$ could in principle be concentrated on a parametrically smaller frequency window, of $\d = o(1)$. If that behavior persists out to $\l\rar\i$, parametrically (exponentially!) large spikes in $A_{\zz}(N)$ will spoil convergence. These two scenarios are best understood visually, as depicted in Figure \ref{fig:nc}.

An extreme example of clustering is when the Cardy density within a given $O(1)$ energy band is localized at a single frequency. This is a property of all chiral CFT partition functions when treated as \textit{Virasoro} partition functions; of the $J$-completion $Z_p^{(J)}(\t)$; and of any partition function of a CFT with an extended chiral algebra $\cA \supset \text{Virasoro}$, which, due to the extra $\eta(\t)$ factors, have a pointwise infinite \textit{Virasoro} primary density at $\l\rar\i$.

\ssec{Convergence as random gap statistics}\label{s54}

The fact that \textit{Convergence $\Longleftrightarrow$ No Clustering} is very much welcome, because we are ultimately interested in studying CFTs whose spectra do not cluster: these are CFTs deep in the heart of theory space, exhibiting chaotic spectral statistics. This subsection is devoted to explaining how the convergence condition follows from random matrix behavior of $\cZ(\t)$. (The reader interested only in formal aspects of $L$-functions may freely move on.)

It is important that the reader understand the difference between this subsection and the previous one. Here, we are providing a \textit{physical interpretation} of the previous \textit{mathematical condition} for convergence; that condition, which we nicknamed ``No Clustering'', made no use of any CFT principles or other assumptions. What we will observe below is that this condition \textit{also} follows from the imposition of random matrix dynamics among CFT primary states\thinspace---\thinspace in particular, the \textit{random statistics of extreme gaps}. This will naturally raise some questions about random matrix universality in 2d CFTs which, being tangential to the convergence question, we defer to Section \ref{s7}.   

\sssec{Level spacings in RMT}

\begin{figure}[t]
\centering
{
\subfloat{\includegraphics[scale=.6]{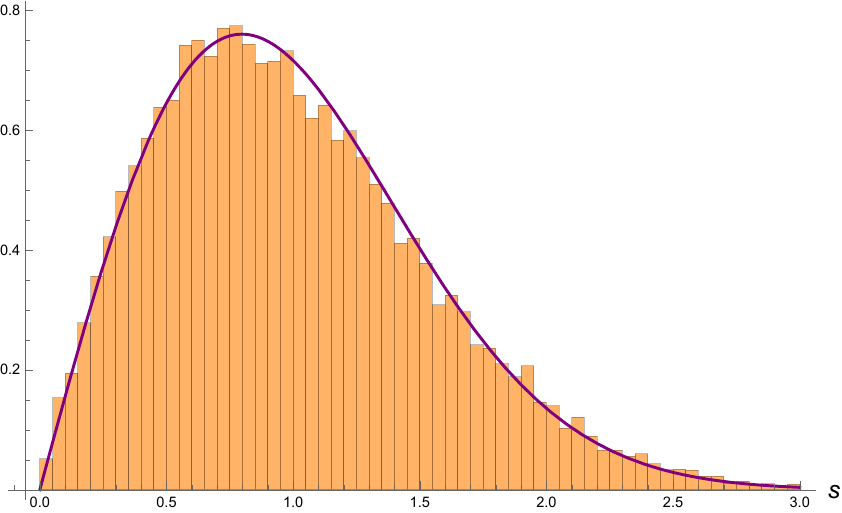}}
}
\caption{Distribution of unfolded nearest-neighbor spacings $s$ for 1000 draws of 50\,$\x$\,50 matrices from a GOE random matrix ensemble, plotted against $P_{\rm Wigner\, GOE}(s) = {\pi \o 2} s e^{-\pi s^2/ 4}$.}
\label{fig:goe}
\end{figure}

An $N\x N$ random matrix with eigenvalues $\{\l_n\}$ has nearest-neighbor spacings $\d_n = \l_{n+1}-\l_n$, with $n=1,\ldots,N-1$. Introduce the ``unfolded'' nearest-neighbor spacings of unit mean, conventionally denoted
\e{}{s := {\d\o \<\d\>}}
where $\<\d\>$ is the inverse local mean density; in the bulk eigenvalue spectrum, $\<\d\> \approx 1/N$. Define
\e{}{P_{\rm RMT}(s) := \text{Probability distribution for unfolded nearest-neighbor spacing $s$}}
%
where ``RMT'' refers to Gaussian random matrices. At short range, Gaussian random matrices enjoy level repulsion:
\e{lrsrrmt}{\textsf{Level Repulsion:}\qquad \lim_{s\rar 0}~\,P_{\rm RMT}(s)=0\qquad\qquad\qquad}
The probability distribution $P_{\rm RMT}(s)$ is known to be well-approximated for matrices of arbitrary size by Wigner's surmise:
\e{wigner}{P_{\rm RMT}(s) \approx P_{\rm Wigner}(s) =  c_\b s^\b e^{-n_\b s^2}}
where $\b=1,2,4$ selects the symmetry class (GOE, GUE, GSE, respectively), and $c_\b$ and $n_\b$ are positive constants \c{mehta04}. See Figure \ref{fig:goe}. In particular, note the power law suppression of small gaps, and the Gaussian suppression of large gaps, reflecting long-range correlations \`a la spectral rigidity. 

Whereas $P_{\rm RMT}(s)$ gives the probability that a typical unfolded spacing is $s$, asking for the minima and maxima over all spacings in a given draw is a different question. At large $N$, \textit{extreme gaps}\thinspace---\thinspace atypically small or large nearest-neighbor spacings, living in the tails of $P_{\rm RMT}(s)$\thinspace---\thinspace are more likely to occur. Let $\{s_n\}$ be the set of unfolded nearest-neighbor gaps, and let $s_{\rm min}$ and $s_{\rm max}$ be its minimum and maximum, respectively. A rigorous computation of their $N$-scaling (for eigenvalues in the bulk spectrum) was done in a set of nice papers spanning multiple random matrix universality classes \c{vinson, BAB,feng2019small,feng2020smallgapscircularbetaensemble, fengsmallgapsgse, bourgade2021extreme,fenglargegapsgue}, with the result
\e{sminsmax}{s_{\rm min} \sim N^{-{1\o \b+1}}\,,\quad s_{\rm max} \sim \sqrt{\log N}}
in probability as $N\rar\i$. In fact, they prove much more than this: they derive the limiting \textit{distributions} of the set of extreme gaps. 

Let us first focus on the small gaps. To understand their statistics, first define the further rescaling $\tau_n := N^{{1\o \b+1}}s_n$ to zoom into the extreme small gap region.\foot{We have now done two rescalings of the physical spacings $\d_n$: one on the scale of typical spacings $\sim N$ of the entire set (``unfolding''), and one on the scale of the extreme small unfolded gaps $\sim N^{{1\o \b+1}}$.} The result of \c{BAB, feng2019small,feng2020smallgapscircularbetaensemble, fengsmallgapsgse} is that at large $N$, the distribution of the gaps $\{\t_n\}$ converges to a Poisson point process, uniformly across the bulk of the eigenspectrum. Denoting $\t^{(k)}$ as the $k$'th smallest gap among all $\{\t_n\}$, the local size distribution is\foot{In a slight abuse of notation we henceforth use $P_{\rm RMT}(X)$ to mean ``the probability of $X$ happening'', not just the nearest-neighbor spacing distribution.}
\e{smallgapdist}{\lim_{N\rar\i} P_{\rm RMT}(\t^{(k)} \in I) \propto {1\o (k-1)!}\int_I dx \,x^{k(\b+1)-1}e^{-x^{\b+1}}}
up to an overall $\b$-dependent constant. So in a large $N$ random matrix, the set of extreme small gaps $s_n\sim s_{\min}$ forms a randomly-scattered $O(N^0)$ subset of the $N-1$ total gaps, with Poisson-distributed count and independent positions within the spectrum.

What is more, \c{BAB,feng2019small} prove the \textit{absence of successive extreme small gaps}. They phrase this neatly by first defining two point processes: one for nearest-neighbor gaps, and one for gaps among all levels, not just nearest neighbors. They then prove that within any bounded interval, their difference converges in law to zero (cf. \c{BAB} Lemma 2.4, \c{feng2019small} Lemma 8). Let
\e{}{\t_{n,k} := N^{{\b+2\o \b+1}}(\l_{n+k}-\l_n)}
be the rescaled $k$'th nearest neighbor spacings with respect to $\l_n$, where $k=1,\ldots,N-n$ and $\t_{n,1} = \t_n$. Then defining the two measures
\e{}{\tilde \chi^{(N)} := \sum_{n=1}^{N-1}\sum_{k=1}^{N-n} \delta_{\t_{n,k}}\,,\quad \chi^{(N)} := \sum_{n=1}^{N-1} \delta_{\t_n}}
the result is that $\tilde \chi^{(N)} - \chi^{(N)} \rar 0$ in probability as $N\rar\i$: in other words, the only gaps that are parametrically of extreme size $\sim s_{\rm min}$ are between nearest neighbors ($k=1$). This is precisely a statement of No Clustering. Indeed, it is a very strong version thereof: the likelihood of having even two adjacent extreme small gaps is zero (not to mention an $N$-dependent number of them). We can phrase this informally as 
\e{NCinformal}{\textsf{No Clustering:}\qquad \lim_{N\rar\i} P_{\rm RMT}(\text{successive extreme small gaps}) = 0}
and more quantitatively as
\e{}{\,\textsf{No Clustering:}\qquad \lim_{N\rar\i} P_{\rm RMT}(\d_{n}+\d_{n+1} \sim \d_{\min}) = 0~~\forall\,n \qquad\qquad }
where the extreme small gap scale is 
\e{}{\d_{\rm min} \sim N^{-{\b+2\o \b+1}}}
in physical units. 

There are also results in \c{vinson, BAB, fenglargegapsgue} for extreme large gap distributions, though we will not make use of them here; the distribution of large gaps is not Poissonian because it mixes with the long-range correlations (spectral rigidity), but is universal nevertheless. The only statement we wish to highlight at large scales is simply the scaling: in \textit{absolute} terms, random matrices have
\e{}{\textsf{No Large Gaps:}\qquad \d_n  \lesssim \d_{\rm max} \sim {\sqrt{\log N}\o N}}
Extreme large gaps are enhanced with respect to the average gaps, but only by a measly log. In particular, there are \textit{no} gaps of $\d_n = O(1)$, or even gaps suppressed by a smaller power of $1/N$: every gap decays parametrically. 

Altogether, if we define an entropy $S := \log N$, bulk physical gaps obey the parametric probabilistic bounds
\e{smalllargeS}{\qquad e^{-{\b+2\o \b+1}S}\lesssim \d_n \lesssim {\sqrt{S} e^{-S}}~~\forall\,n}
in the large $N$ limit, with no clustering at extreme scales.\foot{The results quoted above do not technically rule out clustering at intermediate scales between the extreme smallest gaps and the average gap. A heuristic interpolation between extreme gap statistics and spectral rigidity suggests suppression at large $N$.}

\sssec{Back to convergence}

We now pass back to the convergence of the spectral zeta series $\zz(s)$. The essential connection is hopefully obvious: the No Clustering condition required for convergence is an emergent property of random matrix dynamics. Conversely, a sufficient condition for convergence of the series representation of $\zz(s)$ is that its frequency spectrum obeys random extreme gap statistics. More precisely, recall that $\cZ(\t)$ has scalar states with degeneracies of both signs:
\e{}{\{\l_n\} = \{\l_n\}_+ \cup \{\l_n\}_-\,,\quad \text{where}\quad \{\l_n\}_\pm := \{\l_n : \text{sgn}(a_n) = \pm\}\,.}
Denote their associated nearest-neighbor gaps as $\{\d_n\}_\pm$. Then a sufficient condition for convergence is that the $\{\d_n\}_{\pm}$ exhibit No Clustering. The sign grading is necessary to guarantee the cancellations, as in the top of Figure \ref{fig:nc}. 

Let us make a closing comment. What one would really like is to view this property of the CFT level spacings $\{\d_n\}$ not just as a statistical statement about a sequence of numbers, but as a physical statement about the underlying theory. In other words, we want to properly embed random extreme gap statistics in a 2d CFT context. This subsequently forms a bridge to an interesting open question: what is the definition of random matrix universality in 2d CFT? We return to this topic in Section \ref{s7}.

\sec{Square root cancellation and AdS EFT: What's left in the UV?}\label{s6}
In this section we interpret some of our formal results so far in physical terms for 2d CFT and AdS$_3$ quantum gravity, in the language of UV-IR relations and effective field theory (EFT). When we subtract two CFT partition functions with the same light spectra, how much of the heavy spectra remain? In gravitational terms, we are asking about the ``wiggle room'' in the black hole Hilbert space: upon fixing the gravitational path integral to generate a given EFT at low energies, what's left in the UV? 

Two properties of the spectral zeta function $\zz(s)$ indicate very strong constraints. 

First, the trivial zeros of $\zz(s)$,
\e{}{\zz\Big(\half-p\Big)=0~~\forall\,p\in\Z_{\geq 0}\,,}
imply that all regularized polynomial moments of the spectral density \textit{vanish}:
\e{vanmoments}{\sum_\l a_\l \l^p\Big|_{\rm reg.}=0~~\forall\,p\in\Z_{\geq 0}\,.}
We have indicated explicitly that this sum is to be understood in a regularized sense, familiar from computations of Casimir energy in quantum field theory. In simple terms, the regularized spectral density of $\zz(s)$ \textit{vanishes on average:} when we subtract partition functions with the same light spectra, there is almost nothing left. 

Second, analyticity of $\zz(s)$ for $\s>1$ means
\e{src}{\sum_\l {a_\l\o \l^{\half+\eps+it}}<\i~~\forall\,\eps>0\,.}
Recall that, by theorems for generalized Dirichlet series, this is equivalent to
\e{}{\left|\sum_{n\leq N} a_n\right| \sim \sqrt{\l_N}\qquad (N\rar\i)\,.}
This is the statement that the $\{a_\l\}$ exhibit \textit{square root cancellation}: the sum enjoys a square root savings due to fluctuations of the coefficients, which apparently behave as independent and identically distributed random variables. Square root cancellation is a central concept in analytic number theory, observed many times over and often invoked to heuristically justify apparent-but-unproven cancellations. To convey its importance, one need only note that both the Lindel\"of and Riemann Hypotheses are equivalent to forms of square root cancellation, the former as
\e{}{\sum_{n\lesssim \sqrt{t}} n^{-\half+it} = O(t^{\eps})~~\forall\,\eps>0\qquad (t\rar\i)}
and the latter as 
\e{rhmobius}{{1\o \z(s)} = \sum_{n=1}^\i {\mu(n)\o n^s} <\i~~\forall\,\s>\half\,,}
equivalently,
\e{}{\left|\sum_{n\leq N} \mu(n)\right| = O(N^{\half+\eps})~~\forall\,\eps>0\qquad (N\rar\i)}
where $\mu(n)$ is the M\"obius function. This is sometimes called pseudorandomness, and forms part of the M\"obius Randomness Conjecture due to Sarnak \c{mobiusLecturesIAS}: this is the idea that $\mu(n)$ is sufficiently (pseudo)random so as to be orthogonal to any deterministic sequence, in a sense that can be made precise.

\begin{figure}[t]
\centering
{
\subfloat{\includegraphics[scale=.6]{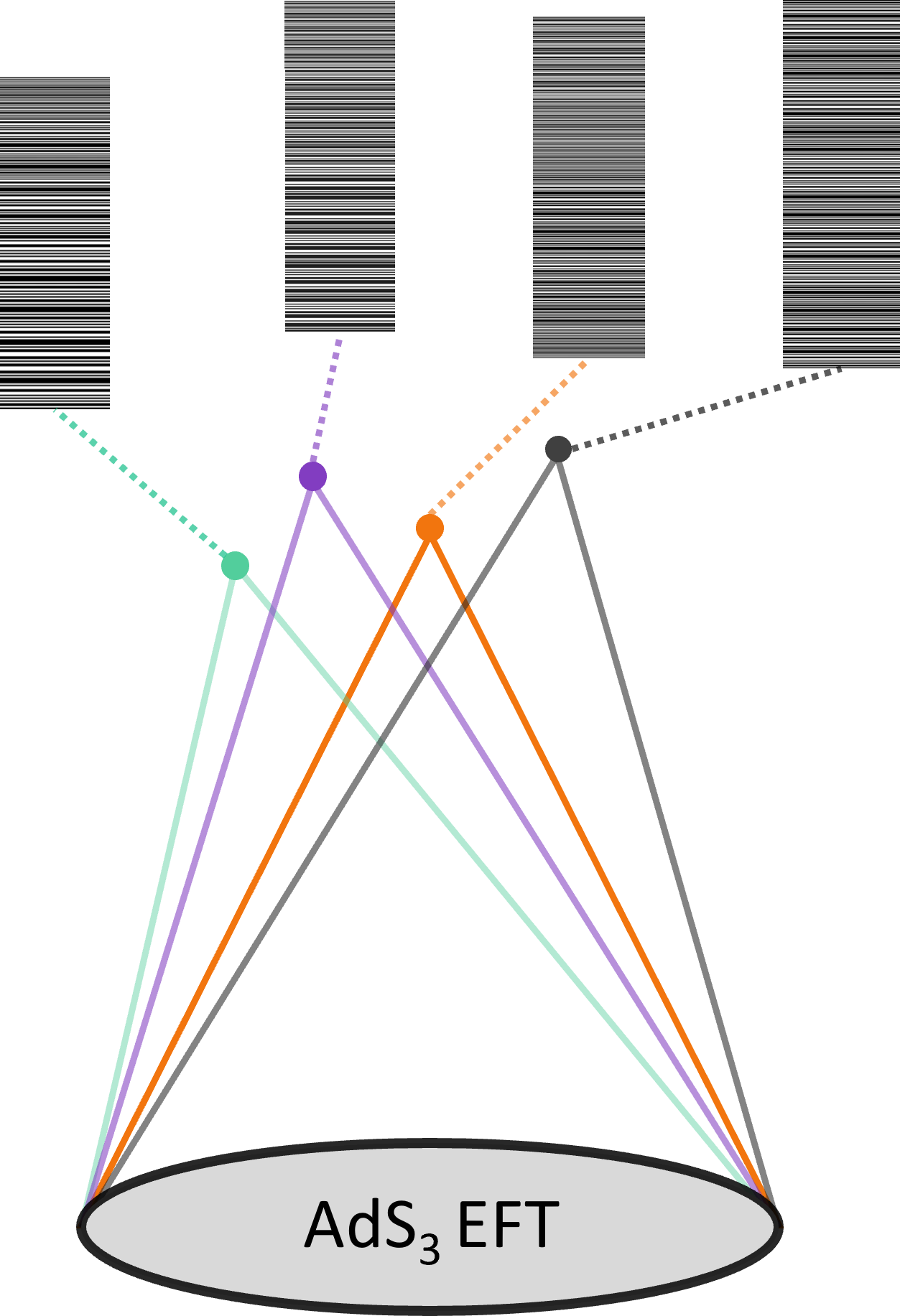}}
}
\caption{A given AdS$_3$ EFT may have multiple UV completions, characterized by their distinct black hole microstate spectra. To each spectrum is attached an $L$-function.}
\label{fig:ads3}
\end{figure}

This parallel can be made sharper. Not only do the degeneracies $\{a_\l\}$ exhibit square root cancellation, but they are actually structurally rather similar to $\mu(n)$: while $a_\l\in\ZZ$ in general, absent any accidental degeneracies one has $a_\l = \pm 1$, just like $\mu(n)$. And indeed, recalling that $a_\l = d_\D - \tilde d_\D$ is borne of a difference of partition functions, it is even clear what is cancelling. So we see that the $\{a_\l\}$ furnish a certain extension of the M\"obius function to non-integer arguments, while retaining its central analytic properties; they are, in the above precise sense, pseudorandom. It would be interesting to ask whether notions of determinism can be applied to these data in the spirit of \c{mobiusLecturesIAS}.\foot{As an aside, let us point out the analogous statement for $\qz(s)$. Recall that $\qz(s)$ has a pole at $s=3/4$, not $s=1$. This means that its expansion coefficients $\{a_\mu\}$ actually exhibit square root cancellation not in frequency $\l$, but in \textit{Liouville momentum $P$}. This follows upon recalling the usual definition of Liouville momenta $(P,\Pb)$ as $h={c-1\o 24}+P^2$ and $\hb={c-1\o 24}+\Pb^2$, such that a scalar state has $\l=P^2+\Pb^2 = 2P^2$.}

Returning to our question, then: what \textit{is} left in the UV? On the one hand (cf. \eqr{vanmoments}), the answer is ``not much''. On the other (cf. \eqr{src}), what \textit{is} left is highly fluctuating, chaotic and pseudorandom. In AdS$_3$ language, these are quantifications of the  freedom in the black hole microstate spectrum (see Figure \ref{fig:ads3}). The fact that we need to match light spectra to construct $\cZ(\t)$ makes this, at heart, a \textit{swampland} statement: given an AdS$_3$ low-energy EFT, the space of consistent UV completions, i.e. consistent black hole microstate spectra, contains the space of $L$-functions $\lz(s)$\thinspace---\thinspace self-dual of degree-4, with gamma factor $\g_\cZ(s)$ and root number $\vareps=1$. These spectra obey strict finiteness conditions: low-energy matching is such a strong constraint that the remaining UV degrees of freedom must almost cancel, exhibiting square root cancellation. Conversely, one can view this as a quantification of the ``miracle'' needed for chaotic 2d CFTs, and their AdS$_3$ quantum gravity duals, to exist at all!

\sec{Interlude: On random matrix universality in CFT}\label{s7}

Our work so far has been irrespective of any properties of the space of 2d CFTs. In this section, safely cloistered from the rest of the paper\thinspace---\thinspace and with no direct relation to $L$-functions\thinspace---\thinspace we take a slight detour.

In understanding the physical meaning of convergence of the spectral zeta series in Sections \ref{s53}--\ref{s54}, the statistics of extreme gaps in RMT naturally arose. In this section we want to draw a line between these statistics and 2d CFT per se. In other words, we want to embed random extreme gap statistics in a 2d CFT context, by formulating their appropriate avatars for primary operator spectra. Precisely what would it mean for a 2d CFT to exhibit these statistics? In what sectors of the theory ought these statistics to hold, and to what degree? It is not well-recognized that these questions are quite open, even for the ``mainstream'' spectral statistics of RMT: most work on the subject has focused on two-point statistics, but other cornerstones of RMT\thinspace---\thinspace e.g. nearest-neighbor spacings, $n$-point correlations, number variance\thinspace---\thinspace have not yet been transmuted into sharp properties of 2d CFT spectra.

One of the main motivations for doing so is that it would allow for a more robust definition of random matrix universality in 2d CFT. But we emphasize that this is a two-step process: only upon having formulated what ``random matrix statistics in 2d CFT'' ought to mean (what are the properties?) does the concept of random matrix universality (for what theories do they hold?) become contentful.\foot{We are being meticulous in making these logical distinctions, in an attempt to counter some hazy language that is often used in the literature (present company included).}

Having distilled the problem, we discuss each of these questions in turn. We focus on the narrow purview of extreme gap statistics for nearest neighbors. 

A 2d CFT avatar of random matrix statistics must incorporate the following two features. First, it is only concerned with the high-energy spectrum, asymptotically as $\D\rar\i$: this is a semiclassical limit in which one expects an effective random description to apply. Second, it must account for the local conformal symmetry of 2d CFTs. In particular, to the extent that they do apply, random matrix statistics should apply to the \textit{conformal symmetry-resolved} density of states\thinspace---\thinspace in other words, to the spectra of primaries $\{\D_n^{(j)}\}$ within every sector of fixed spin $j$ (see similar remarks in \c{CotlerJensen2021}).\foot{We point out the recent work \c{Pal:2025yvz} deriving nice results for spectral densities in 2d CFT in the reverse regime of finite twist and asymptotically large spin, as opposed to the random matrix regime.}

Let 
\e{}{\d_n^{(j)} := \D^{(j)}_{n+1}-\D^{(j)}_n}
be the physical level spacing between neighboring spin-$j$ states indexed by $n\in\Z_+$, with conformal dimensions ordered as $\D_{n}^{(j)} < \D_{n+1}^{(j)}$ for all $n$ and with $\D_{n\rar\i}^{(j)} \rar\i$. Let 
\e{}{s_n^{(j)} := {\d_n^{(j)}\o\<\d_n^{(j)}\>}}
be the unfolded level spacing. At $n\rar\i$, this becomes 
\e{}{\qquad \qquad \quad s_n^{(j)} \approx \d_n^{(j)}\rho_{\rm Cardy}^{(j)}(\D_n) \qquad (n\rar\i)}
with the Cardy density given in \eqr{cardydef}. It is to these fixed-spin primary level spacings that we are entitled to apply random matrix ideology.

We now carry this out for level repulsion and extreme gap statistics, following Section \ref{s54}. There are, of course, strong and weak forms. The strongest form would be that the full distributions of extreme gaps match those of large $N$ RMT: as $n\rar\i$ for every fixed $j$, the $\{\d_n^{(j)}\}$ obey random extreme gap distributions up to small corrections. In this scenario, the extreme small gaps in CFT would be distributed as \eqr{smallgapdist}, where the notion of probability is understood as statistical sampling within a sliding energy window out to infinity.

A weaker, and more plausibly universal, form is simply to incorporate random extreme gap statistics at the level of scaling: as $n\rar\i$ for every fixed $j$, the level spacings obey
\es{extcft}{\textsf{Level Repulsion:}&\qquad P_{\rm CFT}(\d_n^{(j)}=0) = 0\\
\textsf{No Clustering:}&\qquad P_{\rm CFT}(\d_n^{(j)}+ \d_{n+1}^{(j)} \sim \d_{\min}^{(j)}) = 0\\
\textsf{No Large Gaps:}&\qquad P_{\rm CFT}(\d_n^{(j)} = O(1)) = 0}
where $P_{\rm CFT}$ is defined by the statistical sampling within an $O(1)$ energy window centered at some $\D^{(j)}\rar\i$. By design, this form of random extreme gap statistics, in which we only apply the properties at the smallest and largest scales, allows for the possibility that the distribution of level spacings at intermediate scales $s_n^{(j)} = O(1)$ might deviate from RMT behavior in a more complicated, possibly theory-dependent, way. Note that in phrasing the No Clustering condition we could have opted to use the informal phrasing \eqr{NCinformal} instead; the scale $\d_{\min}^{(j)}$ in \eqr{extcft} stands for a scale exponentially smaller than the mean level spacing.\foot{A stronger, intermediate version would identify $\d_{\min}^{(j)}$ with the specific scale \eqr{smalllargeS} from RMT with $S \mapsto S_{\rm Cardy}^{(j)}(\D_n)$, presumably with $\b=1$ for GOE statistics, but this seems somewhat ad hoc. Likewise for the extreme large gap scale.}

In spectral analysis, a spectrum with no degenerate eigenvalues is said to be \textit{simple}. So the conditions \eqr{extcft} may altogether be paraphrased in plain terms as follows: as $\D^{(j)}\rar\i$ in sectors of fixed spin $j$, the Virasoro primary spectrum is asymptotically simple, with no clustering or large gaps.

We can now turn to the question of universality: having formulated these conditions in 2d CFT, when do they hold?

\ssec{A simple extreme conjecture}\label{s71}

As noted earlier, a full formulation of random matrix universality in 2d CFT is lacking. Work so far has mostly focused on the two-point spectral statistics, and in particular the linear ramp of the two-point spectral form factor, a hallmark of spectral rigidity. (We will return to the ramp in Section \ref{s9} vis-\`a-vis $L$-functions.) 

In \c{BAB}, it was observed that even the extreme gap statistics of large $N$ RMT are well-realized by Riemann zeta zeros high on the critical line, demonstrating that universality can indeed extend to the extreme regime.

The robustness \c{BGS, haake, richter} of random matrix universality across quantum chaotic systems suggests that \textit{all} irrrational CFTs might be expected to obey the fundamental tenets of RMT in the appropriate semiclassical regime, with possible deviations at fine scales. For 2d CFTs, we have just formulated a few such tenets in \eqr{extcft}, focusing on scaling properties of extreme gaps in particular. That Gaussian statistics, not Poissonian statistics characteristic of arithmetic systems, are the relevant universality class for 2d CFT is supported by the study of black holes in AdS gravity: their dynamics are random matrix-like \c{Cotler:2016fpe} (maximally so for Einstein gravity), and highly non-integrable. So this gives us a large $c$ data point for the above suggestion. 

In this spirit of universality, and on the back of \eqr{extcft}, we are led to formulate the following

\result{Conjecture:}{In a unitary, compact Virasoro CFT with $c>1$, in every sector of fixed spin $j$, the Virasoro primary spectrum is asymptotically simple, with no clustering or large gaps, as $\D\rar\i$.}

\noindent We call this the \textit{Simple Extreme Conjecture}. Recall that a Virasoro CFT has no extra conserved currents. 

It is underappreciated that asymptotic simplicity would be implied just by the onset of level repulsion. This perspective stands in contrast to similar statements sometimes made on the basis of ``typicality'', which imagine a measure on the space of Virasoro CFTs that, one might expect, suppresses putative theories with degenerate spectra. However, our knowledge of the space of irrational CFTs being primitive as it is, we are far from defining such a measure from first principles.\foot{See \c{Belin:2025qjm} for a recent bottom-up proposal of a measure on the space of \textit{all} CFTs, without any condition on the chiral algebra. A lesson of that work is that, absent such a condition, Virasoro CFTs are swamped by what we would call ``composite CFTs'', constructed from lower-$c$ theories by simple operations (e.g. tensor product and orbifolding), which have a great many extra conserved currents.} We have instead posed a stronger form of the conjecture\thinspace---\thinspace that \textit{all} $c>1$ Virasoro CFTs obey \eqr{extcft}\thinspace---\thinspace in part to avoid conditioning on a definition that does not yet exist (and is bound to be subtle). 

At the same time, the conjecture is conservative: as we emphasized in formulating \eqr{extcft}, it only pertains to the spectrum of nearest-neighbor spacings at the largest and smallest scales. It is intentionally robust against deviations from pure RMT nearest-neighbor spacings at intermediate scales, where level repulsion and spectral rigidity cross over (relevant for dynamical processes at timescales of order the Heisenberg time), which have been observed in various quantum systems (see e.g. \c{Winer:2023btb} for a recent study). At such scales, larger deviations from random matrix behavior may well be possible in Virasoro CFTs.

It has been shown that the maximal possible spectral gap between neighboring fixed-spin primaries in any unitary, modular-invariant 2d CFT obeys $\d_{\rm max}\leq1$ as $\D\rar\i$ \c{baursasha,baursridip}. Our conjecture, if correct, would imply that this bound is not tight for $c>1$ Virasoro CFTs. Indeed, in RMT, large gaps $\d = O(1)$ are \textit{doubly-exponentially suppressed} in the entropy! This can be seen heuristically via Wigner's surmise \eqr{wigner}; the same applies for any gap $\d \sim e^{-\a S}$ with $\a<1$. The picture for Virasoro CFTs is that, at ever-larger $\D$, large gaps appear increasingly sporadically until, as $\D\rar\i$, they disappear.

To summarize, the crux of the Simple Extreme Conjecture is an explicit definition of level repulsion and extreme gap statistics in 2d CFT via \eqr{extcft}, and the appeal to the robustness of random matrix universality as a first principles argument for its total applicability. A feasible probe of the conjecture would be the computation of level splittings among degenerate high-energy states in conformal perturbation theory near rational points of conformal manifolds. If, in fact, the conjecture fails to hold, those CFTs that violate it should have a ``reason'' to do so; we would expect such theories, by analogy with certain phenomena in number theory, to be in some sense \textit{arithmetic}\thinspace---\thinspace whatever that may mean. (Perhaps arithmeticity of CFT data will prove easier to define than a measure on the space of Virasoro CFTs.)

\sec{An approximate functional equation}\label{s8}

In this section we derive an approximate functional equation (AFE) for CFT $L$-functions. In spite of the name, this is an \textit{exact} representation of the $L$-function, as a generalized Dirichlet series modified by the insertion of a certain smoothing function which converges even in the critical strip. The smoothing function acts effectively like an exponential cutoff above some critical frequency, which is the origin of the word ``approximate'': analytic number theorists typically use AFEs to truncate the series with controlled errors.\foot{For example, the Riemann-Siegel formula is an asymptotic expansion of the error term in the AFE for Riemann zeta, known to remarkable accuracy; for clear introductions, see \c{BerryKeating1999, berry}. A widespread number theory application of AFEs is in the computation of moments of $L$-functions, or other mean-value theorems. Often the truncation is used without full accounting of subsequent errors, so as to give a heuristic explanation of an observed phenomenon, a famous example of that approach being the ``recipe'' of \c{cfkrs} for arbitrary moments of primitive $L$-functions on the critical line: via random matrix theory heuristics, this led to a conjecture for arbitrary $2k$'th moments of $|\z(\half+it)|$, the apparent veracity of which relies on various off-diagonal cancellations that remain to be proven rigorously.} Besides convergence in the critical strip, an AFE makes subconvexity (see Section \ref{s92}) of standard $L$-functions highly plausible, and allows excellent approximation of the locations of non-trivial zeros. 

For standard $L$-functions, the derivation of an AFE is straightforward, and we reproduce it in Appendix \ref{appf}. In our setting in which simple, rather than absolute, convergence is present, we have to start from scratch. Nevertheless, we will derive an AFE for $\lz(s)$ which converges for all central charges $c$, and moreover converges \textit{absolutely} for $c<7$. This will lead to our last, and best, form of the zero sum rule in \eqr{sumrule3}, in which the zero sum equals a \textit{manifestly convergent} sum over CFT frequencies. 

\ssec{Review: AFE for standard $L$-functions}\label{s81}
We begin by stating the AFE for standard $L$-functions, whose derivation we present in Appendix \ref{appf} following the canonical treatment of \c{IKtext}. Relative to \c{IKtext} our version is very slightly generalized, and adapted to our present notation. 

\result{AFE for standard $L$-functions.}{Let $L(s)$ be an $L$-function with completion $\L_L(s) = q^{s/2}\g(s)L(s)$, where $\L_L(s)$ is meromorphic with poles at most at $s=0,1$, and an absolutely convergent series representation
\e{}{L(s) = \sum_\D {a_\D\o \D^s}\,,\quad \s_a(L)=1}
where the sum runs over an infinite set $\{\D_n\}$ with $n\in\Z_+$. Let $G(u)=G(-u)$ be any function which is holomorphic and bounded in the strip $|\Re(u)| < \b+1$ for some $\b\geq 1$ and normalized as $G(0)=1$. Then 
\e{AFEIK}{L(s) = \sum_\D {a_\D\o \D^s} V_s\({\D\o X \sqrt{q}}\) + \vareps(s) \sum_\D  {a_\D\o \D^{1-s}} V_{1-s}\({\D X\o  \sqrt{q}}\)-R(s)}
converges absolutely for all $s$ in the strip $1-\b< \s < \b$, where $V_s(y)$ is a smooth function
\e{Vdef}{V_s(y) = {1\o 2\pi i}\int\limits_{~(\b)} {du\o u} y^{-u} G(u) {\g(s+u)\o \g(s)}}
and
\e{}{\vareps(s) = q^{\half-s} {\g(1-s)\o \g(s)}}
and $R(s)$ is a remainder from poles of $\L_L(s+u)$,
\e{}{R(s) = {1\o q^{s/2} \g(s)} \(\Res_{s+u=0} + \Res_{s+u=1}\) X^u \L_L(s+u) {G(u)\o u}}
Equivalently, in terms of the completed $L$-function,
\e{AFEIK2}{\L_L(s) = \sum_\D {a_\D\o \D^s} Y_s\({\D\o X \sqrt{q}}\) +  \sum_\D {a_\D\o \D^{1-s}} Y_{1-s}\({\D X\o\sqrt{q}}\)-\widehat R(s)}
with completed smoothing function
\e{}{Y_s(y) = q^{s/2} \g(s) V_s(y) = {q^{s/ 2}\o 2\pi i}\int\limits_{~(\b)}{du\o u}  y^{-u} G(u) {\g(s+u)}}
and
\e{}{\widehat R(s) = q^{s/2} \g(s) R(s)}
}

\sssec*{Comments}
\enumcom

\item The symmetrized remainder $\widehat R(s)$ simplifies when $G(u)=X=1$:
\es{}{\widehat R(s) &= {1\o s(1-s)}\Res_{s=1}\L_L(s)}

\item In typical treatments \c{IKtext}, the choice $\b=3$ is made. The actual value of $\b$ is not terribly important because there already exists an absolutely convergent series for $\s>1$, so the real value of the AFE is to furnish a convergent series \textit{inside} the critical strip, for which any $\b>1$ will do. However, with an eye toward our extension to the conditionally convergent setting, the generalization to arbitrary $\b>1$ (which costs nothing in the derivation) is useful. 

\item The choice $G(u)=1$ allows us to take $\b$ arbitrarily large, by Cauchy's theorem and the absence of poles of $\L_L(s)$ for $\s>1$. 

\item $V_s(y)$ and its derivatives are of rapid decay in $y$ for fixed $s$: it acts as a sharp cutoff function, with crossover scale
\e{}{y_* \sim \sqrt{\mathfrak{q}_\i(s)}}
where 
\e{}{\mathfrak{q}_\i(s) := \prod_{i=1}^d (|s+\kappa_i|+3)}
is the analytic conductor $\mathfrak{q}(s)$ divided by the conductor $q$ (cf. \eqr{aconddef}). In the $t$-aspect, this crossover scale at degree $d$ is
\e{}{y_*^{(d)}(t) \sim t^{d/2}\,.}
Accordingly, one can approximate $L(s)$ by truncating the AFE to terms with ${\D/ X} \lesssim \sqrt{\mathfrak{q}(s)}$ in the first sum and ${\D /X^{-1}} \lesssim \sqrt{\mathfrak{q}(s)}$ in the second. On the critical line, for example, the completed AFE, with the simplifying choice $X=1$, is thus well-approximated by
\es{AFEapprox2}{\L_L\(\half+it\) &\approx \sum_{\D \lesssim \sqrt{\mathfrak{q}(\half+it)}} {a_\D\o \sqrt{\D}}\(\D^{-it}Y_{\half+it}\({\D\o \sqrt{q}}\)+\cc\)-\hat R\(\half+it\) + \mathsf{E}(t)}
with small errors $\mathsf{E}(t)$. In well-studied cases like Hardy-Littlewood or Riemann-Siegel for $\z(s)$, $\mathsf{E}(t)$ is power law-suppressed as $t\rar\i$, with asymptotic expansions amenable to resurgence.

\end{enumerate}

\sssec{Degree-4 smoothing functions}\label{s811}

Our CFT $L$-functions $\lz(s)$ are degree-4, with conductor $q=4$ and a gamma factor $\g_\cZ(s)$ given in \eqr{gcft}. Taking $G(u)=1$, we find 
\e{VG}{\qquad \qquad Y_s(y) = 8\sqrt{\pi} \(8\pi^2\)^{\half-s} \,\Gamma \left(2 s-1,4 \pi  \sqrt{y}\right)\qquad (G(u)=1)}
where
\e{}{\Gamma(n,z) = \int_z^\i du \,u^{n-1} e^{-u}}
is the incomplete gamma function. This will be our choice of smoothing function below. We record the value at $s=1$, 
\e{}{Y_1(y) = 2\sqrt{2\o \pi}e^{-4\pi\sqrt{y}}}
and the asymptotics at large $y$, 
\e{3145}{\qquad Y_s(y) \approx \sqrt{2\o\pi}\,2^sy^{s-1}e^{-4\pi\sqrt{y}} \qquad (y\rar\i)\,.}
For smoothing functions of general degree-$d$ gamma factors, see \c{dokchitser}.

\ssec{AFE for CFT $L$-functions}\label{s82}
We now turn to our $L$-functions $\lz(s)$. Our goal is to show that \eqr{AFEIK} holds even without absolute convergence. 

\sssec*{Take I: Absolute convergence for $c<7$}
The proof actually goes through unmodified for $\lz(s)$ when $c<7$, leading to an absolutely convergent AFE for all $\s \in\mathbb{C}$. The reason is simple: at large $y$, the exponent in \eqr{3145} beats the Cardy growth of states \eqr{cardydef}--\eqr{coarsecardy} when $c<7$. The first sum in \eqr{AFEIK} is absolutely convergent when
\e{}{4\pi \sqrt{\D\o 2X} > 2\pi \sqrt{{c-1\o 3}\D}\quad \Longrightarrow \quad X^{-1} > {c-1\o 6}}
and similarly for the reflected sum with $X \rar X^{-1}$. For any central charge $c$ we can achieve absolute convergence of either sum by choosing $X$ appropriately, but absolute convergence of \textit{both} sums holds when 
\e{Xc}{ {c-1\o 6} < X < {6\o c-1}\,.}
This implies $c<7$.\foot{See \c{Benjamin:2025kvm} for a similar $c<7$ phenomenon, which we suspect can be explained with the AFE.}

\sssec*{Take II: Conditional convergence and general $c$}

In fact, upon assuming that $\lz(s)$ admits a convergent series representation, the AFE \eqr{AFEIK} holds for \textit{all} $c$. The logic here is simple: since it is legal to swap sums and integrals for any finite sum, one can do so for a conditionally convergent sum provided that the error, i.e. the difference between the partial sum and the full sum, gives a vanishing contribution in the limit. That is, we replace $\lz(s)$ by its partial sum up to some cutoff $\nu<\chi$, then show that the error in the AFE vanishes in the limit $\chi\rar\i$.\foot{We thank Anshul Adve for suggesting this strategy.} Define the error
\e{}{\textsf{E}(s;\chi) := \lz(s)-\lz^{<\chi}(s)\quad \text{where}\quad \lz^{<\chi}(s) := \sum_{\nu<\chi}{a_\nu\o \nu^{s-\half}}}
By convergence,
\e{errorlimitmain}{\lim_{\chi\rar\i}\textsf{E}(s;\chi) = 0~~\forall\,\s>1\,.}
The AFE follows if
\e{errorlimitint}{\lim_{\chi\rar\i} {q^{s/2}\o 2\pi i} \int\limits_{~(\b)}{du\o u}(\sqrt{q}X)^u \textsf{E}(s+u;\chi) \g(s+u) G(u) =0\,.}
In view of \eqr{errorlimitmain}, the goal is to show that the limit can be moved inside the integral. We prove that it can in Appendix \ref{appg}, by using the uniform growth bound \eqr{bound} to demonstrate the existence of an integrable dominating function and applying the dominated convergence theorem. 

\sssec*{Statement of AFE}
Altogether, we have derived the following

\result{AFE for CFT $L$-functions.}{Assume $\lz(s)$ admits a convergent $L$-series for $\s>1$. Then 
\ebox{AFECFT}{\L_\cZ(s) = \sum_{\nua} {a_\nua\o \nua^{s-\half}} \Gamma \left(2 s-1,\sqrt{\nua}\right) + \sum_{\nua}{a_\nua\o \nua^{\half-s}} \Gamma \left(1-2 s,\sqrt{\nua}\right)-{r_0\o s(1-s)}}
converges for all $s\in\mathbb{C}$, where
\e{nuadefs}{\nua := 8\pi^2\nu\,,\quad a_\nua := 8\sqrt{\pi}a_\nu}
and the residue $r_0$ is defined in \eqr{l0r0}. For $1<c<7$, the series \eqr{AFECFT} converges absolutely, even without assuming that the $L$-series converges.}

 We highlight the specialization to the critical line:
\e{AFEcrit}{\L_\cZ\(\half+it\) = \sum_\nua {a_\nua\o \nua^{it}} \Gamma \left(2it,\sqrt{\nua}\right) + \text{(c.c.)}-{r_0\o {1\o4}+t^2}\,.}
In terms of an exponential integral,
\e{AFEcrit}{\L_\cZ\(\half+it\) = 2\sum_\nua a_\nua\int_1^\i {du\o u}e^{-\sqrt{\nua}u}\cos(2t\log u)-{r_0\o {1\o4}+t^2}\,.}
This representation makes manifest that motion along the critical line is implemented by periodic modulation of the Fourier coefficients $a_\nua$. 

\ssec{Comments}\label{s83}

\enumcom

\item We have chosen parameters $G(u)=X=1$, with $Y_s(y)$ given by \eqr{VG} with $y=\l/2$. Other choices would give other AFEs besides \eqr{AFECFT}, so there is really an infinite family of AFEs.

\item Our assumption of convergence of the $L$-series helps to phrase things in generality, but it is clear that \eqr{AFECFT} would still hold (for any $c$) even without that assumption as long as the growth of the series' partial sums is slow enough to be dominated by the decay of the smoothing function \eqr{3145} (indeed, this is what always happens for $c<7$).

\item One naturally wonders whether it is possible to increase the exponent in the smoothing function $Y_s(y)$ by choosing $G(u)\neq 1$ appropriately: if so, this would increase the range of $c$ for which an absolutely convergent AFE exists. However, this is not possible: parameterizing
\e{}{\log Y_s(y) \sim -4\pi \a_G \sqrt{y}\qquad (y\rar\i)}
to leading order, we prove in Appendix \ref{apph} that for any admissible choice of $G(u)$, 
\e{aG}{\a_G \leq 1\,.}

\item At the central point,
\e{AFEcrit}{\L_\cZ\(\half\) = 2 \sum_\nua a_\nua \Gamma \left(0,\sqrt{\nua}\right) -4r_0\,.}
Compare this with the modular integral representation \eqr{bsd} (where $\L_\cZ(\thalf) = 4\sqrt{\pi}\lz'(\thalf)$). 

\item An important avenue for future work is to estimate the magnitude of errors in truncation of \eqr{AFECFT}, which would facilitate estimation of low-lying zeros of $\lz(s)$. The question is how long, or rather how short, the sums must be under a controlled approximation \`a la Riemann-Siegel. We revisit this in the Discussion.

\end{enumerate}

\ssec{Zero sum rule revisited}\label{s84}

The AFE leads to an improved version of the zero sum rule. As in \eqr{BLreln4}, we have another expansion of $\L_\cZ(s)$ around $s=1$ in terms of the dimensions, but this time it manifestly converges: the pole is stripped off! Therefore, by expanding \eqr{AFECFT} around $s=1$ and extracting the ratio $\ell_1/\ell_0$, we derive a convergent expression for the zero sum in terms of a sum over frequencies:
\ebox{sumrule3}{\sum_\rho{1\o\rho} ={1\o r_0}\sum_\nua a_\nua\({1\o\sqrt{\nua}} \Gamma \left(1,\sqrt{\nua}\right) +\sqrt{\nua} \Gamma \left(-1,\sqrt{\nua}\right)\)}
We recall from \eqr{nuadefs} that $\{\nua_n\}= \{8\pi^2\nu_n\}$ are the rescaled frequencies of the $L$-series \eqr{zlqser}. In contrast to \eqr{BLreln4}, there is no longer any need for regularization. Indeed, it is useful to recall the exponential integral representation of $\Gamma(\pm 1,\sqrt{\nua})$,
\e{}{{1\o\sqrt{\nua}} \Gamma \left(1,\sqrt{\nua}\right) +\sqrt{\nua} \Gamma \left(-1,\sqrt{\nua}\right) = {e^{-\sqrt{\nua}}\o\sqrt{\nua}} + \int_1^\i {du\o u^2} e^{-\sqrt{\nua}u}\,.}
The high-frequency tail of the summand behaves as
\e{}{\qquad\qquad\quad{e^{-\sqrt{\nua}}\o\sqrt{\nua}} + \int_1^\i {du\o u^2} e^{-\sqrt{\nua}u}\approx {2e^{-\sqrt{\nua}}\o\sqrt{\nua}}\qquad (\nua\rar\i)\,.}
Comparing to \eqr{BLreln4}, we see that the smoothing factor acts precisely as an exponential regulator that renders the sum convergent (absolutely so for $c<7$).

\sec{Random matrix universality and the critical line}\label{s9}

In this section we address the following question, in view of our principal interest in understanding chaotic dynamics of irrational CFTs: \textit{What does a linear ramp in the two-point spectral form factor of a 2d CFT imply for $\lz(s)$?} The linear ramp is one of the most robust hallmarks of random matrix universality (RMU), a statistical signature of spectral rigidity among distant energy levels, well-studied in CFT and holography.\foot{We will not review here the definition of the spectral form factor nor the connection between the linear ramp and spectral rigidity, which has been the subject of much recent literature. See e.g. \c{Cotler:2016fpe} for an introduction.}

The ubiquity of random matrix theory in the study of the non-trivial zeros of standard $L$-functions (partially reviewed in Section \ref{s23}) obliquely suggests that the answer could relate to the behavior of $\lz(s)$ on the critical line. The logic is reversed in the present CFT context\thinspace---\thinspace randomness is an input as a property of the theory, not an output as a property of the $L$-function\thinspace---\thinspace but we will nevertheless find such a connection, building on earlier observations in \c{2307} using $\sl$ spectral decomposition. 

We will find that a linear ramp implies a \textit{linear second moment on the critical line}, for a certain product of $\lz(s)$ with a Riemann zeta function. For reasons to become clear, we call this \textit{Riemann zeta universality}. In turn, using known properties of $\z(1+it)$, the Riemann zeta function on the 1-line, we deduce a subconvexity result for $\lz(s)$ on the critical line in terms of its growth on the 1-line. 

\ssec{Random matrix universality $\Rightarrow$ Riemann zeta universality}\label{s91}

In \c{2307}, an integral condition for a 2d CFT to have a linear ramp in the two-point spectral form factor was derived. Spectral form factors in 2d CFT are computed for conformal symmetry-resolved partition functions, measuring correlations among primaries of fixed spins, so the two-point case carries two spin indices $(j_1,j_2)$; the analysis here is for the scalar sector, with $j_1=j_2=0$. From (4.13), (4.21) and (4.22) of \c{2307}, in the present language the linear ramp condition is
\e{ramp1}{ \int_{-\i}^\i dt |(\cZ,E_{\half+it})|^2 \cosh\(2t\arctan\({T\o \b}\)\) \approx  {\mathsf{C}_{\rm RMT}\o 2} {T\o \b} \qquad \({T\o \b}\rar\i\)}
$T$ is Lorentzian time, $\b$ is the inverse temperature, and $\mathsf{C}_{\rm RMT}$ sets the Gaussian ensemble of random matrices with respect to which the theory exhibits random matrix universality (RMU). In the 2d CFT context, it is the GOE ensemble that appears most natural \c{Yan:2023rjh}, for which $\mathsf{C}_{\rm RMT}=2$. We make this choice henceforth, though $\mathsf{C}_{\rm RMT}$ can be easily restored. 

Our goal is to turn \eqr{ramp1} into a result for the \textit{second moment} of the $L$-function: that is, trading the Lorentzian time $T$ for height $T$ on the critical line. This falls under the purview of what are known as \textit{mean value theorems}, central quantities in the study of $L$-functions (nicely introduced in e.g. Chapter 7 of \c{Titchmarsh1986}).

Henceforth working in units of $\b=1$, our ramp integral \eqr{ramp1} is (twice)
\e{}{ I(T) := \int_{0}^\i dt |(\cZ,E_{\half+it})|^2 \cosh\(2t\arctan\(T\)\)\,.}
It will prove convenient to work in terms of the $L$-quotient $\qz(s)$, remembering as always that one may trade $\qz(s)$ for $\lz(s)$ via \eqr{qdef2}. In a slight abuse of notation, we use
\e{}{\qz(t) := \qz(\thalf+it)}
to denote the critical line specialization of $\qz(s)$. Following the definitions of Section \ref{s3}, we have
\e{}{|(\cZ,E_{\half+it})|^2 = \qz(t)^2 |\z(1+2it)|^2\cosh(\pi t)^{-1}}
in terms of which the ramp integral $I(T)$ is
\e{Idef}{I(T) =  \int_{0}^\i dt \,\qz(t)^2 |\z(1+2it)|^2{\cosh\(2t\arctan\(T\)\)\o \cosh(\pi t)}\,.}
On the other hand, we introduce the moment integral
\e{Jint}{J(T) := \int_1^T dt \,\qz(t)^2 |\z(1+2it)|^2\,.}
\begin{figure}[t]
\centering
{
\subfloat{\includegraphics[scale=.62]{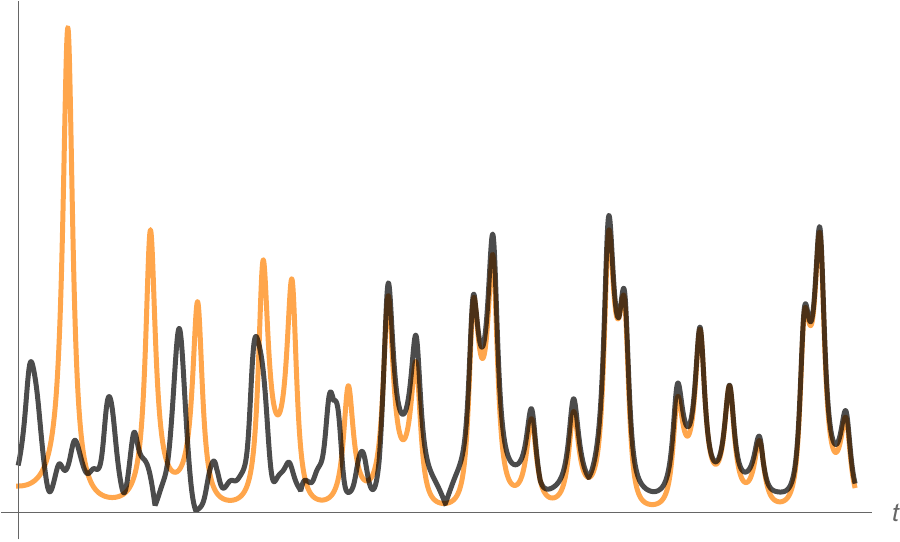}}
}
\caption{\textit{Riemann zeta universality:} $\qz(t)^2$ (drawn heuristically, in black) approaches $|\z(1+2it)|^{-2}$ (drawn exactly, in orange) on average as $t\rar\i$.}
\label{fig:zetauniv}
\end{figure}

In Appendix \ref{appi} we prove the necessary Tauberian result:
\e{IJres}{I(T) \approx {T\o 2} \quad \Longrightarrow \quad J(T)  \approx  T}
In other words, introducing a height average notation
\e{}{\qquad\qquad~~ \<f(T) \> := \int_1^{T} dt \, |f(\thalf+it)|\qquad (T\rar\i)}
we have shown that
\ebox{qav}{\textsf{RMU}\quad \Longrightarrow \quad \<\qz(T)^2 |\z(1+2iT)|^2 - 1\> = o(T) }
On average as $T\rar\i$, the (squared) $L$-quotient approaches the (squared) reciprocal Riemann zeta function on the 1-line! We call this phenomenon \textit{Riemann zeta universality}. The equivalent statement for $\lz(t) := \lz(\thalf+it)$ is that it approaches Riemann zeta on the 0-line,
\e{}{\<\lz(T)^2 |\z(2iT)|^{-2} - 1\> = o(T) }
One can shift to the more conventional 1-line using the functional equation, $|\z(2iT)| \approx |\z(1+2iT)|\sqrt{T\o \pi}$ at large $T$. See Figure \ref{fig:zetauniv}. 

\ssec{Reviewing subconvexity}\label{s92}
An immediate corollary of Riemann zeta universality is a subconvexity result, given some known facts about the Riemann zeta function. Let us first review the notion of subconvexity.\foot{For a nice overview of the state-of-the-art in the context of standard $L$-functions, see \c{Michel2022Bourbaki}.} 

The quantity of interest is the \textit{growth exponent} of the $L$-function, defined via the tightest possible upper bound on the growth along vertical lines: for some function $f(s)$,
\e{}{\mu_f(\s) := \text{inf}\(\xi:f(\s+it) = O(|t|^\xi)\)\,.}
By Phragmen-Lindel\"of (cf. Appendix \ref{appb}), $\mu(\s)$ is a convex, continuous function of $\s$.\foot{$\mu_f(\s)$ is sometimes called the ``order'', or ``order function'', of $f(s)$, but we find ``growth exponent'' less likely to be conflated with other notions of ``order'' in complex analysis.} In what follows we use $\mu_L(\s)$ and $\mu_\Q(\s)$ to denote the growth exponents of $\lz(\s+it)$ and $\qz(\s+it)$, respectively.

The logic of convexity bounds is simple; we spell it out with reference to $\Lamz(s)$. Recall that $s(1-s)\Lamz(s)$ is entire, of degree-4, with functional equation \eqr{LZFE}. Given some behavior of $\L_\cZ(s)$ in a right-half plane $\s>\s_*$, the functional equation fixes the behavior in the left half-plane $\s<1-\s_*$. To bound the behavior in the strip $\s\in[1-\s_*,\s_*]$, one uses that $(1-s)\lz(s)$ is entire and polynomially bounded along vertical lines for $\s\geq \half$: then Phragmen-Lindel\"of for strips implies that $\mu_L(\s)$ is upper-bounded across the strip by the linear function, and is non-increasing as $\s$ increases (see Appendix \ref{appb}). 

\begin{figure}[t]
\centering
{
\subfloat{\includegraphics[scale=.55]{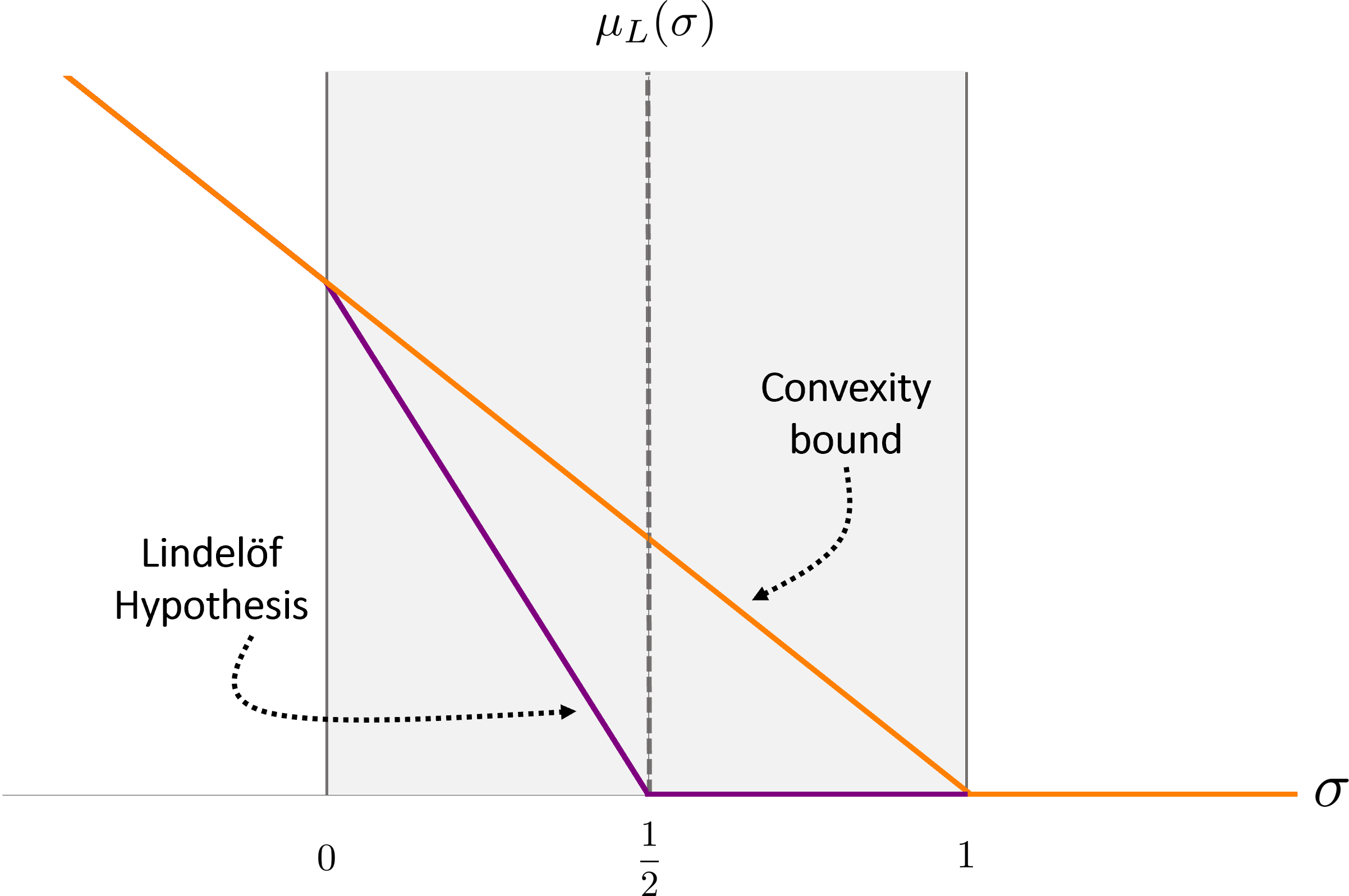}}
}
\caption{Convexity vs. Lindel\"of.}
\label{fig:conlind}
\end{figure}

For a standard $L$-function of degree $d$, where the series representations converges absolutely for $\s>1$ and $\mu_L(\s\geq 1)=0$, the tightest bound (without further assumptions) comes from taking $\s_*=1$ as the ``reference point''. Stirling's approximation of the gamma factor then yields
\e{stdconvex}{\mu_L(\s) \leq {d\o 2}(1-\s) ~~\forall\,\s\in[0,1]\,.}
For a function like $\lz(s)$ for which $\mu_L(1)= 0$ is not guaranteed,\foot{If we assume that the series representation of $\lz(s)$ converges, then the uniform growth bound \eqr{bound} implies $\mu_L(1) \leq 1$, but otherwise $\mu_L(1)$ is unconstrained.} \eqr{stdconvex} generalizes straightforwardly by adding a shift by $\mu_L(1)$. On the lower end, a bound is supplied by Phragmen-Lindel\"of. Altogether, taking $d=4$, we thus have\foot{In principle an even tighter bound can come from choosing $\s_*>1$, if $\mu_L(\s)$ drops off precipitously: the best bound on the critical line comes from, $\mu_L(\thalf) \leq \min \(\mu(\s) +2(\s-\thalf)\)$ where one minimizes over $\s\geq 1$.}
\e{plc}{\textsf{Phragmen-Lindel\"of + Convexity:}\quad 0 \leq \mu_L(\s) - \mu_L(1) \leq 2(1-\s)\quad~}
for all $\s\in[0,1]$. \textit{Subconvexity} is any behavior that beats the upper bound. On the critical line, 
\e{plc}{\textsf{Phragmen-Lindel\"of + Convexity:}\quad 0 \leq \mu_L\(\half\) -\mu_L(1)\leq 1\qquad\qquad  }

For standard $L$-functions, saturation of the lower bound is the Lindel\"of Hypothesis: $\mu_L(\thalf)=0$. The statement of Lindel\"of is that $\mu_L(\s)$ is flat for $\s\in[\thalf,1]$\thinspace---\thinspace that is, as small as possible consistent with complex analysis and the $\s=1$ boundary condition. (See Figure \ref{fig:conlind}.) By that definition, saturation of the lower bound of \eqr{plc} is Lindel\"of scaling for CFT $L$-functions. 

In view of approximate functional equations like \eqr{AFEIK} and \eqr{AFECFT}, in which $L$-functions admit convergent series representations on the critical line, \textit{subconvexity is phase incoherence}. Conversely, an approximate functional equation makes (at least some degree of) subconvexity highly plausible. Much of the work on improving subconvexity bounds comes down to proving that the exponential sums do indeed enjoy phase cancellation. For context, for the Riemann zeta function\thinspace---\thinspace of degree-1, hence $\mu_\z(\half)\leq {1\o4}$ by convexity\thinspace---\thinspace the current best bound is $\mu_\z(\half) = {13\o 84}$; for standard $L$-functions of $GL(n)$, see \c{Nelson2021}. That these are well short of Lindel\"of scaling, despite over a century \c{weyl16, ingham26, littlewood26} of concerted effort, indicates the profundity of the problem.

\ssec{Subconvexity and moments of CFT $L$-functions}\label{s93}

We now return to the CFT $L$-function. Its asymptotic growth on the critical line is strictly bounded by \eqr{qav}, which follows upon recalling some elementary facts about the Riemann zeta function on the 1-line.

\sssec{Riemann zeta on the 1-line}
It is well-known that $|\z(1+it)|$, the magnitude of Riemann zeta on the 1-line, is uniformly bounded by logarithms, unconditionally.

For starters, note that on RH,
\e{}{{1\o \log\log t}\ll |\z(1+it)|^{-1} \ll \log\log t\,.}
Explicit uniform bounds are known. The upper bound on the magnitude is 
\e{sound}{|\z(1+it)|^{-1} \leq {12 e^{\g_E}\o \pi^2}\(\log\log t +O(1)\)}
for all $t\geq 10^{10}$, originally derived by Littlewood \c{littlewood26, littlewood28} and extended in \c{soundetal} (cf. Theorem 1.5). 

Unconditionally, there are various explicit upper and lower bounds, all logarithmic and uniform for $t>t_*$. In our context it is the lower bounds on $|\z(1+it)|$ that are most relevant. A simple unconditional bound from \c{hly} is
\e{zeta1}{1.731 {\log\log t\o \log t} \leq |\z(1+it)|^{-1} \leq 431.7 {\log t\o \log\log t} ~~\forall\,t\geq 3}
with upper coefficient subsequently improved to 107.7 in \c{cully2024explicit}. Other works \c{ford2002vinogradov,trudgian} have established a logarithmic scaling of
\e{zetaford}{{c_-\o\log^{2/3}t} \leq |\z(1+it)|^{-1} \leq c_+ \log^{2/3}t(\log\log t)^{1/3}~~\forall\,t\geq t_*}
for some $(c_\pm,t_*)<\i$. 

Note that although $|\z(1+it)|$ is known to take arbitrarily large values on the 1-line, and does not have a proper asymptotic due to unbounded variation ($\z'(1+it)$ is not uniformly bounded by a constant, but rather by a power of $\log t$ \c{hly}), all of its moments are exactly linear. This is in contrast to moments on the critical line, which scale (conjecturally) as $T\log^{k^2} T$ for the $2k$'th moment \c{cfkrs}. Defining\foot{Moment integrals on the 1-line start above zero to avoid the pole.}
\e{}{I_{2k}(T) := \int_1^T dt\,|\z(1+2it)|^{2k}}
one has
\e{zeta1moments}{I_{2k}(T) = a_k^{(\z)}T + o(T)}
where
\e{}{a_k^{(\z)}= \sum_{n=1}^\i{d_k(n)^2\o n^2}\,,\quad \text{with}\quad {\z(s)^k} = \sum_{n=1}^\i {d_k(n)\o n^s}\,.}
At low $k$ (e.g. Theorems 7.2 and 7.5 of \c{Titchmarsh1986}),
\es{}{a_1^{(\z)} = \z(2) \,,\quad a_2^{(\z)}= {\z^4(2)\o \z(4)}\,.}
\eqr{zeta1moments} holds for $k\notin \Z_+$, and in particular for $k<0$ (see e.g. \c{negmom} for work on negative moments). 

\sssec{Subconvexity}

Plugging any one of the aforementioned logarithmic bounds for Riemann zeta into \eqr{qav} implies that the second moment of $\qz(t)$ scales no faster than linearly times logarithms. To wit, the second moment, call it $M_2(T)$, is
\es{}{M_2(T) &:= \int_1^T dt\,\qz(t)^2 \\&= \int_1^T dt \big(\qz(t)|\z(1+2it)|\big)^2|\z(1+2it)|^{-2}\,.}
Recognizing the first term of the integrand as the integrand of $J(T)$, we have the obvious bound
\e{M2bounds}{J(T) \x\underset{t\in[1,T]}{\text{inf}}|\z(1+2it)|^{-2}\leq M_2(T) \leq J(T)\x \underset{t\in[1,T]}{\text{sup}}|\z(1+2it)|^{-2}\,.}
The RMU result $J(T) \approx T$, together with the logarithmic upper and lower bounds for $|\z(1+2it)|^{-1}$, implies the stated result. 

In turn, the second moment implies a bound on the growth exponent $\mu_\Q(\thalf)$, 
\ebox{muGrmu}{{\textsf{RMU}\quad \Longrightarrow \quad 0 \leq \mu_{\mathrm{Q}}\(\half\) \leq \half}}
The upper bound is the ``worst-case'' scenario, in which $\qz(t)$ has large spikes of square root amplitude that persist out to arbitrarily large $t$, but of sufficiently small measure as to preserve linearity of $M_2(T)$. The lower bound is the scaling obeyed by $\z(1+it)$\thinspace---\thinspace and, conjecturally, by $\z(\thalf+it)$\thinspace---\thinspace in which the amplitude of fluctuations of $\qz(t)$ is smaller than any power of $t$. Using\foot{Potential mollification of $\qz(t)$ by the Riemann zeta factors\thinspace---\thinspace by way of alignment between maxima of the former and minima of the latter\thinspace---\thinspace cannot lower $\mu_L(\thalf)$, because the magnitude of Riemann zeta minima is only logarithmic.}
\e{muqlcrit}{\mu_{\mathrm{Q}}\(\half\)= \mu_L\(\half\)-\half\,,}
the bound for $\mu_L(\half)$ is read off from \eqr{muGrmu} as
\ebox{muLRMU}{{\textsf{RMU}\quad \Longrightarrow \quad \half \leq \mu_L\(\half\) \leq 1}}
Notice that it is $\qz(t)$, and not $\lz(t)$, that is allowed to exhibit sub-polynomial growth. 

This result has a few corollaries:

\begin{itemize}

\item Combining \eqr{muLRMU} with \eqr{plc} implies $\mu_L(1)\leq 1$. By Phragmen-Lindel\"of, this extends everywhere to the right of $\s=1$ (i.e. $\mu_L(\s)$ is non-increasing), so
\e{rmus1}{\textsf{RMU}\quad \Longrightarrow \quad \mu_L(\s) \leq 1~~\forall\,\s\geq 1\,.~~~~}
This non-trivially affirms our earlier interpretation of convergence as random statistics. In particular, \eqr{rmus1} is precisely the same condition implied by demanding convergence of the series representation for $\lz(s)$ for $\s>1$, because convergence implied the uniform growth bound \eqr{bound}, which implies \eqr{rmus1}; but convergence was \textit{not} assumed in deriving \eqr{rmus1}. It is well-understood that spectral rigidity is the origin of a linear ramp, which led to \eqr{rmus1}, so the physical picture here holds together nicely. 

\item If $\mu_L(1)>0$, then RMU implies subconvexity: $\mu_L(\thalf) < \mu_L(1)+1$.

\item If $\mu_L(1) = 1$, then RMU implies Lindel\"of-type scaling: $\mu_L(\thalf)=\mu_L(1)$.

\end{itemize}

\sssec{Moments}

In order to probe the value of $\mu_\Q(\thalf)$ or $\mu_L(\thalf)$ within their respective allowed ranges, we can study higher moments, a typical approach used for standard $L$-functions. If all moments of $\qz(t)$ are linear (times logs), then $\mu_\Q(\thalf)=0$ (cf. \c{Patterson1988}, Section 5.3).\foot{This behavior is known as ``Lindel\"of-on-average'' scaling for standard $L$-functions. The moment approach to standard $L$-functions on the critical line has established unconditional bounds only for some low moments: even for Riemann zeta, Lindel\"of-on-average scaling beyond the fourth moment \c{ingham26} has not been rigorously derived. See \c{heathbrown82} for a Weyl-strength bound from the twelfth moment of $|\z(\thalf+it)|$.} On the other hand, just the fourth moment will determine definitively whether subconvexity must hold.  

We choose to phrase the moment problem using $\qz(t)$. Define 
\e{}{M_{2k}(T) := \int_1^Tdt\, \qz(t)^{2k}\,,~~k\in\mathbb{R}\,.}
In terms of $\lz(t)$, 
\e{}{M_{2k}(T) := \int_1^Tdt\, \({\pi\o t \tanh(\pi t)}\)^k\left|{L_\cZ(t)\o \z(1+2it)\z(1-2it)}\right|^{2k}  \,,~~k\in\mathbb{R}\,.}
The $\tanh(\pi t)$ factor rapidly asymptotes to unity, and can be ignored for the purposes of large $T$ asymptotics. We can parameterize the leading asymptotic as
\e{}{M_{2k}(T) \approx a_k T^{b_k} \log^{f_k}T}
for some $(a_k,b_k,f_k)\in \RR$. For standard $L$-functions, $(b_k,f_k)$ are monomials in $k$, while $a_k$ is a coefficient conjecturally known via RMT heuristics and the CFKRS recipe \c{cfkrs}. 

Let us review what we know about these moments. We studied the second moment ($k=1$) earlier, with the result 
\es{M1M2sum}{ M_2(T) &= 
    \begin{cases}
      O(T\log^{4/3}T (\log\log T)^{2/3}) &\quad \textsf{(unconditional)}\\
      O(T(\log\log T)^2) &\quad \textsf{(on RH)} 
    \end{cases}}
So $b_1=1$, while $f_1$ depends on whether one does ($f_1=0$) or does not ($f_1\leq 4/3$) assume RH. By H\"older's inequality, this linear-times-logs behavior holds for all positive smaller $k$:
\e{}{M_{2k}(T) \lesssim T^{1-k}M_2(T)^k\qquad  (0\leq k \leq 1)\,.}
So $b_k=1$ for $k$ in the above range, with $f_k$ implied by $f_1$. Moreover, the first moment ($k=\thalf$) can be shown to obey a slightly stronger upper bound, without any logarithms, using Cauchy-Schwarz and $J(T) \approx T$:
\es{}{M_1(T) &= \int_1^T dt \big(\qz(t)|\z(1+2it)|\big)|\z(1+2it)|^{-1} \\&\leq \sqrt{J(T) I_{-2}(T)}\approx \sqrt{a_{-1}^{(\z)}}\, T}
where $a_{-1}^{(\z)} = 15/\pi^2$, so in fact $f_\half=0$. 

Now, the higher moments admit variations on two main possibilities:

\begin{enumerate}[label*=  \bf \arabic*)]

\item Only some moments have linear scaling, i.e. $\exists~ k>1$ such that $b_{k>1}>1$.

\item All moments have linear scaling, i.e. 
\e{Loa}{b_k = 1 \quad \Leftrightarrow \quad M_{2k}(T) \approx a_k T \log^{f_k}T~~\forall\,k\in\R_+}
This would imply that $\mu_\Q(\thalf)=0$. Different behaviors for $f_k$ are possible:

\bul $f_k=k^2$ (log normal): this is the behavior of standard (primitive) $L$-functions on the critical line, proven for $0\leq k \leq 2$ \c{radziwill12, heap19} and conjectured for higher $k\in\Z_+$ \c{cfkrs} (with a precise asymptotic coefficient $a_k$). 

\bul $f_k=k$ (normal): this is the behavior of Hurwitz zeta functions $\z(s,\a)$ on the critical line for transcendental $0<\a\leq 1$ (with irrationality exponent $\mu(\a)=2$), proven for $0\leq k \leq 2$ and conjectured for higher $k\in\Z_+$ \c{heap}.

\bul $f_k=0$: this is the behavior of standard $L$-functions on the 1-line. (This could also include $\log\log T$ terms in $M_{2k}(T)$ in principle.)

\bul Something else.
    
\end{enumerate}

\ni The natural suggestion here is that \eqr{Loa} indeed holds for all $k\in\Z_+$, and in turn, that $\mu_\Q(\thalf)=0$. It is plausible that this proposal can actually be determined with current methods, for the same reason that our derivation of linearity of $M_2(T)$ came from RMU of the CFT data.\foot{At risk of repetition, this is of a fundamentally different nature than the use of RMT in the moment conjectures for standard $L$-functions: in that setting, the presence of RMT statistics is based on a set of heuristic\thinspace---\thinspace albeit extremely compelling and highly robust\thinspace---\thinspace  techniques and numerical evidence. In contrast, our starting point is the presumption that CFTs obeying spectral rigidity exist.} Proof of $\mu_\Q(\thalf)=0$ could thus proceed by proving that $M_{2k}(T) \sim T$ follows from RMU. We sketch a possible argument along these lines. The idea is to relate $M_{2k}(T)$ to a $2k$-point spectral form factor,
\e{}{K_{2k}(T) := \<|\cZ(iT)|^{2k}\>}
written here at $\b=0$ for simplicity, where $T$ is Lorentzian time. For the case $k=1$, the linear scaling $M_2(T) \sim T$ derived above followed from the linear ramp of $K_2(T)$. The same simple manipulations that led to \eqr{M2bounds} gives its $2k$-fold generalization
\e{int937}{M_{2k}(T) \sim \int_1^T dt \big(\qz(t)|\z(1+2it)|\big)^{2k} \x (\text{logs})\,.}
It is known that spectral rigidity implies $K_{2k}(T) \sim T^{k}$ for late times $T$ (before the Heisenberg time) \c{Cotler:2017jue,Liu:2018hlr,Legramandi:2024ljn}. However, those leading contributions come from certain \textit{disconnected} terms in the $2k$-fold sum over energies: there are other contributions, linear in $T$, coming from connected terms. (See e.g. Appendix C of \c{Cotler:2017jue} and Section 2 of \c{Liu:2018hlr}.) So the goal is to show that the integral \eqr{int937} captures these specific connected terms of $K_{2k}(T)$. We leave a proper analysis for future work. If this strategy succeeds, it further implies that $f_k=0$\thinspace---\thinspace the moments $M_{2k}(T)$ have no logarithms\thinspace---\thinspace indicating that $\qz(t)$ fluctuates very weakly at large $t$. 

\ssec{Comments}\label{s94}

\enumcom

\item Let us restate the logic of the result \eqr{qav}. The framework of RMT$_2$ effectively studies correlators of $\cZ(\t)$ in spectral space \c{2307, 2503}. Upon factoring out some gamma functions, these are correlators of $\lz(s)$. The two-point function $\<\cZ\cZ\>$ contains the linear ramp in a theory obeying RMU, so one expects that this can be couched as a statement about the two-point function of $L_\cZ(s)$. This is what the second moment \eqr{Jint} is computing. 

\item The RMU lower bound $\mu_L(\half)=\half$ is reminiscent of general Epstein zeta functions of degree $d\geq 4$, which obey subconvexity but not Lindel\"of scaling \c{blomer2020epstein}. Moreover, the $d=4$ Epstein zeta series has $\mu(\half)=\half$. 

\item It is $\mu_\Q(\thalf)$, not $\mu_L(\thalf)$, that is allowed to vanish by RMU. Indeed, in general by \eqr{muqlcrit}, $\mu_\Q(\thalf)$ is less than $\mu_L(\thalf)$. Evidently, in their respective frequency sums evaluated on the critical line, $\qz(t)$ has more phase cancellation. 

\ni To understand what this might signify, consider an instructive discussion of the properties of the Hurwitz zeta function $\z(\thalf+it;\a)$ in \c{heap}. The upshot is that for sufficiently\foot{See the paper for the precise meaning of ``sufficiently''.} irrational $\a$, its fourth moment scales as $T\log^2 T$, which is slower than Riemann zeta by two logarithms. A heuristic explanation for the enhanced cancellations is given from the AFE point of view: expanding out the product of four series, Hurwitz zeta has more cancellation in the off-diagonal terms because the phases $(n+\a)^{-it}$ act more like independent and identically distributed random variables when $\a$ is transcendental. 

\item It would obviously be interesting if Riemann zeta universality holds \textit{locally}, not just on average, such that $\qz(t) \rar |\z(1+2it)|^{-1}$ either pointwise or almost everywhere as $t\rar\i$. (We have no real reason to expect this.) This would require proving that the variation of $g(t) := \qz(t)^2|\z(1+2it)|^2$ is sufficiently bounded, as 
\e{lmvc}{\int_T^{T+1} dt |g(t)-g(T)| = o(1) \qquad (T\rar\i)}
in the almost everywhere case (the average deviation across a unit interval vanishes, but pointwise maxima could be unbounded for a Lebesgue measure zero set of ``bad points''), or 
\e{}{\underset{t\in[T,T+1]}{\text{sup}}|g(t)-g(T)| = o(1)\qquad (T\rar\i)}
in the pointwise case (there are no unbounded oscillations in any bounded interval). 

\end{enumerate}

\sec{Generalizations}\label{s10}

\ssec{Extended chiral algebras}\label{s101}

We already explained in Section \ref{s3} how to construct $L$-functions for CFTs with extra holomorphic currents generating an extended chiral algebra $\mathcal{A}$, but here we register some formulas. We take
\e{}{N = \#\{\text{strong generators of $\cA$}\}}
where a ``strong generator'' cannot be written as a normal-ordered product of lower-spin currents ($N=1$ is the Virasoro, or $U(1)$, case). We consider irrational CFTs with respect to $\cA$, which necessarily have $c>N.$\foot{In irrational CFT, $N = c_{\rm currents}$, the effective central charge of the $\cA$ vacuum module.} The $\mathcal{A}$-primary partition function
\e{ZAP}{Z_p(\t) = (\sqrt{y}|\eta(\t)|^2)^{N}Z(\t)}
is once again modular-invariant, now with $q$-expansion 
\e{}{Z_p(\t) = y^{N/2} \,{\sum_{h,\hb}}' \,d_{h,\hb} \,q^{h-{c-N\o 24}}\qb^{\hb-{c-N\o 24}}\,.}
Following the same subtraction logic as in the Virasoro case, we construct $\cZ(\t)\in L^2(\cF)$, whose spectral overlap with the Eisenstein series defines a scalar spectral zeta function:
\e{}{(\cZ,E_{1-s}) := (2\pi)^{-s_N} \G(s_N)\zz^{(N)}(s)}
where 
\e{}{s_N := s+{N-2\o 2}}
and the $L$-function is
\e{}{\lz^{(N)}(s) = \z(2s)\zz^{(N)}(s)\,.}
The completed $L$-function $\Lamz^{(N)}(s)$ and its functional equation are
\es{LamzNdef}{\L^{(N)}_\cZ(s) &:= 2^s \g_{\cZ}^{(N)}(s) \lz^{(N)}(s)\\& = \L^{(N)}_\cZ(1-s)}
where
\e{gN}{\g_{\cZ}^{(N)}(s) = \pi^{-2s}\prod_{i=1}^4 \G\({s+\k_i\o 2}\)\,,\quad \{\k_i\} = \{0,1,\tfrac{N}{2}-1,\tfrac{N}{2}\}\,.}
Once again, this defines a degree-4, self-dual $L$-function with root number $\vareps=1$, now with $N$-dependent spectral parameters $\{\kappa_i\}$. All of these properties follow from the Rankin-Selberg integral representation
\e{LamzNint}{\Lamz^{(N)}(s) =  4\pi^{N\o2}\int_\cF{dx dy\o y^2} \cZ(\t) E^*_s(\t)\,.}
Developing the generalized Dirichlet series as
\e{GDSN}{\zz^{(N)}(s) = \sum_\l {a_\l \o \l^{s_N}}}
yields, by inverse Mellin transform, precisely the required form of the $y\rar\i$ expansion of the scalar mode $\cZ^{(0)}(y)$. Note that the pole at $s=1$ occurs at $s_N = \tfrac{N}{2}$.

The analyticity properties of $\Lamz^{(N)}(s)$ are essentially the same as the Virasoro case\thinspace---\thinspace with one interesting difference. The Rankin-Selberg integral \eqr{LamzNint} again implies that its only pole is a simple pole at $s=1$, and hence that $s(1-s)\Lamz^{(N)}(s)$ is entire; there are non-trivial zeros scattered throughout the complex $s$-plane, for which one can once again write down sum rules that directly generalize the $N=1$ case; and trivial zeros occur at $s_N=0,-1,-2,\ldots$, which again correspond to the vanishing positive moments of the density $a_\l$ via \eqr{GDSN}. However, when $N>1$, the $L$-function does not necessarily vanish at the central point $s=\half$:
\e{}{\underset{s=\half}{\text{ord}}\,\lz^{(N>1)}(s) \geq 0\,.}
The reason is that $\G(s_N)<\i$ at $s=\half$ when $N>1$.\foot{This special feature of the Virasoro case deserves further study in view of possible relations to elliptic curves or generalizations thereof.}

Finally, we may form the $N$-current generalization of the $L$-quotient $\qz(s)$. Removing both the $\sl$ and local conformal gamma factors, we construct
\e{qNdef}{\qz^{(N)}(s) = {\Lamz^{(N)}(s)\o 4 \pi^{N/2}\L(s)\L(s_N)} = {\lz^{(N)}(s)\o 2^{s_N}\z(2s)\z(2s_N)}}
where
\e{}{(\cZ,E_{1-s}) = \L(s_N) \qz^{(N)}(s)\,.}
The functional equation is
\e{}{\qz^{(N)}(s) = {\L(1-s)\o \L(s)}{\L({N\o2}-s)\o \L(s+{N\o2}-1)}\qz^{(N)}(1-s) \,.}
When $N=1$, this reduces to $\qz(s)$, which is reflection-symmetric. 

\ssec{Explicit example: Narain $L$-functions}\label{s102}

We now present the explicit $L$-functions for $U(1)^c$ primaries of Narain CFT at central charge $c$. As usual, the fact that Narain CFTs are theories of free bosons is what makes such explicit results possible. What follows uses previous computations of \c{2107}, to which we refer the reader for original proof of certain statements below. 

Before presenting the $L$-functions, let us note a special consequence of the free-ness of these theories. A theory of $N$ free bosons has $c=N$. Consequently, the asymptotic growth of $U(1)^c$ primaries is power law, not exponential, and likewise for the primary partition function $Z_p^{(c)}(\t) \sim y^{c/2}$ near the cusp at $y\rar\i$. Due to the fact that the renormalized Rankin-Selberg transform of an Eisenstein series vanishes \c{zagier} (this is a precise version of the statement that $(E_r,E_{1-s}) = 0$ for $r\neq s$), one can actually attach $L$-functions to \textit{individual} Narain CFTs, as opposed to differences thereof. We continue to use the $\cZ$ subscript notation for Narain CFTs (i.e. $\cZ$ stands for the unsubtracted $U(1)^c$-primary Narain partition function). 

The $U(1)^c$-primary spectrum of a Narain CFT is encoded in a quadratic form $M_{n,w}(m)^{2}=2\D_{n,w}(m)$, where $n$ and $w$ are momentum and winding vectors, respectively, living in a $c^2$-dimensional integral lattice, and $m\in\mathcal{M}_{\rm Narain}$ are the moduli with
\e{}{\mathcal{M}_{\rm Narain} := O(c,c;\ZZ)\backslash O(c,c;\RR)/O(c;\RR)\x O(c;\RR)\,.}
The scalar spectrum $\{\D^{(0)}_{n,w}(m)\}$, determined by the subspace of orthogonal vectors $n\cdot w=0$, may be packaged to form the constrained Epstein series
\e{}{\E^{(c)}_s(m) = \sum_{(n,w)\,\in\, \ZZ^c \x \ZZ^c\backslash\{(0,0)\}}{\delta_{n\cdot w,0}\o M_{n,w}(m)^{2s}}\,.}
This is a sum over orthogonal vectors spanning the Narain lattice. Its functional equation is most easily stated in terms of a completed Epstein zeta function
\e{}{\widehat \E^{(c)}_{s}(m) := \pi^{-s} \G(s) \L\({s-{c-2\o 2}}\)\E^{(c)}_{s}(m)}
whereupon, at $s=s_c$, this becomes invariant under $s \rar 1-s$:
\e{eq1225}{\widehat \E^{(c)}_{s+{c-2\o 2}}(m) = \widehat \E^{(c)}_{1-s+{c-2\o 2}}(m)\,.}

The scalar spectral zeta function for Narain CFT is\foot{Here and in a few places below, we suppress the $m$-dependence in our $L$-function notations.}
\e{zznarain}{\zz^{(c)}(s) = 2^{s_c} \E^{(c)}_{s_c}(m)\,.}
By \eqr{LamzNdef}, the completed $L$-function for Narain CFT is
\e{Lamznarain}{\L_\cZ^{(c)}(s) = 4\pi^{c/2}\widehat \E^{(c)}_{s_c}(m)\,.}
This is manifestly invariant under $s \rar 1-s$, as required, thanks to \eqr{eq1225}. 

Actually, this presentation obscures an essential physical feature of $\Lamz^{(c)}(s)$. Because Narain CFT spectra are governed by lattices, they enjoy a scaling symmetry which, moreover, preserves the orthogonality condition $n\cdot w=0$: given a scalar primary of dimension $\D^{(0)}_{n,w}$, the spectrum also contains primaries generated by taking
\e{scalmult}{(n,w) \rar k(n,w) \quad \Longrightarrow \quad \D^{(0)}_{n,w} \rar k^2\D^{(0)}_{n,w}}
for all $k\in\ZZ$. Without loss of generality, we can factor these rescalings from the Epstein zeta function as
\e{eq1217}{{\E^{(c)}_{s_c}(m) = \z(2s_c) \mathsf{E}^{(c)}_{s_c}(m)}}
where $\mathsf{E}^{(c)}_{s_c}(m)$ is the sum over all pairs of orthogonal vectors which are not related by integral scalar multiplication \eqr{scalmult}:
\e{}{\mathsf{E}^{(c)}_{s}(m) := \sum_{(n,w)\,\in \,\cV_{\rm core}}{(2\D_{n,w}(m))^{-s}}}
with
\e{}{\cV_{\rm core} := \left\{(n,w)\in \ZZ^c \x \ZZ^c\backslash\{(0,0)\}:n\cdot w=0\,,~(n,w)_i \neq k (n,w)_{j\neq i} ~~\forall\,k\in\Z\,,~\forall\,(i,j)\right\}}
comprising the set of ``core'' lattice vectors of the theory (indexed by $i$ and $j$). Accordingly, $\mathsf{E}^{(c)}_{s}(m)$ is the core scalar spectral zeta function of a Narain CFT. We may alternatively write this in terms of a set of core frequencies $\mu \in \cS_{\rm core}$, rather than lattice vectors $(n,w) \in \cV_{\rm core}$, as
\e{Ecoresum}{\mathsf{E}^{(c)}_{s}(m) := \sum_{\mu\,\in \,\S_{\rm core}}{\mu^{-s}}}
with
\e{}{\cS_{\rm core} := \left\{2\D_{n,w}(m): (n,w)\in \cV_{\rm core}\right\}\,.}
Now the completed $L$-function may be written cleanly as
\e{}{\L_\cZ^{(c)}(s) = 4\pi^{c/2}\L(s)\L(s_c)\mathsf{E}^{(c)}_{s_c}(m)\,.}
What this reveals is that, from \eqr{qNdef}, 
\e{}{\qz^{(c)}(s) = \mathsf{E}^{(c)}_{s_c}(m)\,.}
So $\qz^{(c)}(s)$ is not just a convenient quotient: it is the \textit{core scalar spectral zeta function} of a Narain CFT, from which the full scalar spectrum is generated by the union of discrete lattice dilatations. The frequency set $\cS_{\rm core}$ is the frequency set $\{\mu_n\}$ of $\qz^{(c)}(s)$. This reinforces our earlier remarks regarding $\qz^{(N)}(s)$ as a core object more generally.

\sssec{Analytic structure}

The analytic structure of the $L$-function is easily determined using the analysis of \c{Angelantonj:2011br} of the constrained Epstein zeta series, and provides a check of our general analysis of Section \ref{s101} with $N=c$.

The poles of $\widehat\E^{(c)}_{s_c}(m)$, and their implications for the $L$-function and series representations, are as follows: 
\begin{itemize}

\item $\widehat\E^{(c)}_{s_c}(m)$ is meromorphic with simple poles at $s=1,{c\o2}$ and their reflections at $s=0,1-{c\o 2}$.\foot{See (2.25) and (3.17) of \c{Angelantonj:2011br}, with $\mathcal{R}^*(s) \mapsto \widehat \E^{(c)}_{s_c}(m)$.} By \eqr{Lamznarain}, these are the poles of $\Lamz^{(c)}(s)$. For $c=1,2$, familiar degenerations occur (see the next subsection).

\item The series representation of $\E^{(c)}_{s_c}(m)$ is absolutely convergent for $\Re(s_c)>c-1$.
This can be ascertained from the known asymptotic Cardy density of $U(1)^c$ scalar primaries \c{Cohn}\foot{The scalar Cardy density is down by $1/\D$ relative to the full Cardy density, which is important here.}
\e{}{a^{(0)}_{\rm Cardy}(\D)\approx {2\pi^c \z(c-1)\o \z(c) \G({c\o2})^2}\D^{c-2} \qquad (\D\rar\i)}
So the zeta function
\e{}{\zz^{(c)}(s) = 2^{s_c} \E^{(c)}_{s_c}(m) = \sum_\D {a_\D\o \D^{s_c}}}
converges absolutely when $\Re(s_c)>c-1$. In terms of $s$, this corresponds to $\s={c\o 2}$, matching the location of the right-most pole of $\E^{(c)}_{s_c}(m)$ in agreement with \eqr{GDSN}. 

\item We note also that, from \c{BC},
\e{BCres}{\E^{(c)}_0(m)=-1\,.}
This implies that the residue of the pole of $\widehat\E^{(c)}_{s_c}(m)$ at $s={c\o 2}$ is moduli-independent, as seen using the functional equation:
\e{}{\Res_{s={c\o2}}\widehat \E^{(c)}_{s_c}(m) = -\L\(1-{c\o2}\)\,.}
This means that in the \textit{difference} of two Narain partition functions, the right-most pole moves to $s=1$ (i.e. $s_c = {c\o 2}$), and the completed $L$-function has simple poles only at $s=0,1$. This matches the result from the general analysis of Section \ref{s101} for CFTs with $N$ generating currents, providing both a useful check on that and on the residue result \eqr{BCres} itself. 

\end{itemize}

As for the zeros, the subject of Epstein zeta zeros is a rich one. In general, Epstein zetas do not obey a Generalized Riemann Hypothesis and, lacking an Euler product, can even have zeros in the half-plane of absolute convergence. For a sample of work on this subject, see \c{stark, ribeiro-yakubovich22, rezvyakova24}.

\sssec{Low $c$}

Specializing to $c=1,2$ allows us to give yet-more-explicit formulas using \c{2107}.
 
At $c=1$, 
\e{}{\qz^{(c=1)}(s) = 2(r^{2s-1}+r^{1-2s})\,.}
We read off the core frequency set
\e{corec1}{\cS_{\rm core}^{(c=1)} = \left\{{r^2},{r^{-2}}\right\}\,.}
The $L$-function and its completion are
\es{}{L_\cZ^{(c=1)}(s) &= 2^{s+\half}(r^{2s-1}+r^{1-2s})\z(2s)\z(2s-1)\\
\L_\cZ^{(c=1)}(s) &= 4\sqrt{\pi} 2(r^{2s-1}+r^{1-2s})\L(s)\L(1-s)\,.}
The meaning of these Riemann zeta factors was originally unclear in \c{2107}, but the $L$-function perspective gives their appearance a certain synergistic explanation: the ``simplest'' CFT has the ``simplest'' $L$-function!

A few remarks on analyticity. First, $\L_\cZ^{(c=1)}(s)$ has simple poles at $s=0,1$ and a double pole at $s=\half$, as expected from degeneration of the $c>2$ case. Second, we point out two notable features of $\qz^{(c=1)}(s)$: the absence of Riemann zeta poles, and that $\cS_{\rm core}^{(c=1)}$ obeys {\it reciprocity}, i.e. invariance under inversion of every element. As explained in Appendix \ref{appd}, these two features are one and the same: convergence of the series down to the critical line implies reciprocity. Whereas either feature would be forbidden in an irrational Virasoro CFT, the $c=1$ free boson is not irrational. Indeed, here we verify the manifest interplay of these two properties. Finally, we can also explicitly verify the zero density bound \eqr{gdb}. At $c=1$, the gap is bounded as $\l_1\leq \half$, with saturation at the self-dual radius $r=1$. Indeed, the self-dual boson saturates the density bound: at $r=1$, $\qz^{(c=1)}(s)$ is constant.

At $c=2$,
\e{}{\qz^{(c=2)}(s) = \mathsf{E}_s^{(c=2)}(m) = 2E_s(\boldsymbol\rho)E_{s}(\boldsymbol\s)}
where $\boldsymbol\rho$ and $\boldsymbol\s$ are (in an unfortunate clash of standard notational conventions) modular parameters for $\sl$ fundamental domains, $\cM_{\rm Narain}^{(c=2)} = \cF_{\boldsymbol\rho}\x \cF_{\boldsymbol\s} / \Z_2 \x \Z_2$. From the series representation \eqr{Esgds} of the Eisenstein series as a sum over coprimes, we read off the core frequency set
\e{}{\cS_{\rm core} ^{(c=2)}= \left\{ {|a\boldsymbol\rho+b|^2\o \Im(\boldsymbol\rho)}{|c\boldsymbol\s+d|^2\o\Im(\boldsymbol\s)}:(a,b)=1\,,(c,d)=1\right\}\,.}
The $L$-function and its completion are 
\es{}{L_\cZ^{(c=2)}(s) &= 2^s\z(2s)^2E_s(\boldsymbol\rho)E_s(\boldsymbol\s)\\
\L_\cZ^{(c=2)}(s) &= 8\pi E^*_s(\boldsymbol\rho)E^*_s(\boldsymbol\s)\,.}
$\L_\cZ^{(c=2)}(s)$ has double poles at $s=0,1$, as expected from degeneration of the $c>2$ case. 

\ssec{Correlator $L$-functions}\label{s103}

$L$-functions may be attached to other CFT observables besides torus partition functions. 

Torus one-point functions of Virasoro primaries with dimensions $(h,\hb)$ are modular-covariant, with weight $(h,\hb)$. Accordingly, we can attach $L$-functions to scalar one-point functions dressed by a power of $y^h$, which renders them modular-invariant:
\e{}{y^h \<\cO\>_\t \quad \longrightarrow \quad L_{\<\cO\>}(s)\,.}
$L_{\<\cO\>}(s)$ will again be of degree-4, self-dual with root number $\vareps=1$, now with spectral parameters
\e{}{\{\kappa_i\} = \{0,1,h,h-1\}\,.}
This is the same analytic structure as the $L$-function attached to a CFT partition function with $N = 2h$ currents, cf. \eqr{gN}. The torus one-point conformal blocks for general $h$ and $c$ are unknown analytically, so the $L$-function frequencies are not just primaries. In the special case $h=1$, the blocks are simply characters, so we can append a further $\sqrt{y}|\eta(\t)|^2$, and the attached $L$-function will be primary, now with $\{\kappa_i\} = \{0,1,\tfrac{3}{2}, \thalf\}$. 

A similar construction exists for scalar four-point functions on the sphere after mapping to the pillow metric \c{Maldacena:2015iua}. Consider a four-point function of identical scalar primaries $\cO$ of dimension $\D=2h$, written as 
\e{}{\<\cO(0)\cO(z,\zb)\cO(1)\cO(\i)\> = \left|\theta_3(q)^{{c\o2}-16h}(z(1-z))^{{c\o 24}-2h}\right|^2 g_\cO(q,\qb)}
where 
\e{}{q:= e^{-\pi {K(1-z)\o K(z)}}\,,\quad K(z) = \half\int_0^1 {dt\o \sqrt{t(1-t)(1-zt)}}\,.}
Like torus one-point functions, $g_\cO(q,\qb)$ is covariant under modular transformations. The modular-invariant product, and its associated $L$-function, are 
\e{}{y^{{c\o2}-16h}g_\cO(q,\qb)\quad \longrightarrow \quad L_{\<\cO\cO\cO\cO\>}(s)}
$L_{\<\cO\cO\cO\cO\>}(s)$ will again be of degree-4, self-dual with root number $\vareps=1$, now with spectral parameters
\e{}{\{\kappa_i\} = \left\{0,1,{c\o2}-16h,{c-2\o2}-16h\right\}\,.}
This is the same analytic structure as the $L$-function attached to a CFT partition function with $N = c-32h$ currents, cf. \eqr{gN}. It is curious that the Virasoro case $N=1$ corresponds to the case $h={c-1\o 32}$, which is a special threshold: it is the conformal weight above which there are no light composite operators in Virasoro Mean Field Theory \c{VMFT,Kusuki:2018wpa}. In addition, note that taking $h>{c\o 32}$, which includes all heavy operators, drives two of the spectral parameters negative. 

We close with a couple of technical points. First, for $L_{\<\cO\>}(s)$, subtraction may not be necessary because the vacuum module never appears as a channel of $\<\cO\>_\t$, due to conformal invariance. Second, convergence properties of their $L$-series now rely on the Cardy OPE asymptotics, which introduce exponentially small damping factors from heavy correlations \c{Collier:2019weq}.

\sec{Dirichlet, Maass, Riemann: Toward spinning $L$-functions}\label{s11}

A natural question is whether and how the $L$-function framework presented here generalizes to spinning sectors of CFT primaries. We take a step in this direction using $\sl$ spectral decomposition, in the notation of \c{2307}. In doing so, we will point out some intriguing interrelations between the Riemann zeta function and Maass cusp forms for $\sl$, implied by the existence of Virasoro CFTs with discrete spectra. 

Any modular-invariant $\cZ(\t)\in L^2(\cF)$ admits a spectral decomposition into a complete eigenbasis of $\sl$-invariant functions, comprised of the constant function, the non-holomorphic Eisenstein series with $s=\half+it$, and the Maass cusp forms $\{\phi_n(\t)\}$ with $n\in\Z_+$. These functions have Fourier expansions\foot{For simplicity we restrict our attention to even Maass cusp forms here, relevant for parity-invariant CFT observables.}
\es{Ephifourier}{
E^*_s(\tau)&= \sum_{j= 0}^\i(2-\delta_{j,0})\sfa_j^{(s)} \cos(2\pi j x)  \sqrt{y} K_{s-\half}(2\pi j y)\\ 
\phi_n(\tau)&= \sum_{j= 1}^\i 2\sfb_j^{(n)} \cos(2\pi j x) \sqrt{y} K_{s_n-\half}(2\pi j y)
\,.}
The Eisenstein Fourier coefficients are 
\e{}{\mathsf{a}_j^{(s)} = {2\s_{2s-1}(j)\o j^{s-\half}} = 2\sum_{d|j} \({d^2\o j}\)^{s-\half}}
whereas the $\sfb_j^{(n)}$ are unknown analytically. Likewise for the parameters $s_n := \half+iu_n$, where
\e{}{\D_\t\phi_n(\t) = \({1\o4}+u_n^2\) \phi_n(\t)\,,\quad u_n\in\RR_+\,.}

In the microcanonical ensemble, the spin-$j$ Virasoro primary eigendensity is \c{2307}
\e{rhosj}{\rho^*_{s,j}(\l) = \mathsf{a}_j^{(s)}{\Theta(\l-j)\o \sqrt{\l^2-j^2}}\cosh\(\(s-\thalf\)y_j(\l)\)}
for the completed Eisenstein series, where\foot{In this section $\rho_j(\l)$ really stands for $\rho_{|j|}(\l)$, i.e. it is the density for both spins $\pm j$. This is clear from \eqr{Ephifourier} and the derivation of \eqr{rhosj} in Appendix \ref{appj}. We continue to use the notation $\l=\D-{c-1\o 12}$.}
\e{yjldef}{y_j(\l) := \cosh^{-1}\({\l\o j}\)\,,\quad {\p y_j(\l)\o \p \l} = {1\o \sqrt{\l^2 - j^2}}\,.}
and likewise for the cusp forms with the substitution $\mathsf{a}_j^{(s)} \rar \mathsf{b}_j^{(s_n)}$. We re-derive this in Appendix \ref{appj}. By spectral decomposition of $\cZ(\t)$, its Virasoro primary density of spin-$j$ is thus
\e{rhojintt}{\rho_j(\l) = {\Theta(\l-j)\o \sqrt{\l^2-j^2}}\Bigg({1\o4\pi i}\int\limits_{~\big(\thalf\big)} ds \,\mathsf{a}_j^{(s)} \qz(s) \cosh\(\(s-\thalf\)y_j(\l)\) + \sum_n \mathsf{b}_j^{(s_n)} (\cZ,\phi_n) \cos\(u_n\,y_j(\l)\)\Bigg)\nonumber}

Consider the Eisenstein term of \eqr{rhojintt}, integrated over the critical line. To make contact with the scalar frequency spectrum, we shift the contour rightward to the region where $\qz(s)$ can admit a convergent series representation \eqr{zlqser}. If $\qz(s)$ has Riemann zeta poles\thinspace---\thinspace recall from Section \ref{s4} that these are its only possible poles\thinspace---\thinspace we will pick their residues up along the way. So the total spin-$j$ density is
\e{spinj}{{\rho_j(\l) = \rho_j^{\rm (D)}(\l) + \rho_j^{(\z)}(\l) + \rho_j^{(\phi)}(\l)}}
The first term $\rho_j^{\rm (D)}(\l)$, where ``D'' stands for Dirichlet, is the one we will analyze below:
\es{}{\rho_j^{\rm (D)}(\l) &= {2\Theta(\l-j)\o \sqrt{\l^2-j^2}} \x{1\o 4\pi i}\int\limits_{~(\s)} ds \sum_\mu a_\mu \sum_{d|j} \({d^2\o j\mu}\)^{s-\half}\cosh\(\(s-\thalf\)y_j(\l)\)}
with $\s$ to the right of the non-trivial zero of $\z(2s-1)$ with largest real part (on RH, $\s>\tfrac{3}{4}$). The other two terms come from the Riemann zeta and Maass cusp terms, respectively: 
\es{rhorem}{\rho_j^{(\z)}(\l)  &=  {\Theta(\l-j)\o 2\sqrt{\l^2-j^2}} \sum_{m}\Res_{s\,=\,\tfrac{1+\rho_m}{2}}\(\mathsf{a}_j^{(s)}{\qz(s)}\)\cosh\(\tfrac{\rho_m}{2}y_j(\l)\) + (\text{c.c.})\\
\rho_j^{(\phi)}(\l) &= {\Theta(\l-j)\o \sqrt{\l^2-j^2}} \sum_n \mathsf{b}_j^{(s_n)}(\cZ,\phi_n)\cos\(u_n\, y_j(\l)\)}
where we recall that $\rho_m$ stands for a non-trivial Riemann zeta zero. 

\ssec{A discrete spinning density}
We now derive a more transparent expression for the Dirichlet part, $\rho_j^{\rm (D)}(\l)$. The idea is to ask: what does a given scalar frequency $\mu$ contribute to the spinning density? The surprise is that the resulting density is \textit{discrete}\thinspace---\thinspace a feature unique to the Virasoro case, as we will see. 

By holomorphy of the integrand, we are free to choose any convenient $\s$ to the right of the Riemann zeta poles. We choose $\s=1$. Now, let us swap the integral with the sum over $\mu$, keeping in mind that convergence of the series for $\qz(s)$ is a necessary condition (though possibly not sufficient) for this to hold. We need to perform the integral
\e{}{\mathcal{I}(x,y) := {\sqrt{x}\o 4\pi}\int_{-\i}^\i dt\, x^{it}\cosh\(\(\thalf+it\)y\)}
in terms of which the Dirichlet density is
\e{rhoixy}{\rho_j^{\rm (D)}(\l) = 2\Theta\(\l-j\){\p y\o \p \l} \sum_\mu a_\mu \sum_{d|j} \mathcal{I}\({d^2\o j\mu},y_j(\l)\).}
The integral is a Fourier transform:
\e{Ixyfourier}{\mathcal{I}(x,y) = {\sqrt{x}\o 4}\(e^{-y/2}\d(\log x-y) + e^{y/2}\d(\log x +y)\)\,.}
Viewing $\log x \pm y$ as a function of $\l$, 
\e{}{\d(\log x\pm y) = {\d(\l-\l_0)\o \left|{\p y\o \p \l}\right|_{\l_0}}\,, \quad \l_0 = {j\o2}(x+x^{-1})}
where the zero $\l_0$ exists for the delta function with $-$ sign when $x>1$, and $+$ sign when $x< 1$. The case $x=1$ receives an equal contribution from both (these are $\l=j$ states, exactly at the twist threshold $\D-j={c-1\o 12}$). So altogether,
\e{}{\mathcal{I}(x,y) = {(1+\delta_{x,1})\o 4\left|{\p y\o \p \l}\right|_{\l_0}}\d(\l-\l_0)}
for all $x$, where delta functions let us replace $\sqrt{x} e^{\pm y/2} \to 1$ in \eqr{Ixyfourier} upon viewing the density of states as living under an integral over $\l$. Plugging into \eqr{rhoixy} gives
\e{}{{\rho_j^{\rm (D)}(\l) = {1+\delta_{\l,j}\o 2}\Theta\(\l-j\)\sum_\mu a_\mu \sum_{d|j} \d\(\l - {j\o2}\({d^2\o j\mu}+\({d^2\o j\mu}\)^{-1}\)\)}\,.}
For $\mu\in\mathbb{R}$, which follows by unitarity of the CFT spectrum,
\e{}{\underset{d|j}{\text{min}} \[{j\o2}\({d^2\o j\mu}+\({d^2\o j\mu}\)^{-1}\)\] \geq j}
so the Heaviside theta is automatically turned on; and in the usual convention $\Theta(0)=\thalf$, the factor ${1+\delta_{\l,j}\o 2}\Theta\(\l-j\)$ equals $\thalf$ for both threshold and above-threshold states. So we can simply yet further to arrive at the final result 
\ebox{rhojD}{{\rho_j^{\rm (D)}(\l) = \half\sum_\mu a_\mu \sum_{d|j} \d\(\l - {j\o2}\({d^2\o j\mu}+\({d^2\o j\mu}\)^{-1}\)\)}}

It is a pleasant surprise that this is a discrete density. As we show in Appendix \ref{appj}, the analogous computation for a CFT with an extended chiral algebra does \textit{not} give a discrete density, so this is to be regarded as yet another special feature of Virasoro CFT $L$-functions in particular. It is also notable that it is the frequency spectrum $\{\mu_n\}$ of $\qz(s)$, not that of $\lz(s)$ or $\zz(s)$, that directly determines the spinning spectrum. 

The contribution of $\rho_j^{(\rm D)}(\l)$ to the spin-$j$ frequency spectrum $\l = \D^{(j)}- {c-1\o 12}$ is therefore given by the set 
\e{specjD}{\left\{{j\o2}\({d^2\o j\mu}+\({d^2\o j\mu}\)^{-1}\):\mu\in\{\mu_n\}\,,\,d|j\right\}}
For a fixed $\mu\in\{\mu_n\}$ the spin-$j$ operator has multiplicity $a_{\mu_n}/2$. The gap to the spin-$j$ state of smallest conformal dimension (call it $\D_1^{(j)}$) is determined by the minimum of \eqr{specjD} over both $\mu$ and $d|j$. We can conveniently write this as a twist gap bound,
\e{jgap}{\D_1^{(j)} - j \leq  {c-1\o 12} + {j\o2}\,\min\limits_{\mu,\,d|j}\(\sqrt{d^2\o j\mu}-\sqrt{j\mu\o d^2}\)^2\,.}
Closure of the spin-$j$ twist gap to ${c-1\o 12}$ occurs if and only if there exists an $n$ such that $\mu_n=d^2/j$ for some divisor $d|j$. (In the $c=1$ free boson context, these would be rational points of symmetry enhancement by emergent conserved currents.) As a corollary, if $\mu_n\notin\QQ$ for all $n$, the twist gap above ${c-1\o 12}$ must be nonzero.

\ssec{Riemann vs. Maass}\label{s112}
The \textit{full} spin-$j$ density is given in \eqr{spinj}, which augments the Dirichlet term \eqr{rhojD} with the Riemann and Maass terms. So the statements of the previous paragraph hold for the actual spinning spectrum of $\cZ(\t)$ only if the Riemann and Maass terms do not somehow \textit{cancel} these terms; for example, absent such cancellation, \eqr{jgap} would be a robust upper bound on the spin-$j$ twist gap. Whether such cancellation is possible is a fascinating question that involves complicated aspects of sums over Maass cusp forms and Riemann zeta zeros, well beyond the scope of this work.\foot{We thank Nathan Benjamin for discussions on this topic.}

However, even without asking whether \textit{cancellation} is possible, one can already infer the need for some remarkably intricate structure solely from imposing \textit{discreteness}. In particular, the Riemann and Maass terms must together preserve discreteness of the total density. The condition is
\e{rhojdisc}{\rho_j^{(\z)}(\l) + \rho_j^{(\phi)}(\l) \stackrel{!}{=} \sum_n c_j^{(n)} \d(\l-\l_j^{(n)})}
for some set of spin-$j$ frequencies $\{\l_j^{(n)}\}\in\RR$ and coefficients $\{c_j^{(n)}\}\in\R$. The power of this observation is that neither $\rho_j^{(\z)}(\l)$ nor $\rho_j^{(\phi)}(\l)$ is manifestly discrete on its own (see \eqr{rhorem}). This makes \eqr{rhojdisc} a non-trivial condition, raising several interesting questions. Perhaps the main point is the following: \textit{\eqr{rhojdisc} implies relations between the Riemann zeta zeros $\{\rho_n\}$ and the Maass cusp form spectral parameters $\{u_n\}$.} This can be turned into a quantitative, not just a qualitative, constraint as an infinite set of sum rules polynomial in the Riemann and Maass data. 

Needless to say, $\{\rho_n\}$ and $\{u_n\}$ are two famously enigmatic sets of chaotic number-theoretic data. No element of either set is known analytically. But by \eqr{rhojdisc}, they are bound together by sum rules that follow from enforcing discreteness. It would be very exciting to make this observation fully rigorous, and seek a functional algorithm for the mutual determination of these data. The pursuit of this possibility, and a complete analysis of spinning $L$-functions for CFTs, is left for the future.

\sec{Discussion}\label{s12}

Analytic number theory should provide a natural language for resolving black hole spectra and semiclassical asymptotia of conformal field theory data. We have initiated an exploration of $L$-function formalism in general two-dimensional conformal field theories. There appear to be many stimulating connections between central notions in these two subjects. We close by discussing a handful of the most interesting conceptual directions. 

\textbf{Families of CFTs:} A pressing issue for abstract CFT (in any dimension) is that there is no obvious non-perturbative meaning of a ``family of CFTs''. Apart from certain examples, such as those which happen to admit a Lagrangian formulation with an index $N$ defined by the presence of a gauge field, we do not understand how to group CFTs in any formal, or even natural, sense.\foot{Reference to ``a theory labeled by $N$'' is not well-defined from an abstract CFT point of view: what does it mean to distinguish a ``single'' theory from an interpolation among ``different'' theories?} Can we define a ``family of CFTs'' using ideas from $L$-functions, in some adaptation of the Katz-Sarnak philosophy?

In the $L$-function context, the notion of a family $\mathsf{F}$ has been deeply influential both taxonomically and pragmatically, as we briefly reviewed in Section \ref{s2}. $L$-functions are usually grouped into infinite families such that the analytic conductor varies over $\mathsf{F}$. On the other hand, the CFT $L$-functions constructed herein have a \textit{fixed} analytic conductor once we specify the chiral algebra $\mathcal{A}$. This means that the more literal CFT analog involves grouping theories with increasing number of currents $N$, the latter playing the role of the conductor $q$.\foot{One concrete suggestion is to understand symmetric orbifold CFTs as a family of $L$-functions. These are built from a reference (seed) modular function, in a similar fashion as families of modular forms twisted by Dirichlet characters.} It also means that the question we would especially like to answer\thinspace---\thinspace how do we group $c>1$ Virasoro CFTs into families?\thinspace---\thinspace still appears puzzling. 

We emphasize that the question of CFT families is important not only formally, but for questions currently under research investigation in AdS/CFT. Let us highlight two: first, for understanding the emergence of semiclassical spacetime in AdS/CFT, which requires grappling with the meaning of ``large $N$ limits'' (e.g. \c{Schlenker:2022dyo,Liu:2025cml,Gesteau:2025obm}); and second, for giving an abstract definition of ``fortuity'' in BPS state counting, which so far relies on the existence of an $N$ in a sense only understood in gauge theories with trace relations \c{Chang:2024zqi}.

\textbf{Finding and understanding the non-trivial zeros:} We derived several constraints on the non-trivial zeros of CFT $L$-functions, but they were rather general, and did not probe the physical nature of the zeros. It seems possible that in chaotic CFTs, the zeros of $\lz(s)$ are actually simply distributed, without random matrix statistics: since the frequencies $\{\l_n\}$ are chaotic,  perhaps $\lz(s)$ ``inverts'' which quantities are complex/random/chaotic and which are not relative to standard $L$-functions.\foot{We thank Steve Shenker for a discussion about this possibility.} As for where the zeros may be, the tools presented here should be capable of more sharply constraining their locations. One might expect zeros in the half-plane of convergence based on precedent from $L$-functions lacking Euler products. We should emphasize that while we are not aware of a possible Euler product representation of $\lz(s)$, we have not yet shown that there \textit{cannot} be one. Of course, it would be absolutely remarkable if $\lz(s)$ had an Euler product, as it would imply that CFT scalar spectra are constructible multiplicatively from a set of atomic dimensions. Clearly, future work should address this (in our view, remote) possibility. 

\textbf{A Riemann-Siegel formula for CFT $L$-functions:} We derived an approximate functional equation in Section \ref{s8}. It would be useful to develop a truncation algorithm with controlled errors. If possible, this would lead, following methods from standard $L$-functions, to an effective estimator for the low-lying zeros of $\lz(s)$ in terms of its low-lying frequencies $\{\l_n\}$. The key question is whether the errors will be sufficiently small so as to shorten the sum in a useful way; in particular, the asymptotic Cardy level spacings among the $\{\l_n\}$ could lead to an effective cutoff on the index $n$ that is larger than power law, in contrast to \eqr{AFEapprox2}. If, on the other hand, the cutoff is parametrically below the scale where Cardy growth kicks in, that would be especially nice.  

\textbf{Bootstrapping the zeros:} The $L$-function is constructible from its zeros. This suggests an alternative to the usual conformal or modular bootstrap in which it is the zeros, not the dimensions, that are being bootstrapped. We envision a hybrid approach: first, ``seed'' the $L$-function by low-lying spectral data, then read off the approximate low-lying zeros, then feed this back into the $L$-function (e.g. integrating the Hadamard product against an exponential factor \`a la Schnee's formula). The idea here is similar to what has been effectively implemented in the large spin lightcone bootstrap \c{Fitzpatrick:2012yx,Komargodski:2012ek,Simmons-Duffin:2016wlq,Caron-Huot:2017vep}, in which one seeds the algorithm with low-twist states, uses crossing to deduce high-spin asymptotic data, then feeds that data back into the crossed channel to derive new refinements, etc. An effective truncation scheme for the approximate functional equation would be useful for this algorithm. 

\textbf{``Elliptic curves for CFTs'':} It would be interesting to ask whether there is some generalized lattice structure, or elliptic curve, that can be associated to generic 2d CFTs. (See \c{Kondo} for a more literal version of this.) We remark that the $L$-quotient $\qz(s)$ has the functional shape of (the recpirocal of) the Hasse-Weil zeta function. We also noted earlier some properties of $\lz(s)$ near the central point $s=\half$, in particular its vanishing, and it would be worth asking if there is a BSD-type interpretation. Further on, it would be of clear interest to ask whether a ``BSD for CFT'' conjecture could be sensibly formulated.\foot{We thank Sean Hartnoll for an early conversation about the BSD conjecture.}

\textbf{$J$-completion for AdS$_3$ pure gravity:} In Section \ref{s32} we introduced $J$-completion as a procedure by which to render a given light spectrum fully $\sl$-invariant, to be subtracted from a partition function of interest. But it is interesting to consider the constructive point of view on $J$-completion: under what conditions on the central charge $c$ is the $J$-completion of a given light spectrum {\it unitary and integral}? The $J$-completion of the Virasoro vacuum module is a natural target: it is a compact alternative to the non-unitary, non-compact MWK construction \c{MW,MK} of a partition function of AdS$_3$ pure gravity. By enforcing unitarity and integrality on this object, one extracts number-theoretic constraints on the allowed values of $c$ \c{yiannis}. 

\textbf{Black hole zeros and the $L$-function of AdS$_3$ pure gravity:} What is the $L$-function of AdS$_3$ pure gravity? This could be particularly clean analytically\thinspace---\thinspace an \textit{extremal $L$-function}. Do the ``BTZ black hole zeros'' have special structure? As per our previous remarks, there may be multiple $L$-functions which have a gap to the black hole threshold, parameterizing the freedom in the scalar microstate spectrum.  

\textbf{String universality of $L$-functions:} String/M-theory have not yet arisen in this work. It is interesting to ask how the analytic structure of the CFT $L$-functions would reflect the existence of an AdS$_3$ string/M-theory dual.\foot{Whatever the answer, it seems that we should call these objects $M$-functions.} Perhaps this could help identify sharp stringy signatures on black hole microstate spectra. It is also natural to revisit string universality, the broad idea that every consistent quantum gravity theory is a string theory. What is string universality from an $L$-function point of view (if anything)? 

\textbf{OPE randomness as critical behavior:} We briefly introduced correlator $L$-functions in Section \ref{s10}, whose systematic understanding would be good to pursue, for instance, as a possible probe of OPE statistics. Generalized spectral form factors for one-point functions have been introduced in \c{Belin:2021ibv}. In a chaotic CFT, the two-point generalized spectral form factor has a linear ramp and three-point coefficients are believed to obey the ``OPE Randomness Hypothesis'' \c{Belin:2020hea}. Does OPE randomness imply subconvexity of $L_{\<\cO\>}(s)$ on the critical line, or other bounds?

\textbf{Toward a full definition of random matrix universality in 2d CFT:} In Section \ref{s7} we laid out the problem of random matrix universality of 2d CFT, and developed a local conformal avatar of extreme gap statistics, leading us to explore the Simple Extreme Conjecture in Section \ref{s71}. The RMT results \c{vinson, BAB,feng2019small,feng2020smallgapscircularbetaensemble, fengsmallgapsgse, bourgade2021extreme,fenglargegapsgue} for extreme gaps appear not to have previously made their way to the CFT literature; it would be worthwhile to interpret the full slate of results in CFT terms. There are many other basic aspects of random matrix universality in 2d CFT worth studying, among them: the behavior of nearest-neighbor spacings $\{\d_n^{(j)}\}$ at intermediate scales (the approaches of \c{Altland:2020ccq,TD} to this crossover in RMT might be useful for addressing this); the gap ratio \c{Huse, gap1, gap2}; the rate of approach to RMT statistics as $\D^{(j)}\rar\i$; and relations to notions of chaos, in the spirit of the BGS conjecture. It would also be interesting to determine the asymptotic extreme gap scales $\d_{\min}^{(j)}$ and $\d_{\rm max}^{(j)}$ in $c>1$ Virasoro CFT, analogous to their RMT counterparts \eqr{smalllargeS}. A restricted version of this question would be to determine these scales assuming the presence of a linear ramp in the spectral form factor, thereby seeding the theory with a random statistic.\foot{We thank Sridip Pal for long-term discussions on this topic.}

Application of the $L$-function approach to bootstrap questions will be presented elsewhere. 

\sec*{Acknowledgments}

It is a pleasure to thank Amina Abdurrahman, Anshul Adve, Jan Albert, Nathan Benjamin, Dustin Clausen, Tom Hartman, Dalimil Maz\'a\v{c}, Hirosi Ooguri, Sridip Pal, Steve Shenker, Douglas Stanford and Yiannis Tsiares for discussions. We thank the organizers of Strings 2025, where this work was first presented. This work was made possible by Institut Pascal at Universit\'e Paris-Saclay with the support of the program ``Investissements d'avenir'' ANR-11-IDEX-0003-01. This work was performed in part at the Aspen Center for Physics, which is supported by National Science Foundation grant PHY-2210452. This work was supported in part by ERC Starting Grant 853507. We thank the IHES for providing an environment especially conducive to the development of this work.  

\begin{appendix}

\sec{Non-holomorphic Eisenstein cheat sheet}\label{appa}

Here we give three representations of the non-holomorphic Eisenstein series $E_s(\t)$ for $\sl$, a modular-invariant eigenfunction of the hyperbolic Laplacian
\e{Eslaplace}{\D_\t E_s(\t) = s(1-s)E_s(\t)}
where $\D_\t = -y^2(\p_x^2+\p_y^2)$. This obeys a functional equation
\e{}{E_s^*(s) = E_{1-s}^*(\t)}
where $E_s^*(\t) := \L(s) E_s(\t)$ is the completed Eisenstein series and $\L(s)$ was defined in \eqr{Lamdef}.

The Poincar\'e series representation is a sum over group elements: 
\e{Espoincare}{E_s(\t) = \sum_{\g\,\in \,\G_\i\backslash PSL(2,\ZZ)} \Im(\g \t)^s}
where $\G_\i$ is the subgroup of $PSL(2,\ZZ)$ that fixes $y$. 

The generalized Dirichlet series representation, derivable directly from \eqr{Espoincare}, is a sum over coprimes: 
\e{Esgds}{E_s(\t) = \half\sum_{(m,n)=1} \({y\o |m\t+n|^2}\)^s\,.}

The Fourier representation meromorphically continues $E_s(\t)$ from $\s>1$, where the above representations converge absolutely, to all $s\in\CC$:
\e{Esfourier}{E_s(\t) = y^s + {\L(1-s)\o \L(s)} y^{1-s} + \sum_{j=1}^\i 4\cos(2\pi j x) {\s_{2s-1}(j)\o j^{s-\half}\L(s)} \sqrt{y}K_{s-\half}(2\pi j y)\,.}
$E_s(\t)$ has poles at the locations of the zeros of $\L(s)$, which are the non-trivial zeros of $\z(2s)$. 

\sec{Phragmen-Lindel\"of for strips}\label{appb}

We quote essentially verbatim from \c{IKtext}, Appendix A.2, and \c{Titchmarsh1939}, eq. (5.65):

\ni \textit{Let $f(s)$ be a function that is regular in the strip $\s\in[\s_-,\s_+]$ where $\s_-<\s_+$, such that $|f(\s+it)| = O(e^{\eps t})$ for every $\eps>0$ everywhere in the strip. Let
\e{}{|f(\s_\pm+it)| = O\Big(|t|^{\mu_\pm}\Big)}
%
for $t\in\RR$, for some growth exponents $\mu_\pm$. Let $M(\s)$ be the linear function connecting the two boundary values $\mu(\s_\pm) = \mu_\pm$,
\e{}{M(\s) := {\mu_+\o \s_+-\s_-}(\s-\s_-)+{\mu_-\o \s_--\s_+}(\s-\s_+)\,.}
Then
\e{}{|f(\s+it)| = O\(|t|^{\mu(\s)}\)}
uniformly in $\s$, where $\mu(\s)$ is a convex function obeying $\mu(\s) \leq M(\s)$ everywhere in the strip.}

\sec{Uniform growth bound on vertical lines for convergent series}\label{appc}

In this appendix we recall a useful result about the behavior of convergent series high on vertical lines ($|t|\rar\i$ for fixed $\s$), which we quoted in \eqr{bound}. Take $\s_c$ to be the abscissa of convergence of a generalized Dirichlet series
\e{}{f(s) = \sum_n {a_n \l_n^{-s}}\,.}
Also define the partial sums,
\e{}{f_N(s) = \sum\limits_{n=1}^N a_n\l_n^{-s}}

\result{Uniform growth bound:}{For every fixed $\s>\s_c$, there exists some $C(\s)<\i$ such that
\e{boundapp}{|f(\s+it)| \leq C(\s)(1+|t|) ~~\forall\,|t|\geq 0\,.}
}We use this result in a few places in the text. Note that, in the language of the growth exponent, \eqr{boundapp} implies that $\mu_f(\s)\leq 1$ for all $\s>\s_c$. 

The proof combines two theorems from the theory of generalized Dirichlet series \c{hrtext}. The first is a ``local'' bound: conditional convergence for $\s>\s_c$ implies uniform convergence on compact subsets for $\s>\s_c$. In particular, this implies boundedness on every fixed vertical segment:
\e{}{\underset{|t|\leq T}{\text{sup}} \,|f(\s + it)| < \i~~\forall\, T<\i\,,~\s<\s_c\,.}
The second, more non-trivial, theorem is a ``global'' bound, which restricts the behavior on \textit{non-compact} domains (cf. \c{hrtext}, Theorem 12): along every vertical line within the half-plane of convergence, 
\e{}{|f(\s + it)| = o(|t|)~~\text{as}~~|t|\rar\i}
uniformly in $\s>\s_c$, and likewise for every partial sum $f_N(s)$. Equivalently, for every $\d>0$ there exists a $t_0=t_0(\d,\s)$ such that
\e{}{\left|{1\o t}f_N(\s+it)\right| < \d\,,~~\forall\,\s>\s_c\,,~~|t|\geq t_0\,.}
This tells us that, while the function can grow high on vertical lines $|t|\rar\i$, the growth must be sub-linear, uniformly in the distance to the abscissa of convergence (and in $N$). Putting these two together yields the uniform $o(|t|)$ bound on the growth rate along vertical lines (the shift $|t| \rar 1+|t|$ is to avoid vanishing at the origin). \qed

In words, for $|t|\leq T<\i$ the supremum is bounded, so there is always some large enough constant $C<\i$ for which \eqr{boundapp} is true. Then for $|t|>T$, we make the RHS linear in $|t|$ to satisfy the linear growth bound. An explicit construction is simple: take $C = \text{max}(M,1)$, where $M := \text{sup}_{|t|\leq T}|f(t)|$ for $T$ large enough that $|f(t)|\leq |t|$ for all $|t|\leq T$.

\sec{Modularity and the Riemann zeta poles of $\qz(s)$}\label{appd}

Recall that, from \eqr{qpoles}, $\qz(s)$ is allowed \`a priori to have simple poles at the locations of non-trivial zeros of $\z(2s-1)$. These poles would imply power-law behaviors in the high-temperature limit $y\rar 0$ of the scalar partition function $\cZ^{(0)}(y)$. To see this, we write the latter as
\e{}{\cZ^{(0)}(y)=\<\cZ\> + {1\o 2\pi i} \int\limits_{~(\half)} ds \,\L(s) \qz(s) y^s}
and develop the $y \rar 0$ limit by deformation into the RHP, 
\e{z0falloff}{\cZ^{(0)}(y \rar 0) \approx \<\cZ\> - \sum_n \Res\limits_{s\,=\,s_n} \qz(s)\L(s_n) y^{s_n} + \cc\,,\quad s_n := {1+\rho_n\o 2}}
plus terms non-perturbative in $y$. This expansion was independently observed in \c{Benjamin:2025kvm} in different language. We would like to understand whether there must be nonzero residues, or whether, say, an infinite subset must not vanish.

What we wish to point out, first of all, is that the behavior \eqr{z0falloff} is somewhat mysterious from the point of view of modularity. A generalization of the Cardy argument implies exponential decay of the \textit{full} $\cZ(iy)$ as $y\rar 0$: this follows from the exponential decay $\cZ(iy)\sim \sqrt{y}e^{-2\pi \l_* y}$ as $y\rar\i$, where $\l_*$ is the frequency of the lightest state (of any spin) in $\cZ(\t)$; and $S$-invariance, which acts as $y \rar y^{-1}$. Since
\es{Zexpdecay}{\sum_{j\in\Z} \cZ^{(j)}(y) = \cZ(iy) \stackrel{!}{\sim} {1\o \sqrt{y}}e^{-2\pi \l_* y^{-1}} \,,}
the falloff \eqr{z0falloff} is only compatible with \eqr{Zexpdecay} if there are intricate cancellations among spin sectors. 

However, we now prove that (with one conservative assumption) if $\qz(s)$ admits a convergent series representation, then there must be at least one Riemann zeta pole. To prove this, assume the absence of all such poles. This would imply that the series representation of $\qz(s)$ converges all the way to $\s=\half$. On the critical line, the invariance $\qz(s) = \qz(1-s)$ would then imply \textit{reciprocity} of the frequency spectrum, i.e. invariance under inversion of every element of the frequency set: $\{\mu_n\} = \{\mu_n^{-1}\}$. However, in the Virasoro CFT context, $\{\mu_n\}$ must comprise a dense set of frequencies at high-energy, reflecting the Cardy growth of the underlying $Z_p(\t)$ and $\tilde Z_p(\t)$.\foot{Here we make the assumption that in forming $\cZ(\t) = Z_p(\t) - \tilde Z_p(\t)$, there is not so much point-wise cancellation in $\l$ that the support of the frequency spectrum is \textit{bounded}, such that the spectrum $\{\mu_n\}$ has finitely many elements. That would obviously be a special case.} Reciprocity would therefore imply an accumulation point at $\l=0$, which is in flat contradiction with finiteness of a torus partition function. 

One therefore concludes that, in irrational Virasoro CFTs, $\qz(s)$ must have at least one Riemann zeta pole as long as its series representation \eqr{zlqser} converges. Apparently, the cancellations among spins required by demanding \eqr{Zexpdecay} must actually occur\thinspace---\thinspace including the $y$-dependent Riemann zeta terms. That is a tantalizing conclusion in the context of $\sl$ spectral decomposition: spinning modes are a function of Maass cusp form data, but the scalar mode is not. This cancellation therefore implies a set of mutual constraints among Riemann zeta zeros and Maass cusp form data. It would be interesting to understand this in conjunction with the Riemann-Maass connections observed in Section \ref{s11}. 

In closing, note that the $c=1$ free boson provides an explicit contrapositive to the above result: $\qz^{(c=1)}(s)$ is pole free, and from \eqr{corec1}, $\{\mu_n\} = \{r^2,r^{-2}\}$, compatible with the absence of Cardy growth. This also gives another definition of T-duality invariance, as reciprocity of the core scalar spectrum.

\sec{Proof of \eqr{oo} (order one condition) for convergent series}\label{appe}

Here we give an alternative proof of the order one condition \eqr{oo}, assuming that $\lz(s)$ admits a convergent generalized Dirichlet series representation. The proof in the main text did not require this assumption, but we find it enlightening to provide this proof all the same. 

Let us sketch the logic. We use general properties of convergent series to establish the order one property in the half-plane of convergence $\s>1$, and its image under the functional equation. Then to cover the strip in between, in which the series may have poles, we multiply it by an order one function which kills its poles\thinspace---\thinspace thus allowing Phragmen-Lindel\"of to extend the order one property across the strip, and hence everywhere. The proof relies on the convergence properties articulated in Section \ref{s5}.

Consider the $L$-quotient $\qz(s)$, defined in \eqr{qdef1}. $\qz(s)$ is analytic for $\s$ larger than the real part of the right-most non-trivial zero of $\z(2s-1)$. We assume $\qz(s)$ admits a convergent series in its right-half-plane of analyticity. Now:

\enumcom
\item Choose any fixed $\s=\b>\s_c(\qz)$. By \eqr{boundapp}, $|\qz(\b+it)|$ is uniformly bounded (sub-linearly) in $t$. 

\item Since this holds for any $\b>\s_c$, and $\qz(s)$ is analytic there, Phragmen-Lindel\"of implies that $|\qz(\b+it)|$ is uniformly bounded in any strip in the half-plane of convergence $\s>\s_c(\qz)$. Combined with the asymptotic $|\qz(\s+it)| \sim \l_1^{-\s}$ as $\s\rar\i$, we have that $\qz(s)$ obeys the order one condition in the half-plane $\s>\s_c(\qz)$.

\item By the functional equation $\qz(s)=\qz(1-s)$, the same holds in the half-plane $\s<1-\s_c(\qz)$. Therefore, $\qz(s)$ is order one everywhere outside the strip 
\e{}{S_\b := \{s:\s\in[1-\b,\b]\}}
for any $\b>\s_c(\qz)$.

\item Form 
\e{}{s(1-s)\L_\cZ(s) =4\sqrt{\pi} s(1-s)\L(s)\L(1-s)\qz(s)\,,}
which is entire and invariant under $s \rar 1-s$. In particular, $s(1-s)\L_\cZ(s)$ is analytic in the strip. The prefactor $s(1-s)\L(s)\L(1-s)$ is order one. Therefore, from {\sf (iii)}, $s(1-s)\L_\cZ(s)$ is also order one for $s \notin S_{\b}$. What remains is to extend this to the strip $s\in S_{\b}$.

\item On the boundary $\p S_\b$, the prefactor $s(1-s)\L(s)\L(1-s)$ is uniformly bounded, with exponential decay along vertical lines dominated by $\Gamma(\b+it)\Gamma(1-\b+it)\sim e^{-\pi |t|}$ at $|t|\rar\i$. Thus, $s(1-s)\L_\cZ(s)$ is the product of two functions uniformly bounded on $\p S_\b$, and is therefore itself uniformly bounded on $\p S_\b$: in particular, the prefactor exponentially decays and $\qz(s)$ is sub-linear, so $s(1-s)\L_\cZ(s)$ is (easily) uniformly polynomially bounded on $\p S_\b$. 

\item The result now follows: $s(1-s)\L_\cZ(s)$ is analytic for $s\in S_\b$ and uniformly polynomially bounded for $s\in \p S_\b$, so by Phragmen-Lindel\"of it is so bounded for all $s \in S_\b$. Therefore, $s(1-s)\L_\cZ(s)$ is order one inside the strip $s\in S_\b$, and hence, by our earlier result, everywhere. 

\ni Therefore, $s(1-s)\L_\cZ(s)$ is order one. \qed

\end{enumerate}

\sec{Proof of \eqr{AFEIK} (AFE for standard $L$-functions)}\label{appf}

Here we derive \eqr{AFEIK}, following the treatment of Iwaniec-Kowalski \c{IKtext}. Start with the integral
\e{Idefafe}{I(X,s) := {1\o 2\pi i} \int\limits_{~(\b)}{du\o u} X^u \L_L(s+u) G(u) \,.}
$\L_L(s)$ decays exponentially along vertical lines, so $I(X,s)<\i$. Shifting the contour to $\Re(u)=-\b$, then using the functional equation, then performing a change of variables $u \rar -u$ gives
\es{}{I(X,s) &= \L_L(s) + \widehat R(s) + {1\o 2\pi i} \int\limits_{~(-\b)}{du\o u}X^u \L_L(s+u) G(u)\\
&= \L_L(s) + \widehat R(s) + {1\o 2\pi i} \int\limits_{~(-\b)}{du\o u}X^u \L_L(1-s-u) G(u)\\
&= \L_L(s) + \widehat R(s) - {1\o 2\pi i} \int\limits_{~(\b)}{du\o u}X^{-u} \L_L(1-s+u) G(u) }
where $\widehat R(s)$ picks up possible poles of $\L_L(s)$ at $s=0,1$. 

Now we evaluate $I(X,s)$ in the form \eqr{Idefafe}. By absolute convergence of the series expansion of $L(s)$ for $\s>1$, we can exchange sum and integral when $\Re(s+u) >1$, i.e. when $\s > 1-\b$, which gives
\es{IVint}{I(X,s) &= q^{s/2}\sum_\D {a_\D \o \D^s} {1\o 2\pi i} \int\limits_{~(\b)}{du\o u}\({\D\o X \sqrt{q}}\)^{-u} \g(s+u) G(u) \\
&= \sum_\D {a_\D \o \D^s}Y_s\({\D\o X \sqrt{q}}\)}
We can treat $I(X^{-1},1-s)$ similarly when $\Re(1-s+u)>1$, i.e. when $\s < \b$, which gives
\e{}{I(X^{-1},1-s) = \sum_\D {a_\D \o \D^{1-s}}Y_{1-s}\({\D X \o \sqrt{q}}\)\,.}
Putting things together yields \eqr{AFEIK2}. Dividing through by $q^{s/2}\g(s)$ gives \eqr{AFEIK}, the AFE for $L(s)$. \qed

\sec{Proof of \eqr{errorlimitint} (AFE for conditionally convergent series)}\label{appg}

To prove\foot{We thank Anshul Adve for showing us how the desired result follows as a special case of a more general theorem. What follows is a more direct, more pedestrian proof.} \eqr{errorlimitint}, from which follows our approximate functional equation \eqr{AFECFT}, the starting point is \eqr{Idefafe}, on which we perform the following manipulations:
\es{}{I(X,s) &= {q^{s/2}\o 2\pi i} \int\limits_{~(\b)} {du\o u}(\sqrt{q}X)^u L(s+u) \g(s+u) G(u)\\
&= \lim_{\chi\rar\i} {q^{s/2}\o 2\pi i} \int\limits_{~(\b)} {du\o u}(\sqrt{q}X)^u \(L^{<\chi}(s+u)+\textsf{E}(s+u;\chi)\) \g(s+u) G(u) \\
&= \sum_\D {a_\D\o \D^s} Y_s\({\D\o X\sqrt{q}}\) + \lim_{\chi\rar\i} {q^{s/2}\o 2\pi i} \int\limits_{~(\b)} {du\o u}(\sqrt{q}X)^u \textsf{E}(s+u;\chi) \g(s+u) G(u) }
where
\e{}{\textsf{E}(s;\chi) := L(s)-L^{<\chi}(s)\,,\quad \text{where}\quad L^{<\chi}(s) := \sum_{\D<\chi}{a_\D\o \D^s}\,.}
By convergence,
\e{errorlimit}{\lim_{\chi\rar\i}\textsf{E}(s;\chi) = 0~~\forall\,\s>\s_c(L)\,.}
So if we can move the limit inside the integral, the AFE holds for all $\Re(s+u)>\s_c(L)$.

We can bound the error by its supremum over intervals $I\subseteq [\chi,\i)$:
\e{}{|\textsf{E}(s;\chi)| \leq M(s;\chi) := \supr{I\subseteq [\chi,\i)} L_{I}(s)\,,\quad \text{where}\quad L_{I}(s) := \sum_{\l\in I} {a_\D\o \D^s}}
For $\s>\s_c(L)$, $M(s;\chi)<\i$ by convergence. Moreover:

{\sf i)} by the uniform growth bound  \eqr{boundapp}, $M(\s+it;\chi) = o(|t|)$ uniformly in $t$ for fixed $\chi$;

{\sf ii)} by definition, $M(s;\chi)$ is monotonically non-increasing as a function of $\chi$ for fixed $s$.

\noindent Therefore, for every $\s>\s_c(L)$, there exist constants $\a_{\chi}(\s)$ and $\b_{\chi}(\s)$, maximized by $\a_{\l_1}(\s)$ and $\b_{\l_1}(\s)$, respectively, such that
\e{}{M(\s+it;\chi) \leq \a_\chi(\s)+\b_\chi(\s)|t|  \leq \a_{\l_1}(\s)+\b_{\l_1}(\s)|t| ~~\forall\,t\,.}
Putting things together, the magnitude of the error is dominated\thinspace---\thinspace in both $\chi$ and $t$\thinspace---\thinspace by the following function:
\e{}{|\textsf{E}(s;\chi)| \leq D(s) := \a_{\l_1}(\s)+\b_{\l_1}(\s)|t|~~\forall\,(\chi,t)\,,~~\s>\s_c(L)\,.}
Moreover, the dominating function is integrable for $\s+\b>\s_c(L)$,
\e{}{{q^{s/2}\o 2\pi i} \int\limits_{~(\b)}{du\o u}(\sqrt{q}X)^u D(s+u) \g(s+u) G(u)  < \i\,,}
because the rapid decay of $\g(s+u) G(u)$ along vertical lines swamps the linear growth of $D(s+u)$. Therefore, by the dominated convergence theorem and \eqr{errorlimit},
\es{}{\lim_{\chi\rar\i} \,&{q^{s/2}\o 2\pi i} \int\limits_{~(\b)}{du\o u}(\sqrt{q}X)^u \textsf{E}(s+u;\chi) \g(s+u) G(u)  \\=  &~{q^{s/2}\o 2\pi i} \int\limits_{~(\b)}{du\o u}(\sqrt{q}X)^u \lim_{\chi\rar\i}\textsf{E}(s+u;\chi) \g(s+u) G(u) \\= &~0}
as claimed. \qed

\sec{Proof of \eqr{aG} (bounded exponential decay of smoothing functions)}\label{apph}

We start with \eqr{Vdef}, which is an inverse Mellin transform. We will perform a saddle-point analysis. To warmup, set $G(u)=1$. Take $\g(s)=\g_\cZ(s)$, the degree-4 gamma factor in \eqr{gcft}, which we can write as
\e{gfactor2}{\g_\cZ(s) = 2^{9/2}\pi^{3/2}(16\pi^2)^{-s} \G(2s-1)}
so
\e{}{\g_\cZ(s)V_s(y) = 2^{9/2}\pi^{3/2}(16\pi^2)^{-s}{1\o 2\pi i}\int\limits_{~(\b)} du \,e^{\Phi(u,y,s)}\,,}
where
\e{}{\Phi(u,y,s) = -u\log(16\pi^2 y) + \log\G(2u+2s-1) - \log u\,.}
Anticipating a real saddle $u_*\gg1$, we solve $\Phi'(u,y,s)=0$ at $u\gg1$ for fixed $s$, where $' := \p_u$. The derivative is
\e{}{\Phi'(u,y,s) = -\log(16\pi^2 y) + 2{\G'\o \G}(2u+2s-1) - u^{-1}\,.}
Stirling gives the leading-order equation
\e{}{\log(16\pi^2 y) = \log (4u_*^2)+O(u_*^{-1})}
with leading-order solution
\e{}{u_* \approx 2\pi\sqrt{y}\,.}
This consistently captures the leading-order $y\gg1$ asymptotics of $\g_\cZ(s)V_s(y)$ because $u_*\gg1$ in that limit, validating our earlier approximation. Plugging back in gives the saddle point approximation
\es{Vsaddle}{\g_\cZ(s)V_s(y\gg1) &\approx {2^{9/2}\pi^{3/2}(16\pi^2)^{-s}\o 2\pi}\sqrt{2\pi\o\Phi''(u_*,y,s)}e^{\Phi(u_*,y,s)}\Bigg|_{y\gg1} \\&\approx \sqrt{2\o \pi}y^{s-1}e^{-4 \pi  \sqrt{y}} }
Comparing to the exact result \eqr{3145}, we observe the correct asymptotic (including the subleading $y$-dependence, apparently). 

Now we allow $G(u)\neq 1$. Recall that our goal is to ask whether an admissible (i.e. compatible with the assumptions stated in the AFE \eqr{AFEIK}) choice of $G(u)$ exists for which $V_s(y\gg1)$ decays even faster than \eqr{Vsaddle}, by increasing the magnitude of the exponent. 

First, note that when shifting the contour to pick up a saddle at $u_*\gg1$, we would pick up any poles of $G(u)$ to the right of the original contour $\Re(u)=\b$. These would give power-law contributions $y^{-\s_j}$ to the inverse Mellin transform\thinspace---\thinspace which decay much \textit{slower} than \eqr{Vsaddle}\thinspace---\thinspace so we need to avoid these. 

Returning now to the saddle point approximation at large $y$, for $G(u)\neq 1$ the saddle point equation picks up a new term,
\e{}{\Phi'(u,y,s) = -\log(16\pi^2 y) + 2{\G'\o \G}(2u+2s-1) - u^{-1} - {G'\o G}(u)\,.}
We now must differentiate three cases of asymptotic behavior for $G(u\gg1)$:

\enumcom

\item \textit{Sub-exponential:} ${G'\o G}(u\gg1) = o(1)$, decaying at $u\gg1$. This leaves the saddle point unchanged. 

\item \textit{Exponential:} ${G'\o G}(u\gg1) \approx 2\eta$ for some constant $\eta$. This rescales the saddle:
\e{}{u_*\approx 2\pi \sqrt{y} e^{-\eta}}
which, after accounting for the determinant factor, leads to
\e{}{\g_\cZ(s)V_s(y\gg1) \approx e^{2\eta(1-s)}\sqrt{2 \o \pi } y^{s-1} e^{-4 \pi  e^{-\eta} \sqrt{y}}\,.}

\item \textit{Super-exponential:} ${G'\o G}(u\gg1) \approx \a u^p$ for some $\a$ and $p>0$. This gives a new, logarithmic saddle
\e{}{u_*^p  \approx \a^{-1}\log(4\pi^2 y)}
and a corresponding saddle point action
\e{}{\Phi(u_*,y,s)|_{y\gg1} \propto -(\log y)^{1+{1\o p}}\,,}
which gives much slower, sub-exponential decay of $\g_\cZ(s)V_s(y\gg1)$. 
\end{enumerate}

Therefore, the only possibility of generating \textit{faster} decay than $\g_\cZ(s)V_s(y) \sim e^{-4\pi \sqrt{y}}$ would come from the exponential case $G(u\gg1) \sim e^{2\eta u}$ with $\eta<0$, decaying at infinity. However, this decay is not possible for our problem. Phragmen-Lindel\"of and Liouville's theorem together imply that a non-constant $G(u)$ cannot simultaneously be {\sf i)} even and holomorphic and bounded in the strip $|\Re(u)|<\b+1$ (definition of admissible), {\sf ii)} analytic for $|\Re(u)|>\b+1$ (to avoid the power law terms mentioned previously), and {\sf iii)} decaying at infinity. In other words, given boundedness in the original strip, we cannot have both analyticity to the right and unboundedness at infinity. We must insist on analyticity to avoid power law terms, whereupon Phragmen-Lindel\"of implies that $G(u)$ remains bounded on all vertical lines with $|\Re(u)|>\b+1$. But then by Liouville and $G(u)=G(-u)$, it must either grow at infinity or be a constant: decay is not an option.\foot{A useful non-example to keep in mind is $G(u) = {1/\cosh u}$, which decays as $e^{-u}$ as $u\rar\i$, but has poles along vertical lines.} \qed

\sec{Proof of \eqr{IJres} (linear ramp as a linear second moment)}\label{appi}
Recall the definition of the integrals $I(T)$ and $J(T)$ in \eqr{Idef} and \eqr{Jint}, respectively. We prove the following result relating their $T\rar\i$ asymptotics:
\e{IJresapp}{ {I(T) \approx {\mathsf{C}_\RMT\o 4} T \quad \Longrightarrow \quad J(T)  \approx {\mathsf{C}_\RMT\o 2} T }}

We begin by introducing some notation
\es{}{{1\o x} &:= T\\
\phi(y) &:= \arctan(y)\\
f(t) &:= \qz(t)^2 |\z(1+2it)|^2 \geq 0}
in terms of which
\e{}{I(x)  =  \int_{0}^\i dt f(t)H(x,t)\,,\quad H(x,t) := {\cosh(2t\phi(x^{-1}))\o \cosh(\pi t)}\,.}
At small $y$, $\phi(y) \approx y-{y^3\o 3} + O(y^5)$. Noting the elementary identities 
\es{}{\phi(x^{-1}) &= {\pi\o 2}-\phi(x)\\
\cosh(A-B) &= \cosh A \cosh B - \sinh A \sinh B}
we can write
\e{}{H(x,t) = \cosh(2t\phi(x)) - \tanh(\pi t)\sinh(2t\phi(x))\,.}
Define 
\es{}{\D(x,t) &:= H(x,t) - e^{-2t\phi(x)}\\
&= 2\sinh(2t\phi(x)){e^{-2\pi t}\o 1+e^{-2\pi t}}}
such that
\e{Iint}{I(x)  =  \underbrace{\int_0^\i dt f(t) e^{-2t\phi(x)}}_{:=\,M(x)} + \underbrace{\int_0^\i dt f(t) \D(x,t)}_{:=\,R(x)}\,.}
The integral $I(x)$ is the sum of a main term $M(x)$ and a remainder term $R(x)$. We are interested in the small $x$ regime, but so far, our treatment is exact.

We now make our first approximation: at small $x$, we can drop the remainder integral $R(x)$, seen as follows. Because $0<\phi(x) \leq x$ for $x\geq 0$, we have
\e{}{0 < \sinh(2t\phi(x)) \leq \sinh(2tx) \leq 2tx e^{2tx}}
and thus 
\e{}{\D(x,t) \leq 4tx e^{2tx}{e^{-2\pi t}\o 1+e^{-2\pi t}}\leq 2tx e^{-2t(\pi-x)}}
for all $t \geq 0$. With this inequality, the remainder integral $R(x)$ is damped compared to the main integral $M(x)$ in \eqr{Iint} for any $\pi - x > \phi(x)$, i.e. for $x\in[0,2.03)$ or so: in that range, and in particular in the small $x$ region we care about, $R(x) = O(x)$,
\e{}{R(x) \leq  2x \int_0^\i dt \,t e^{-2t(\pi-x)}f(t) \leq  \int_0^\i dt f(t) e^{-2tx}}
where the second inequality is the small $x$ expansion of $M(x)$.\foot{The subleading terms in the expansion $\phi(x) \approx x$ would give rise only to subleading terms in $J(T)$ at $T\rar\i$.} 

We can thus drop the remainder term in evaluating the leading-order asymptotics of $I(x \rar 0)$, and the ramp condition is now the following condition on $M(x)$:
\e{Mramp}{M(x\rar 0)  \approx \int_0^\i dt f(t) e^{-2xt}  \approx {\mathsf{C}_\RMT\o 4x}\,.}
We are now in a position to apply a Tauberian theorem to $M(x)$. Recall that $f(t)>0$. From e.g. (7.12.1) of \c{Titchmarsh1986}, for $f(t)>0$ one has 
\e{eq3411}{\int_0^\i dt\,f(t) e^{-\d t} \approx  C\d^{-1} \quad \Longrightarrow \quad \int_0^T dt\,f(t) \approx C T}
with $\d\rar 0$ on the left and $T\rar\i$ on the right. Applying this to $M(\d/2)$ via \eqr{Mramp} yields \eqr{IJresapp}. Taking $\mathsf{C}_\RMT=2$ gives the result \eqr{IJres}.\qed

\sec{Proof of \eqr{rhosj} (Virasoro $\sl$ eigendensity)}\label{appj}

In this appendix we re-derive \eqr{rhosj}, the Virasoro primary density of spin-$j$ states of the $\sl$ eigenbasis elements $E^*_s(\t)$ and $\phi_n(\t)$, first presented in \c{2307}. Their respective spin-$j$ Fourier modes were given in \eqr{Ephifourier}; as they are of identical functional form for $j>0$ up to the coefficient, we henceforth focus on $E^*_s(\t)$.

Consider an $\sl$-invariant function $f(\t)$ with Fourier decomposition
\e{}{f(\t)= \sum_{j= 0}^\i(2-\delta_{j,0})\cos(2\pi j x)  f_j(y)\,.}
To extract a formula for the Virasoro primary density by regarding $f_j(y)$ as the spin-$j$ component of a Virasoro primary partition function, we set it equal to the expression of the latter as a Laplace transform:\foot{The lower bound on the integral is taken to be 0 because when viewed as a partition function, $E_{\half+it}(\t)$ has no support on negative frequencies.}
\es{rhojvir}{f_j(y) &= \sqrt{y}\int_0^\i d\l \,\rho_j^{(f)}(\l) e^{-2\pi\l y}\\
&= \sqrt{y}\mathcal{L}\[\rho_j^{(f)}(\l);\l;2\pi y\]}
where $\l=\D-{c-1\o 12}$ as usual. To apply this to the completed Eisenstein series on the critical line, $f(\t) = E^*_{\half+it}(\t)$, we take $f_j(y)$ to be the Fourier mode in \eqr{Ephifourier} and invert. Notice that the $\sqrt{y}$ cancels very nicely\thinspace---\thinspace a feature unique to Virasoro!\thinspace---\thinspace such that the Virasoro primary basis density of spin $j$ is
\e{basisd2}{\rho^*_{\half+it,j}(\l) = \sfa_j^{(\half+it)} \mathcal{L}^{-1}[K_{it}(2\pi j y);2\pi y;\l]\,.}
The inverse Laplace transform evaluates to
\es{Klaplace}{\mathcal{L}^{-1}\[K_{it}(2\pi j y);2\pi y;\l\] = {\Theta(\l-j)\o \sqrt{\l^2-j^2}}\cos \(t y_j(\l)\)\,.}
To derive this, we use the integral representation
\e{}{K_{it}(2\pi j y) = \half\int_{-\i}^\i dz \,e^{-2\pi j y\cosh z}e^{itz} \,.}
Moving the inverse Laplace transform inside the integral,\foot{If we impose that $z\in\R$ here, then the last line picks up a $\Theta(\l-j)$. Either way, it is enforced in the next step in which we  integrate only over $z\in\R$.}
\e{}{\mathcal{L}^{-1}[e^{-2\pi j y\cosh z};2\pi y;\l] = {1\o \sqrt{\l^2-j^2}}\(\d(z-y_j(\l)) + \d(z+y_j(\l))\)}
where $\l/j = \cosh z$ has two solutions. Now performing the $z$ integral, one obtains \eqr{Klaplace}. One can check \`a posteriori that this holds for complex $t$. This gives \eqr{rhosj} as claimed, with $it \rar s-\half$. \qed

\sssec*{The $\cA$-primary density for $N>1$ is continuous}

To determine the basis densities for a CFT with an extended chiral algebra $\mathcal{A}$ with $N>1$ generating currents, the generalization of our previous derivation would equate
\es{}{E^{*}_{\half+it,j}(y) &:= y^{N\o 2}\mathcal{L}\[\rho^{*,(N)}_{\half+it,j}(\l);\l;2\pi y\]}
where the $y^{N/2}$ is the overall factor in the $\mathcal{A}$-primary partition function \eqr{ZAP}. Hence instead of \eqr{basisd2}, the $\cA$-primary density would be
\e{}{\rho^{*,(N)}_{\half+it,j}(\l) = \sfa_j^{(\half+it)} \mathcal{L}^{-1}[y^{{1-N\o2}}K_{it}(2\pi j y);2\pi y;\l]\,.}
But plugging this into the spectral decomposition to extract the ``Dirichlet part'' of the density, analogous to \eqr{rhojD}, will no longer give a delta function in the spin-$j$ primary spectrum: in particular, using the same $z$-integral method as in the $N=1$ case,
\es{}{\mathcal{L}^{-1}[y^{{1-N\o2}}e^{-2\pi j y\cosh z};2\pi y;\l] = \Theta (\l-j \cosh z)\frac{(\l-j \cosh z)^{\frac{N-1}{2}-1}}{\Gamma \left(\frac{N-1}{2}\right)}\qquad (N>1)}
For $N=1$, we had delta functions instead (indicated here by the singular behavior as $N\rar 1$). Integrating over $z$ and plugging into the spectral decomposition yields a continuous density. It is interesting that the Virasoro case is special.

\end{appendix}

\end{spacing}

\bibliographystyle{JHEP}
\bibliography{L_bib}

\end{document}